\documentclass[a4paper]{article}

\usepackage{anyfontsize} 
\usepackage{amsthm,amstext,amscd,amsbsy,amsxtra,latexsym,amsfonts,amsmath,amssymb}

\usepackage[english]{babel}
\usepackage[latin1]{inputenc}
\usepackage{verbatim} 

\usepackage{graphicx} 
\usepackage[all]{xy} 
\usepackage{caption} 
\numberwithin{figure}{section}  

\usepackage{xcolor}

\usepackage{hyperref}
 \hypersetup{
   colorlinks   = true, 
   urlcolor     = blue, 
   linkcolor    = black, 
   citecolor   = blue 
 }

\usepackage{calc} 
\newlength\dlf

\usepackage{calrsfs} 
\usepackage[mathscr]{euscript} 

\usepackage{bbm} 
\usepackage{mathabx}
\usepackage{txfonts} 
\usepackage{scalerel,stackengine}

\DeclareMathAlphabet{\mathchorus}{OT1}{cmtt}{m}{n}

\def\={\;=&\;}
\def\+{+&\:}
\def\-{-&\:}

\newcommand{\ts}{\textsuperscript}
\newcommand{\nn}{\nonumber}

\newcommand{\initial}{_A}
\newcommand{\final}{_B}

\newcommand\SL{S\hspace{-0.22em}L} 
\newcommand\Sp{S\hspace{-0.22em}p} 

\def\sheaf{\mathcal} 

\def\IM{\text{Im}\:} 

\newcommand\DD{D\hspace{-0.06em}D} 
\newcommand\RR{R\hspace{-0.1em}R} 
\newcommand\cc{\text{c\hspace{-0.02em}c}} 
\newcommand\pa{\text{p\hspace{-0.02em}a}} 
\newcommand\qp{\text{q\hspace{-0.02em}p}} 

\def\C{\mathbb{C}}
\def\CP{\mathbb{P}_{\C}}
\def\N{\mathbb{N}}
\def\Q{\mathbb{Q}}
\def\R{\mathbb{R}}
\def\Z{\mathbb{Z}}

\newcommand{\G}{\mathcal{G}}

\newcommand{\polyhedral}{_{\text{polyh}}}
\newcommand{\cylinder}{_{\G(\ell)}}
\newcommand{\MT}{_{\text{MT}}}
\newcommand{\LN}{_{\text{LN}}}
\newcommand{\flatz}{_{|dz|^2}}

\newcommand{\tr}{\text{tr}}

\newcommand{\rechts}{\:\rightarrow\:}

\newcommand{\D}{\mathcal{D}}

\def\const{\text{const.}}
\newcommand{\eps}{\varepsilon}

\newcommand{\tC}{\widetilde{C}}

\newtheorem{theorem}{Theorem}
\newtheorem{'theorem'}{''Theorem``}
\newtheorem{lemma}[theorem]{Lemma}
\newtheorem{proposition}{Proposition}
\newtheorem{cor}[proposition]{Corollary}
\newtheorem{definition}[proposition]{Definition}

\newtheorem{example}[proposition]{Example}
\theoremstyle{plain}
\newtheorem*{remark*}{Remark}
\newtheorem*{definition*}{Definition}
\newtheorem*{example*}{Example}

\newtheorem{cor'}[proposition]{Corollary'}
\newenvironment{altcor}
{%
 \addtocounter{proposition}{-1}%
 \cor}
{\endtheorem}

\stackMath
\newcommand\reallywidehat[1]{%
\savestack{\tmpbox}{\stretchto{%
  \scaleto{%
    \scalerel*[\widthof{\ensuremath{#1}}]{\kern-.6pt\bigwedge\kern-.6pt}%
    {\rule[-\textheight/2]{1ex}{\textheight}}
  }{\textheight}%
}{0.5ex}}%
\stackon[1pt]{#1}{\tmpbox}%
}
\makeatletter
\newcommand{\ostar}{\mathbin{\mathpalette\make@circled*}}
\newcommand{\make@circled}[2]{%
  \ooalign{$\m@th#1\smallbigcirc{#1}$\cr\hidewidth$\m@th#1#2$\hidewidth\cr}%
}
\newcommand{\smallbigcirc}[1]{%
  \vcenter{\hbox{\scalebox{0.77778}{$\m@th#1\bigcirc$}}}%
}

\begin{document}

\title{The $(2,5)$ minimal model on genus two surfaces}
\author{Marianne Leitner\\
School of Mathematics, 
Trinity College Dublin, the University of Dublin, \\
College Green, Dublin 2, Ireland\\
\& School of Theoretical Physics,
Dublin Institute for Advanced Studies,\\
10 Burlington Road, Dublin 4, Ireland\\
\textit{leitner@stp.dias.ie}
}

\maketitle

\begin{abstract}
In the $(2,5)$ minimal model, the partition function for genus $g=2$ Riemann surfaces 
is given by a $5$-tuple of functions with appropriate transformation under the mapping class group.
These functions generalise the two Rogers-Ramanujan functions for the torus.
Their expansions around a locus of surfaces with conical singularities
in the interior of the $g=2$ Siegel upper half plane are obtained
in terms of standard modular forms.
The dependence on the metric is controlled by a canonical choice of flat surface metrics.
In the alternative case where a handle of the $g=2$ surface is pinched, 
our method requires knowledge of the two-point function of the fundamental lowest-weight vector in the non-vacuum representation of the Virasoro algebra,
for which we derive a $3$\ts{rd} order ODE.
In order to make the paper more accessible to mathematicians, 
the exposition includes a short introduction to conformal field theory on Riemann surfaces,
which may be of independent interest.\\
\\
MSC 11F03 (Primary),\,57M50,\,81T40 (Secondary)
\end{abstract} 

\maketitle

\tableofcontents

\section{Motivation}\label{sec:1}

Two-dimensional conformal field theories (CFTs) are naturally defined on pairs $(\Sigma,\G)$,
where $\Sigma$ is a compact Riemann surface (of genus $g\geq 0$) and $\G$ is a metric in the conformal class of $\Sigma$.
Denote by $\mathscr{X}$ the space of pairs $M=(\Sigma,\G)$. 
A two-dimensional CFT is characterised by its partition function \cite{Friedan-Shenker:1986},
which is a smooth map 
\begin{equation*}
\mathfrak{Z}:\quad\mathscr{X}\quad\rechts\quad\R
\:.
\end{equation*}
(This paper shows that the range of $\mathfrak{Z}$ can in general not be restricted to $\R^+$.)
The image of $M=(\Sigma,\G)\in\mathscr{X}$ will be written $\mathfrak{Z}_{M}$ or $\mathfrak{Z}_{\G}^g$.

Quantum field theories are supposed to yield many of the simplest smooth maps of this kind.
What has been most studied is the restriction of $\mathfrak{Z}$ to flat tori,
where $\mathfrak{Z}^{g=1}\flatz$ is modular on the full modular group.
In the $(2,5)$ minimal model, also called Yang-Lee model, 
one has
\begin{equation}\label{eq: g=1 partition function w.r.t. the flat metric}
\mathfrak{Z}_{\C/(\Z+\tau\Z)}
\;=\;|q^{-1/60}G(q)|^2+|q^{11/60}H(q)|^2
\:,
\end{equation}
where 
\begin{equation}\label{def: Rogers-Ramanujan functions}
G(q)\;=\;\sum_{n\geq 0}\frac{q^{n^2}}{(q;q)_n}
\:,
\quad
H(q)\;=\;\sum_{n\geq 0}\frac{q^{n^2+n}}{(q;q)_n}
\:
\end{equation}
is the pair of Rogers-Ramanujan functions.
Here $q=\exp(2\pi i)$ and \hbox{$(\:;q)_n$} is the~$q$-Pochhammer symbol.
The so-called holomorphic blocks, the functions $q^{-1/60}G$ and $q^{11/60}H$, span a unitary representation of the mapping class group.
Note that fivefold Dehn twists act by multiplication with a common twelfth root of unity.
This fact should generalise to higher genus.
The aim of the present paper is a calculation of the genus two partition function of this model 
and its holomorphic blocks. 
In particular, this requires the choice of suitable metrics.

Here we follow the approach by Graeme Segal \cite{Segal:1988} 
who correctly believed that it ``is equivalent to that used by physicists''.
The author learned the method from Werner Nahm, who had learned it from him.
Segal's concepts were further explored by physicists \cite{Vafa:1987,Sonoda:1988,Moore-Seiber:1989}.

This paper provides a calculation of $\mathfrak{Z}$ for genus two surfaces
but no proof that it yields the required map on the full moduli space.
Thus it does not aim at proving the existence of the corresponding CFT
but provides the data for comparison with known functions, in particular Siegel modular forms.

\section{Short introduction to CFT}\label{subsec:1.1}

\subsection{The partition function}

Let $M=(\Sigma,\G)\in\mathscr{X}$.
Every point $P\in\Sigma$ has a neighbourhood $U\subset\Sigma$ 
together with a local coordinate $z$ on $U$ 
such that the metric on $U$ takes the form $\G=2\G_{z\bar{z}}\,dzd\bar{z}$, 
where in our conventions, $2\G_{z\bar{z}}=e^{\sigma(z)}$ with $\sigma\in C^0(U,\R)$.
We require that the Gauss curvature two form,
\hbox{$\mathcal{K}=-i\,\partial_z\partial_{\bar{z}}\log\G_{z\bar{z}}\,dz\wedge d\bar{z}$},
defines an $\R$-valued linear functional on $C^0(\Sigma,\R)$.

By definition, $\mathfrak{Z}_M$ is invariant under diffeomorphisms of $M$.
For disjoint unions,
\begin{equation*}
\mathfrak{Z}_{M_a\dot{\cup} M_b} 
\;=\;\mathfrak{Z}_{M_a}\,\mathfrak{Z}_{M_b} 
\:.
\end{equation*}
For a CFT of central charge $c\in\R$, 
it is  postulated that \cite{P:1981}
\begin{equation}\label{eq: trace anomaly, formulated without fields}
{\G}_{\mu\nu}\,\frac{\delta\mathfrak{Z}}{\delta\,\G_{\mu\nu}(x)}
\;=\;\frac{c}{24\pi}\:K(x)\,\mathfrak{Z}
\:. 
\end{equation}
Here $K$ is the Gauss curvature of the Levi-Civit\` a connection,
and the non-vanishing is a feature of the conformal anomaly.

As observed by W.~Nahm \cite{Nahm:2020}, 
eq.~(\ref{eq: trace anomaly, formulated without fields}) yields a simple relationship between the values of the partition function 
when $M_A=(\Sigma,\G_A)$ and $M_B=(\Sigma,\G_B)$
project to the same point in the moduli space of conformal structures.
Under a finite Weyl transformation $\G\initial\mapsto \G\final=\varrho_{BA}\,\G\initial$,
the partition functions changes according to
\begin{equation}\label{eq: Nahm's formula}
\log\bigl(\mathfrak{Z}_{M\final}/\mathfrak{Z}_{M\initial}\bigr)
\;=\;\frac{c}{48\pi}\iint_{\Sigma_g}\Delta\sigma\,\bigl(\mathcal{K}\initial+\mathcal{K}\final\bigr)
\:,
\end{equation}
where $\Delta\sigma=\log\varrho_{BA}$ and $\mathcal{K}$ with index $A,B$ is the curvature two-form 
of the corresponding metric. 
In particular,
for $\lambda\in\R^+$,
\begin{equation}\label{eq: behaviour of partition fct under rescaling of metric for arbitrary g}
\mathfrak{Z}_{\lambda\G}^g
\;=\;\lambda^{\frac{c}{6}(1-g)}\mathfrak{Z}_{\G}^g
\:.
\end{equation}

\begin{example}
We consider $\CP^1$ with the metric
\begin{equation*}
\G(\ell)
=
\begin{cases}
\,e^\ell\,|dz|^2&\quad 0\leq |z|\leq e^{-\ell/2}\:,\\
\hspace{0.5cm}\bigl|\frac{dz}{z}\bigr|^2&\quad e^{-\ell/2}\leq |z|\leq e^{\ell/2}\:,\\
\,e^{\ell}\frac{|dz|^2}{|z|^4}&\quad e^{\ell/2}\leq |z|\:,
\end{cases}
\end{equation*}
for $\ell\geq 0$.
$\G(\ell)$ is continous and its curvature is supported on the two circles \hbox{$\{|z|=e^{\pm\ell/2}\}$}.
By eq.~(\ref{eq: Nahm's formula}),
\begin{equation}\label{eq: cylinder partition function}
\mathfrak{Z}^{g=0}\cylinder
\;=\;e^{\,c\ell/12}\,\mathfrak{Z}^{g=0}_{\G(0)}
\:.
\end{equation} 
The individual contributions to formula~(\ref{eq: Nahm's formula}) are listed in table \ref{table: The contributions to Nahm's formula for g=0,  from the cylinder to the double disc}. 
\begin{table}[htb]
\centering
\begin{tabular}{l|lll}
&$|z|=e^{-\ell/2}$&$|z|=1$&$|z|=e^{\ell/2}$\\ 
\hline
\hspace{0.3cm}$\Delta\sigma$&$-\ell$&\hspace{0.1cm}$0$&$-\ell$\\
$\iint\mathcal{K}\initial$&\hspace{0.2cm}$2\pi$&\hspace{0.1cm}$0$&\hspace{0.2cm}$2\pi$\\
$\iint\mathcal{K}\final$&\hspace{0.3cm}$0$&\hspace{0.0cm}$4\pi$&\hspace{0.3cm}$0$\\
\noalign{\smallskip}\hline
\end{tabular}\\
 \caption{
The curvature of the metric $\G_A=\G(\ell)$ for $\ell>0$
is evenly spread over two circles. 
The metric $\G_B=\G(0)$ has all curvature evenly distributed over the single circle \hbox{$|z|=1$}. 
}
\label{table: The contributions to Nahm's formula for g=0,  from the cylinder to the double disc}
\end{table}
\end{example}

We will only have to study the map on finite dimensional subspaces of $\mathscr{X}$.
The moduli spaces relevant here are all complex manifolds.

\subsection{Sewing Riemann surfaces}\label{subsubsection: Sewing Riemann surfaces}

The values of $\mathfrak{Z}$ on higher genus Riemann surfaces can be constructed from data on lower genus Riemann surfaces 
by sewing along circles.
Let $M=(\Sigma,\G)$ be a connected Riemann surface with metric $\G$.
Let $(U,z)$ be a holomorphic coordinate chart, where $U\subset\Sigma$ is mapped
to a neighbourhood of the unit circle $S^1$ in $\C$.
We assign to $\C$ its standard metric and assume that the restriction of $z$ to modulus one yields an isometry
between a curve $\gamma\subset U$ and $S^1$.
Cut along $\gamma$ and let $M^{0}$ be the resulting metric manifold 
with two boundary curves with oriented isometries to $S^1$ and marked points corresponding to $1\in S^1$.
For any manifold $M$ with such a left boundary, 
we denote by $M^{\theta}$ the manifold obtained by rotating the marked point on that boundary by the angle $\theta$.

When $\gamma$ is a separating curve, 
we have  
\begin{equation*}
M^{0}
\;=\;M_L\dot{\cup} M_R
\:,
\end{equation*}
where $M_L$ and $M_R$ define manifolds with $S^1$ boundaries to the left and to the right, respectively, 
with the corresponding marked point. 
In the following, such surfaces $M_L$ and $M_R$ with parametrised boundary $\gamma$ 
will be referred to as left and right surfaces (for $\gamma$), respectively.

Conversely, given such a pair, we can construct $M$ as the connected sum $M(1)$,
where
\begin{equation*}
M(e^{i\theta})
\;=\;
M_L\sqcup_{\gamma}M_R^{\theta}
\:.
\end{equation*}
This is called ${s}$ formalism (where ${s}$ stands for ``separating'').
A special case is where $D_L$ and $D_R$ are that flat unit discs and the double disc 
\begin{equation*}
\DD
\;=\;D_L\sqcup_{\gamma}D_R 
\end{equation*}
has genus $g=0$.

When $\gamma\subset\Sigma$ is a non-separating curve (${q}$ formalism),
then $M^{\theta}$ is connected.
We write
\begin{equation*}
M(e^{i\theta})
\;=\;\sqcup_{\gamma}M^{\theta} 
\:
\end{equation*}
for the self-connected sum of $M^{\theta}$ along $\gamma$.
Clearly the existence of a non-separating curve requires $\Sigma$ to have genus $g\geq 1$.

Our ${s}$ and ${q}$ formalisms were dubbed $\eps$ and $\varrho$ formalisms in~\cite{M-T:2006}.

\subsection{The vacuum module}

Let \hbox{$\mathcal{V}=\oplus_{h\in\N_0}\mathcal{V}^h$} be the Verma module of the Virasoro algebra 
\begin{equation}\label{eq: Virasoro algebra}
[L_n,L_m]
\;=\;(m-n)\,L_{m+n}+\frac{c}{12}\,m(m^2-1)\,\delta_{m+n,0}\:,
\end{equation}
with central charge $c\in\R$, lowest weight vector $v$, modulo the relations 
\begin{equation*}
 L_0v
 \;=\;0\:,
 \quad
 L_1v
 \;=\;0\:.
 \end{equation*}
in addition to $L_nv=0$ for $n<0$.
(Note that our~$L_n$ is~$L_{-n}$ for many others; the convention here is chosen so that \hbox{$L_1=d/dz$}).
The degree $h$ is the eigenvalue of $L_0$.
We consider $\mathcal{V}^h$ as vector spaces over \hbox{$\C\cong\mathcal{V}^0$}, 
where the number $1$ is identified with $v$.
Thus
\begin{equation*}
\begin{split}
\sum_{h=0}^{\infty}\dim(\mathcal{V}^h)\,q^h
\;=\;&\prod_{n=2}^{\infty}(1-q^n)^{-1}\\
\;=\;&1+q^2+q^3+2 q^4+2 q^5+4 q^6+4 q^7+7 q^8+8 q^9+\text{O($q^{10}$)}
\:.
\end{split}
\end{equation*}
Let $\overline{\mathcal{V}}$ be the corresponding complex conjugate space.
This means that there exists an antilinear involution denoted by an overbar that exchanges $\mathcal{V}$ and $\overline{\mathcal{V}}$.
$\bar{L}_n$ is defined accordingly,
so that $\overline{\mathcal{V}}$ has a grading defined by $\bar{L}_0$
(with eigenvalues $\bar{h}$). 
Let $\bar{v}\in\overline{\mathcal{V}^0}$ be the lowest weight vector. 
The overbar operation extends to an antilinear involution on $\mathcal{V}\otimes\overline{\mathcal{V}}$.
Let $R_2(\mathcal{V})\subset\mathcal{V}\otimes\overline{\mathcal{V}}$ be the corresponding real subspace.
We call $R_2(\mathcal{V})$ the vacuum sector and its element $v\otimes\bar{v}$ the vacuum vector.

The Shapovalov form \cite{Sh:1972}
\begin{equation}\label{map: Shapovalov form}
\langle .|.\rangle:\quad\mathcal{V}\times\mathcal{V}\rechts\R 
\end{equation}
is the sesquilinear form defined by \hbox{$L_n^*=L_{-n}$} (where $L_n^*$ denotes the Hermitian adjoint of $L_n$) and 
\hbox{$\langle v|v\rangle=\parallel v\parallel^2=1$}. 
A null vector is one for which the linear functional defined by the Shapovalov form vanishes.

For $c=-22/5$, 
the generic vacuum module $\mathcal{V}$ has a maximal invariant submodule generated by 
$(L_2L_2-\frac{3}{5}L_4)\,v$.
This is a null vector,
so we can replace $(\mathcal{V},\langle .|.\rangle)$ by the corresponding quotient space $\mathfrak{V}$ with the induced Shapovalov form.
We have
\begin{equation*}
\sum_{h=0}^{\infty}\dim(\mathfrak{V}^h)\,q^h
\;=\;H(q)
\:,
\end{equation*}
cf.\ the second of eqs~(\ref{def: Rogers-Ramanujan functions}).
The resulting theory is the $(2,5)$ minimal model (with vacuum sector $ R_2(\mathfrak{V})$) mentioned above.
It is the only model of interest here.

\subsection{The partition function on sewn Riemann surfaces}\label{Intro: The partition function on sewn Riemann surfaces}

We explore ways to compute the partition function for a pair $(\Sigma,\G)\in\mathscr{X}$ using either sewing scheme,
and for generic central charge $c\in\R$.
Let $\mathcal{M}_R$ (resp.~$\mathcal{M}_L$) be the space 
of all finite formal $\R$-linear combinations of right (resp.~left) surfaces of genus zero.
For the double disc given by the pair $D_L,D_R$, 
put
\begin{equation}\label{eq: normalisation of partition function of lense space}
\mathfrak{Z}_{\DD}
\;=\;1
\:.
\end{equation}
Eqs~(\ref{eq: normalisation of partition function of lense space}) and (\ref{eq: Nahm's formula})
determine $\mathfrak{Z}_{M}$ for any $M\in\mathscr{X}$ of genus $g=0$. 

The following proposition is well motivated from physics and probably implied by the work of Graeme Segal 
though the author is not aware of a detailed proof.

\begin{proposition}\label{proposition: map into the completion of the vacuum sector}
There exists a space $\widetilde{R_2(\mathcal{V})}$ 
together with a pair of natural maps 
\begin{align*}
|.\rangle:\quad \mathcal{M}_R&\rightarrow\widetilde{R_2(\mathcal{V})}\:,\\
\langle.|:\quad \mathcal{M}_L&\rightarrow\widetilde{R_2(\mathcal{V})}\:,
\end{align*}
which are related by reversal of orientation and have the following properties:
\begin{enumerate}
 \item 
 $R_2(\mathcal{V})$ is dense in $\widetilde{R_2(\mathcal{V})}$.
 \item
 The Shapovalov form (\ref{map: Shapovalov form}) on $\mathcal{V}$ 
 induces a bilinear form $\langle .|.\rangle$ on $\widetilde{R_2(\mathcal{V})}$.
 \item
  $|D_R\rangle
=v\otimes\bar{v}\in R_2(\mathcal{V})$.
\item
For $(M_L,M_R)\in\mathcal{M}_L\times\mathcal{M}_R$,
\begin{equation}\label{def: bra and ket}
\langle M_L|M_R\rangle
\;=\;
\mathfrak{Z}_{M_L\sqcup_{\gamma}M_R}
\:.
\end{equation}
(In the product, angle brackets and vertical bars of inserted bra- and ket vectors will be omitted
 and we simply write $\langle M_L|.\rangle$ and $\langle.|M_R\rangle$.)
 \item
  Let the cylinder $\gamma\times[0,\ell]$ have the natural (flat) metric.
 Then
 \begin{equation}\label{eq: ket of twisted cylinder sewn with MR}
 |(\gamma\times[0,\ell])^{\theta}\sqcup_{\gamma\times\{\ell\}} M_R\rangle
 \;=\;e^{-\ell(L_0+\bar{L}_0-c/12)}e^{i\theta(L_0-\bar{L}_0)}| M_R\rangle
 \:.
 \end{equation}
 \end{enumerate}
\end{proposition}

Note that eq.~(\ref{eq: cylinder partition function}) is a special case of eq.~(\ref{eq: ket of twisted cylinder sewn with MR}).

A proof should follow from a Taylor expansion of $\mathfrak{Z}$ about ${\DD}$,
as explained in Section \ref{subsubsection: The Virasoro field, and more general fields}.

\begin{cor}
The Shapovalov form is positive on pairs $(M_L,M_R)\in\mathcal{M}_L\times\mathcal{M}_R$. 
\end{cor}

\begin{proof}
By eq.~(\ref{eq: normalisation of partition function of lense space}), 
$\langle D_L|D_R\rangle=1$.
Positivity follows by conformal symmetry from eq.~(\ref{def: bra and ket}).
\end{proof}

Let $c=-22/5$. 
The family of projections 
$\mathfrak{V}\otimes\overline{\mathfrak{V}}\rechts{\mathfrak{V}}^h\otimes{\overline{\mathfrak{V}}}^{\,\bar{h}}$ 
for $(h,\bar{h})\in\N_0\times\N_0$ gives rise to a set 
\begin{equation}\label{def: standard orthogonal basis for the vacuum sector in the (2,5) minimal model}
\{e_i\}_{i}
\;=\;\dot{\cup}_{h,\bar{h}}\,\bigl\{\,\text{orthogonal basis in $\mathfrak{V}^h\otimes\overline{\mathfrak{V}}^{\,\bar{h}}$}\bigr\}
\end{equation}
satisfying the following properties:
\begin{enumerate}
\item
For $i,j\geq 0$,
\begin{equation}\label{def: standard orthogonal}
\langle e_i|e_j\rangle 
\;=\;\eps_i\,\delta_{ij}
\:,\quad
\eps_i\in\{1,-1\}
\:.
\end{equation}
\item 
$\eps_0=1$ and
\begin{equation*}
e_0
\;=\;\mathfrak{v}\otimes\bar{\mathfrak{v}}
\:,
\end{equation*}
where $\mathfrak{v}\otimes\bar{\mathfrak{v}}$ is the vacuum vector in $\mathfrak{V}^0\otimes\overline{\mathfrak{V}}^0$.
\end{enumerate}
A set of vectors satisfying property (\ref{def: standard orthogonal}) will be said to be standard orthogonal.

The set of $\{e_i\}_i$ is useful for computing the partition function (\ref{def: bra and ket}) 
for the $(2,5)$ minimal model,
due to the identity
\begin{equation}\label{eq: Segal's naive s formula}
\mathfrak{Z}_{M_L\sqcup_{\gamma}M_R}
\;=\;\sum_{i}\langle M_L|\,e_i\rangle\,\eps_i\,\langle e_i|\,M_R\rangle
\:.
\end{equation}
There is a variant for the self-connected sum of an oriented metric manifold $M^{\theta}$ with boundaries $\gamma_1,\gamma_2$,
which are isometrically isomorphic to $S^1$. For typographical reasons,
will write $\gamma$ instead of $S^1$.
Denote by $M_L^{(1)}$ and $M_R^{(2)}$ two bounded manifolds 
which are a left and right surface for $\gamma_1$ and $\gamma_2$, respectively.
Thus the connected sum $M^{\theta}\sqcup_{\gamma_2}M_R^{(2)}$ yields a manifold $M_R^{(1)}$ with boundary $\gamma_1$.
We set
\begin{equation*}
\langle M_L^{(1)}|\,M^{\theta}|\,M_R^{(2)}\rangle
\;:=\;\langle M_L^{(1)}|\,M^{\theta}\sqcup_{\gamma_2}M_R^{(2)}\rangle
\end{equation*}
To simplify notations, the upper indices of $M_L^{(1)}$ and $M_R^{(2)}$ will be dropped. 
If $M=M_L\dot\cup M_R$, the corresponding vector is $|M_L\rangle\,\langle M_R|$
and the partition function of the self-connected sum $\sqcup_{\gamma}M=M_L\sqcup_{\gamma}M_R$ is given by 
\begin{equation}\label{eq: Segal's naive q formula}
\mathfrak{Z}_{\sqcup_{\gamma}M}
\;=\;\sum_{i}
\eps_i\,\langle e_i|\,M|\,e_i\rangle
\:.
\end{equation}
We define CFTs by the assumption that eqs~(\ref{eq: ket of twisted cylinder sewn with MR}), (\ref{eq: Segal's naive s formula}) and (\ref{eq: Segal's naive q formula}) 
extend to surfaces of higher genus but with an enlarged range of summation.
The new $e_i$ are orthogonal to $\mathcal{M}_R$ so that they do not contribute for genus zero.
They characterise the corresponding CFT.
In general, the pairs $(h_i,\bar{h}_i)$ will not be integral, 
but one needs $(h_i-\bar{h}_i)\in\Z$ since $\mathfrak{Z}$ is invariant under a $2\pi$ rotation along $\gamma$.
Moreover, the values of $(h_i+\bar{h}_i)$ have to be bounded from below.

\subsection{Example: The (2,5) minimal model in genus one}\label{subsub:minimal model in genus one}

Let $M=\gamma\times[0,\ell]$ be the flat cylinder with boundary $\gamma\times\{0\}\cup\gamma\times\{\ell\}$.
By eqs (\ref{eq: Segal's naive q formula}) and (\ref{eq: ket of twisted cylinder sewn with MR}),
\begin{equation*}
\mathfrak{Z}_{\sqcup_{\gamma}M^{\theta}}
\;=\;\sum_{e_i}\eps_i\,
\langle e_i|\,(\gamma\times[0,\ell])^{\theta}\,|e_i\rangle
\;=\;\sum_{e_i}\eps_i\,\langle e_i|\,q^{\,L_0-c/24}\bar{q}^{\,\bar{L}_0-c/24}|e_i\rangle
 \:,
\end{equation*}
where $q=\exp(i\theta-\ell)$.
For $c=-22/5$ and for $\tau=(\theta+i\ell)/2\pi$,
the contribution of the vacuum sector is
\begin{equation*}
\;\sum_{e_i\in\mathfrak{V}\otimes\overline{\mathfrak{V}}}\eps_i\,\langle e_i|\,q^{\,L_0-c/24}\bar{q}^{\,\bar{L}_0-c/24}|e_i\rangle
\;=\; e^{\,c\ell/12}\sum_{(h_n,\bar{h}_n)}
q^{\,h_n}\bar{q}^{\,\bar{h}_n}
\;=\;|q^{11/60}H(q)|^2
\:,
\end{equation*}
where $(h_n,\bar{h}_n)$ runs over the weights occurring in the vacuum sector.
This function cannot be a complete partition function,
since it is not invariant under the modular group.
To get the partition function from eq.~(\ref{eq: g=1 partition function w.r.t. the flat metric}), 
the representation of the Virasoro algebra has to be extended to include one further sector. 

Let \hbox{$\mathcal{W}=\oplus_{h\in\N_0-1/5}\,\mathcal{W}^h$} be the Verma module of the Virasoro algebra (\ref{eq: Virasoro algebra}) 
with central charge $c=-22/5$, and lowest weight vector $w$ of holomorphic conformal weight ($L_0$ eigenvalue) equal to $h=-1/5$.
Let $\overline{\mathcal{W}}$ be the corresponding complex conjugate space. 
$\bar{L}_n$ is defined accordingly, so that $\overline{\mathfrak{W}}$ has a grading defined by $\bar{L}_0$ (with eigenvalues $\bar{h}$).
Let $\overline{w}\in\overline{\mathcal{W}^{-1/5}}$ be the lowest weight vector.
The Shapovalov form on $\mathcal{W}$ is non-degenerate only on the quotient $\mathfrak{W}$ of $\mathcal{W}$ 
by the invariant subspace generated by the nullvectors $(L_1L_1-\frac{2}{5}L_2\bigr)\,w$ and $(L_2L_1-\frac{1}{5}L_3\bigr)\,w$.
Let $R_2({\mathfrak{W}})\subset{\mathfrak{W}}\otimes\overline{\mathfrak{W}}$ be the real subspace 
w.r.t.\ the overbar operation on ${\mathfrak{W}}\otimes\overline{\mathfrak{W}}$.
We take 
\begin{equation}\label{expression: full representation of Virasoro algebra in the (2,5) minimal model}
 R_2(\mathfrak{V})\oplus R_2(\mathfrak{W})
\:
\end{equation}
as the space of (real) fields in the $(2,5)$ minimal model. 
Complex fields are obtained by extending scalars.
The fact that there is no pairing of fields between different sectors
follows from $(h-\bar{h})\in\Z$.
Since
\begin{equation*}
\sum_{h=0}^{\infty}\dim(\mathfrak{W}^h)\,q^h
\;=\;G(q)
\:,
\end{equation*}
cf.\ the first of eqs~(\ref{def: Rogers-Ramanujan functions}),
we recover the full partition function from eq.~(\ref{eq: g=1 partition function w.r.t. the flat metric}).

We call $R_2(\mathfrak{W})$ the non-vacuum sector of the $(2,5)$ minimal model.
Since there are no other representations of the Virasoro algebra that share the properties of $R_2(\mathfrak{V})$ and $R_2(\mathfrak{W})$,
one expects that (\ref{expression: full representation of Virasoro algebra in the (2,5) minimal model}) will do for any genus.
This assumption characterises the model.

\subsection{The Virasoro field, and more general fields}\label{subsubsection: The Virasoro field, and more general fields} 

Let $M$ be a manifold with Riemannian metric, and let $U\subseteq M$ be an open set with coordinate $x$.
Based on Einstein's work, Hilbert in classical field theory \cite[p.~404]{H:1915}
and Weinberg in quantum field theory~\cite[p.~360]{Wein:1972}
identified the energy-stress tensor $T^{\mu\nu}$ in the chart $(U,x)$ 
with the functional derivative w.r.t.\ the metric tensor $\G_{\mu\nu}(x)$,
\begin{equation*}
T^{\mu\nu}(x)
\;=\;2\,\frac{\delta}{\delta\,\G_{\mu\nu}(x)}
\:.
\end{equation*}
Invariance of the partition function under change of coordinates implies
\begin{equation}\label{eq: invariance under change of coordinates}
\nabla_{\mu}T^{\mu\nu}
\;=\;0
\:, 
\end{equation}
where $\nabla_{\mu}$ is the Levi-Civit\` a derivative. 
Since for Weyl transformations, $\delta\G_{\mu\nu}\,\propto\,\G_{\mu\nu}$, 
these transformations are described by the trace of $T$.
Eq.~(\ref{eq: trace anomaly, formulated without fields}) now reads 
\begin{equation*}
{T_{\mu}}^{\mu}
\;=\;{\G}^{\mu\nu}T_{\mu\nu}
\;=\;\frac{c}{12\pi}\:K
\:. 
\end{equation*}
On a Riemann surface $\Sigma$ with local complex coordinate $z=x^1+ix^2$ and metric $\G(z)=2\G_{z\bar{z}}\,dzd\bar{z}$,
volume preserving (i.e.\ tracefree) variations of the metric decompose into $\delta\,\G_{zz}$ and $\delta\,\G_{\bar{z}\bar{z}}$.
For locally flat metrics, the corresponding functional derivatives yield components of the complexified energy-momentum tensor,
which are written $T$ and $\overline{T}$, respectively.
For general metrics, one defines $T$, $\overline{T}$ so that they commute with Weyl transformations \cite{EO:1987}.

Under linear fractional coordinate change $z\mapsto u$, $T(z)$ transforms according to
\begin{equation}\label{eq: transformation behaviour of T undr linear fractional transformation}
T(u)
\;=\;\biggl(\frac{dz}{du}\biggr)^{h(T)}T(z) 
\:,
\end{equation}
where $h(T)=2$ is called the (holomorphic) conformal weight of $T$.
Analogous equations hold for $\overline{T}$ in terms of the anti-holomorphic coordinate $\bar{z}$ 
and weight $\bar{h}(\overline{T})=2$.
Conformal invariance and eq.~(\ref{eq: invariance under change of coordinates}) 
imply that \hbox{$\partial_{\bar{z}}T=0$} 
(and \hbox{$\partial_{z}\overline{T}=0$}).
One calls $T$ a holomorphic field, which is reflected by the identity $\bar{h}(T)=0$.
More specifically, $T$ is called the Virasoro field.
By holomorphicity of $T$ (and antiholomorphicity of $\overline{T}$),
\begin{equation*}
T(z_k)\ldots T(z_1)\,
\overline{T}(\bar{z}_{k+\ell})\ldots\overline{T}(\bar{z}_{k+1})\,
\mathfrak{Z}_{M_L\sqcup_{\gamma}D_R} 
\end{equation*}
is a holomorpic function in $z_1,\ldots,z_k\in D_R$ (and antiholomorphic in $\bar{z}_{k+1},\ldots,\bar{z}_{k+\ell}\in D_R$)
for any $M_L$ as above, provided $|z_i|<|z_{i+1}|$ (and $|\bar{z}_{k+i}|<|\bar{z}_{k+i+1}|$).  
Its Laurent series coefficients are denoted by
\begin{equation*}
\langle M_L|(L_{n_k}\ldots L_{n_1}v)\otimes(\bar{L}_{n_{k+\ell}}\ldots\bar{L}_{n_{k+1}}\bar{v})\rangle 
\end{equation*}
where $v\otimes\bar{v}$ is the vacuum vector.
In this sense, 
\begin{equation}\label{eq: Virasoro field}
T(z)
\;=\;\sum_{n\in\Z}z^{n-2}L_n 
\:,
\end{equation}
(and the analogous equation for $\overline{T}$) so that $T$ corresponds to $L_2v\in\mathcal{V}$.
By the Virasoro algebra (\ref{eq: Virasoro algebra}),
\begin{equation*}
L_{-2}L_2v
\;=\; \frac{c}{2}\,v\:,\quad
L_{-1}L_2v
\;=\; L_3v\:,\quad
L_0L_2v
\;=\;2\,L_2v\:.
\end{equation*}
For $z$ close to $z_0=0$,
this yields the Virasoro operator product expansion
\begin{equation}\label{OPE of T and T}
T(z)\,T(0)
\;=\;
\frac{c/2}{z^4}\,.1
+\frac{2}{z^2}\,T(0)
+\frac{1}{z}\,T'(0)
+N_0(T,T)(0)
+O(z)\:.
\end{equation}
The field $1$ on the r.h.s.\ of eq.~(\ref{OPE of T and T}) represents the identity mapping,
$T'=\partial_zT$ is the derivative field 
and $N_0(T,T)$ is the so-called normal ordered product  of~$T$, which corresponds to $L_2L_2v$.

More generally, fields are commuting operators acting on smooth functions \hbox{$\mathscr{X}\rechts\R$}.
Given $M\in\mathscr{X}$ with coordinate chart $(U,z)$, each $e_i$ defines a field $\psi_i$ on $U$ as follows:
If $M=M_L\sqcup_{\gamma}M_R$ and if $z$ defines an isometry between $M_R$ and $D_R(z_0)$,
where $D_R(z_0)$ denotes the flat unit disc centred at $z_0$,
then
\begin{equation}\label{eq: Shapovalov form of ML and ei expressed through 1-point function}
\psi_i(z_0)\,\mathfrak{Z}_M
\;=\;\langle M_L|e_i\rangle
\end{equation}
More generally, we let $\psi_i$ commute with Weyl transformations.
$N$-point (or correlation) functions $\phi_1(z_1)\ldots\phi_N(z_N)\,\mathfrak{Z}_M$
are defined analogously by simultaneous insertion of the corresponding vectors at $z_1,\ldots,z_N$.
This makes sense as long as the $z_i$ are pairwise different.
In physics, the usual notation is
\begin{equation*}
\langle\phi_1(z_1)\ldots\phi_N(z_N)\rangle_{M}
\;=\;\phi_1(z_1)\ldots\phi_N(z_N)\,\mathfrak{Z}_{M}
\:.
\end{equation*}
The quotient $\langle\phi_1(z_1)\ldots\phi_N(z_N)\rangle_M/\mathfrak{Z}_M$ is independent of the metric on $M$
in a conformal class.
Note that the insertion of a field requires a coordinate.
By abuse of notation,
one uses the arguments $z_i$ to refer both to the coordinate function and its values.


Given a holomorphic field $\phi$ and any field $\psi$,
the Laurent expansion of the product operator $\phi(z)\,\psi(0)$ for $z$ close to $z_0$, 
\begin{equation}\label{eq: Nk normal ordered products of general fields}
\phi(z)\,\psi(z_0)
\;=\;\sum_{k\in\Z}(z-z_0)^k\,N_k(\phi,\psi)(z_0)
\:,
\end{equation}
defines a sequence of fields $N_k(\phi,\psi)$. 
This is a special case of the operator product expansion, which will be introduced later.
Eq.~(\ref{eq: ket of twisted cylinder sewn with MR}) implies scale invariance of the fields:
\begin{equation}\label{eq: scale invariance}
\tilde{z}
\;=\;\lambda z
\quad\Rightarrow\quad\psi_i(z)
\;=\;\lambda^{h_i}\bar{\lambda}^{\bar{h}_i}\,\widetilde{\psi}_i(\tilde{z})
\:.
\end{equation}
Here $\psi_i$ and $\widetilde{\psi}_i$ refer to the insertion of $e_i$ using the coordinate $z_i$ and $\tilde{z}_i$,
respectively.
For the conformal weights, this yields
\begin{equation*}
h(N_k(\phi,\psi)) 
\;=\;h(\phi)+h(\psi)+k\:,
\quad
\bar{h}(N_k(\phi,\psi)) 
\;=\;\bar{h}(\psi)
\:.
\end{equation*}
Thus the summation in eq.~(\ref{eq: Nk normal ordered products of general fields}) has a lower bound.
For $\phi=T$, one finds \hbox{$N_{-2}(T,\psi)=L_0\psi$}.
As a generalisation of the OPE (\ref{OPE of T and T}),
one verifies (using contour integration) that $L_1\psi$ is the derivative of $\psi$ w.r.t.~$z$.
That is, $N_{-1}(T,\psi)=d\psi/dz$.

Fields in the image of~$L_1$ or $\bar{L}_1$ will be called derivative fields.
The space of quasi-primary fields is the orthogonal complement of the derivative fields 
w.r.t.~the Shapovalov form (\ref{map: Shapovalov form}), 
thus the intersection of the kernels of~\hbox{$L_{-1}=L_1^*$} and of~$\bar{L}_{-1}$.

Let ${\mathfrak{M}_h}$ be any irreducible representation of the Virasoro algebra, with lowest weight vector $\mathfrak{m}_h$.
${\mathfrak{M}_h}$ is spanned by vectors of the form \hbox{$L_{n_k}\ldots L_{n_1}\mathfrak{m}_h$} with \hbox{$n_k\in\Z$, $k\geq 0$}.
The generating function of~${\mathfrak{M}_h}$ is the character 
\begin{displaymath}
\chi_h
\;:=\;\tr\,q^{L_0}
\:.
\end{displaymath}
where the trace is taken in ${\mathfrak{M}_h}$.
Let~${\mathfrak{M}_h}^{\qp}$ be the space of quasi-primary fields in the representation~${\mathfrak{M}_h}$.
For \hbox{$h=0$} and \hbox{${\mathfrak{M}_0}=\mathfrak{V}$}, the generating function of~${\mathfrak{M}_0}^{\qp}$ is
\begin{displaymath}
{\chi_0}^{\qp}
\;=\;(1-q)(\chi_0-1)\:.
\end{displaymath}
For other weights $h$, one has
\begin{displaymath}
{\chi_h}^{\qp}
\;=\;(1-q)\,\chi_h\:.
\end{displaymath}


\subsection{Segal's formulae}

Segal's formula (\ref{eq: Segal's naive s formula})
for the partition function of the connected sum $M_L\sqcup_{\gamma}M_R$ reads
\begin{equation}\label{eq: Segal's s formula}
\mathfrak{Z}_{M_L\sqcup_{\gamma}M_R}
\;=\;\sum_{i}
\frac{\bigl(\psi_i(z_R=0)\,\mathfrak{Z}_{M_L\sqcup_{\gamma}D_R}\bigr)
\bigl(\psi_i(z_L=0)\,\mathfrak{Z}_{D_L\sqcup_{\gamma} M_R}\bigr)}{\psi_i(z_L=0)\psi_i(z_R=0)\,\mathfrak{Z}_{\DD}}\:.
\end{equation}
Here the summation is over the fields defined by eq.~(\ref{eq: Shapovalov form of ML and ei expressed through 1-point function}).
When $\Sigma_L$ has genus $g_1$ and and $\Sigma_R$ has genus $g_2$ , 
then $M_L\sqcup_{\gamma}M_R$ has genus equal to $g_1+g_2$.
Eq.~(\ref{eq: Segal's s formula}) is used in what we call the ${s}$ formalism. 

A special case of the ${s}$ formalism is the operator product expansion (OPE),
since
\begin{equation*}
\langle\phi_1(z_1)\phi_2(0)\ldots\rangle_M
\;=\;\sum_i\,\eps_i\,\langle\phi_1(z)\phi_2(0)\,\psi_i(z_R=0)\rangle_{\DD}\,\langle\psi_i(z_L=0)\ldots \rangle_M
\:.
\end{equation*}
Scale invariance, eq.~(\ref{eq: scale invariance}), implies
\begin{equation*}
\begin{split}
\langle\phi_1(z)\phi_2(0)\,\psi_i&(z_R=0)\rangle\\
\;=\;&z^{-h_1-h_2+h_i}\,\bar{z}^{-\bar{h}_1-\bar{h}_2+\bar{h}_i}
\,\langle\phi_1(1)\phi_2(0)\,\psi_i(z_R=0)\rangle
\:.
\end{split}
\end{equation*}
Thus
\begin{equation*}
\begin{split}
\phi_1(z_1)&\phi_2(0)\\
\;=\;&\sum_i\,\eps_i\,z^{-h_1-h_2+h_i}\,\bar{z}^{-\bar{h}_1-\bar{h}_2+\bar{h}_i}
\,\langle\phi_1(1)\phi_2(0)\,\psi_i(z_R=0)\rangle_{\DD}\,\psi_i(0)
\:.
\end{split}
\end{equation*}
This generalises eq.~(\ref{eq: Nk normal ordered products of general fields}).

When a single manifold $M^{\theta}$ with two boundary curves is self-sewn, 
a new handle is attached to it.
Segal's formula (\ref{eq: Segal's naive q formula}) for the so-called ${q}$ formalism is
\begin{equation}\label{eq: Segal's q formula}
\mathfrak{Z}_{\sqcup_{\gamma}M^{\theta}}
\;=\;\sum_{i}
\frac{\psi_i(z_L=0)\psi_i(z_R=0)\mathfrak{Z}_{D_L\sqcup_{\gamma} M^{\theta}\sqcup_{\gamma}D_R}}
{\psi_i(z_L=0)\psi_i(z_R=0)\mathfrak{Z}_{\DD}}
\:.
\end{equation}
In theories with superselection rules, the ${q}$ formalism requires additional terms, 
but this is irrelevant for the model we consider.


\subsection{Notations and conventions}

By a surface we mean a smooth manifold of real dimension two.
By a (closed or open) disc we mean a simply connected orientable surface bounded by a circle.
All occurring contour integrals are to be read counterclockwise.
$\N_0$ and $\N$ denote the non-negative and positive integers, respectively,
and $\R^*=\R\setminus\{0\}$. 
For a functional $F$ of the metric $\G$, we use the functional derivative
\begin{equation*}
\delta F
\;:=\;\int\frac{\delta F}{\delta\G_{\mu\nu}}\,\delta\G_{\mu\nu}\,d\text{vol}(\G)
\:
\end{equation*}
where $d\text{vol}(\G)$ is the volume form given by $\G$.
We use Einstein's convention, 
i.e.\ repeated (upper and lower) indices are summed over.
The curvature two-form $K\,d\text{vol}(\G)=\mathcal{K}$ where $K$ is the Gauss curvature of the Levi-Civit\`a connection.

We use the conventions from \cite{Z:1-2-3}, in particular
$G_2=\pi^2E_2/6$, $G_4=\pi^4E_4/90$ and $G_6=\pi^4E_6/945$,
where $E_2$ resp.~$E_4$ and $E_6$ are the quasi-modular resp.\ modular Eisenstein series of weight $2$ resp.~$4$ and $6$.

\pagebreak

\section{Correlation functions in the \texorpdfstring{$(2,5)$}{Lg} minimal model}

\subsection{The fields}\label{subsec: The fields of the (2,5) minimal model}

Now we specialise to the $(2,5)$ minimal model,
which is characterised by \hbox{$c=-22/5$} and by the space of (real) fields (\ref{expression: full representation of Virasoro algebra in the (2,5) minimal model}).
In this model, to the lowest weight vector $\mathfrak{w}\otimes\overline{\mathfrak{w}}\in R_2(\mathfrak{W})$
is associated a real field $\Phi$, 
whose holomorphic and non-holomorphic conformal weight equal $h(\Phi)=\bar{h}(\Phi)=-1/5$.
$\Phi$ is said to be primary, which means that \hbox{$L_{-n}\Phi=0$} and \hbox{$\bar{L}_{-n}\Phi=0$} for \hbox{$n>0$}.
To normalise the Shapovalov form on $R_2(\mathfrak{W})$, we use
\begin{equation}\label{eq: normalisation of Phi}
\langle\mathfrak{w}\otimes\overline{\mathfrak{w}}|\,\mathfrak{w}\otimes\overline{\mathfrak{w}}\rangle 
\;=\;\eps_{\Phi}
\:,
\end{equation}
in accordance with eq.~(\ref{def: standard orthogonal}).
Later (in Section \ref{subsec: The holomorphic blocks}) we will see that $\eps_{\Phi}=-1$.
Arbitrarily but for definiteness,
we normalise the Shapovalov forms on ${\mathfrak{W}}$ and $\overline{\mathfrak{W}}$ individually by setting
\begin{equation}\label{eqs: squared Shapovalov norm of w and bar w}
\begin{split}
\langle\mathfrak{w}|\,\mathfrak{w}\rangle 
\;=&\;1\:,\\
\langle\overline{\mathfrak{w}}|\,\overline{\mathfrak{w}}\rangle
\;=&\;\eps_{\Phi}
\:.
\end{split}
\end{equation}

After extending, in eq.~(\ref{eq: Segal's naive q formula}), the summation over 
$\{e_i\in{\mathfrak{V}}\otimes\overline{\mathfrak{V}}\}$
from eq.~(\ref{def: standard orthogonal basis for the vacuum sector in the (2,5) minimal model}) 
to include a standard orthogonal set
\begin{equation*}
\{e_j\in{\mathfrak{W}}\otimes\overline{\mathfrak{W}}\}
\;=\;\dot{\cup}_{h,\bar{h}}\,\bigl\{\,\text{orthogonal basis in $\mathfrak{W}^h\otimes\overline{\mathfrak{W}}^{\,\bar{h}}$}\bigr\}
\:
\end{equation*}
w.r.t.~the Shapovalov form on $\mathfrak{W}$,
we recover the full partition function from eq.~(\ref{eq: g=1 partition function w.r.t. the flat metric}).

In any irreducible representation ${\mathfrak{M}_h}$ of a minimal model, two fundamental linear relations hold.
Recall that for the~$(2,5)$ minimal model and the vacuum representation ${\mathfrak{M}_0}=\mathfrak{V}$
(with lowest weight vector $\mathfrak{v}$), these are 
\begin{equation*}
\begin{split}
L_1\mathfrak{v}
\;=&\;0
\:,
\\
L_2L_2\mathfrak{v}
\;=&\;\frac{3}{5}\,L_4\mathfrak{v}
\:.
\end{split}
\end{equation*}
From the latter equation follows that
\begin{equation}\label{eq: normal ordered product of T in the (2,5) minimal model}
N_0(T,T)
\;=\;\frac{3}{10}\,T''
\:,
\end{equation}
by the Virasoro OPE (\ref{OPE of T and T}). 
The corresponding Fourier components satisfy
\begin{equation}\label{eq: (2,5) minimal model eq. in Fourier components}
\sum_{m\geq 2}L_mL_{N-m}+\sum_{m<2}L_{N-m}L_m
\;=\;\frac{3}{10}\,(N-2)(N-3)\,L_N
\:, 
\end{equation}
by eq.~(\ref{eq: normal ordered product of T in the (2,5) minimal model}) and Cauchy's Theorem.
For later use, we note that for $N\geq 3$ odd, eq.\ (\ref{eq: (2,5) minimal model eq. in Fourier components}) is equivalent to
\begin{equation}\label{eq: derived equation for N odd}
\sum_{m=-\infty}^1L_{N-m}\,L_m
+\sum_{m=2}^{(N-1)/2}L_{N-m}L_m
\;=\;\frac{1}{40}(N^2-9)\,L_N
\:,
\end{equation}
and for $N\geq 4$ even, to
\begin{equation}\label{eq: derived equation for N even}
\frac{1}{2}\,L_{N/2}L_{N/2}
+\sum_{m=-\infty}^1L_{N-m}L_m
+\sum_{m=2}^{N/2-1}L_{N-m}L_m
\;=\;\frac{1}{40}(N^2-4)\,L_N 
\:.
\end{equation}
Apart from ${\mathfrak{M}_0}=\mathfrak{V}$,
the only lowest weight representation of the Virasoro algebra satisfying eq.~(\ref{eq: (2,5) minimal model eq. in Fourier components}) is
${\mathfrak{M}_{-1/5}}=\mathfrak{W}$ (with lowest weight vector $\mathfrak{w}$).
For $N=2$ and $N=3$, eq.\ (\ref{eq: (2,5) minimal model eq. in Fourier components}) yields the fundamental identities
\begin{align*}
L_2\mathfrak{w}
\;=&\;\frac{5}{2}\,L_1L_1\mathfrak{w}
\:,
\\
L_3\mathfrak{w}
\;=&\;\frac{25}{12}\,L_1L_1L_1\mathfrak{w}
\:.
\end{align*}
These two equations are equivalent to the OPE 
\begin{equation}\label{OPE of T and Phi}
T(z)\,\Phi(0)
\;=\;\frac{-1/5}{z^2}\,\Phi(0)
+\frac{1}{z}\,\Phi'(0)
+\frac{5}{2}\,\Phi''(0)
+\frac{25}{12}\,z\,\Phi^{(3)}(0)
+O(z^2)\:,
\end{equation}
where the dash denotes application of $d/dz$.
Note that eq.~(\ref{OPE of T and Phi}) is compatible with the Virasoro OPE (\ref{OPE of T and T})
and eq.~(\ref{eq: normal ordered product of T in the (2,5) minimal model}).

Let ${\mathfrak{M}_h}$ be an irreducible representation with lowest weight vector $\mathfrak{m}_h$, of weight $h$.
For $h'\geq 1$, the set of partitions of $h'$ defines a generating set of the subvector space of vectors of conformal weight $h'+h$:
To every sequence \hbox{$(n_1,\ldots,n_N)\in\N^N$} with \hbox{$1\leq n_1\leq\ldots\leq n_N$} and \hbox{$\sum_{j=1}^Nn_j=h'$} 
is associated the vector \hbox{$L_{n_N}\ldots L_{n_1}\mathfrak{m}_h\in {\mathfrak{M}_h}$}.
Eqs (\ref{eq: derived equation for N odd}) and (\ref{eq: derived equation for N even}) show that 
when ${\mathfrak{M}_h}$ is a representation of the $(2,5)$ minimal model,
we may require $n_i+2\leq n_{i+1}$ for $i\geq 1$.
Thus the generating function for the number of holomorphic fields of a given weight in~$\mathfrak{V}$ and in~$\mathfrak{W}$ 
is the character
\begin{equation*}
\begin{split}
\chi_{\mathfrak{V}}
\;:=\;&\sum_{n\geq 0}\frac{q^{n^2+n}}{(q;q)_n}\\
\;=\;&1+q^2+q^3+q^4+q^5+2q^6+2q^7+3q^8+3q^9+4q^{10}+4q^{11}+6q^{12}+\ldots\:,\\
%
\chi_{\mathfrak{W}}
\;:=\;&q^{-1/5}\sum_{n\geq 0}\frac{q^{n^2}}{(q;q)_n}\\
\;=&\;q^{-1/5}\bigl(1+q+q^2+q^3+2q^4+2q^5+3q^6+3q^7+4q^8+5q^9+6q^{10}+7q^{11}+\ldots\bigr)
\end{split}
\end{equation*}
respectively.

\begin{proposition}\label{proposition: quasi-primary fields and their squared norms in the vacuum rep}
In the $(2,5)$ minimal model, to every conformal weight \hbox{$0<h\le 11$}, 
the space of quasi-primary fields of weight $h$ in~${\mathfrak{V}}$ is at most one-dimensional,
while for $h=12$ it is of dimension two.
For $h\leq 12$, an orthogonal basis is given by
\begin{center}
\begin{tabular}{lll}
$h$&quasi-primary vector in ${\mathfrak{V}}$&squared norm\\
\hline
$\,2$&\hspace{0.3cm}$L_2\mathfrak{v}$&\hspace{0.3cm}$c/2$\\
$\,4$&\hspace{0.3cm}--&\hspace{0.3cm}--\\
$\,6$&$\bigl(7L_4L_2-2L_6\bigr)\,\mathfrak{v}$&$-217\,c$\\
$\,8$&$\bigl(\frac{42}{5}L_5L_3+\frac{12}{5}L_6L_2-\frac{28}{5}L_8\bigr)\,\mathfrak{v}$&$-12\,792\,c/5$\\
$10$&$\bigl(12L_6L_4+3L_7L_3+8L_8L_2-\frac{756}{55}L_{10}\bigr)\,\mathfrak{v}$&$105\,400$\\
$12$&\hspace{0.3cm}$v_{12,1}$&$-111\,587\,133\,c/{13}$\\
&\hspace{0.3cm}$v_{12,2}$&$4\,842\,771\,607\,546\,875\,c$
\end{tabular}\\ 
\label{table: quasi-primary fields in Fv}
\end{center}
where
\begin{align*}
v_{12,1}
\;=\;&\bigl(\frac{495}{4}L_7L_5+\frac{55}{2}L_8L_4+\frac{341}{4}L_9L_3+59L_{10}L_2-\frac{2772}{13}L_{12}\bigr)\mathfrak{v}\\
v_{12,2}
\;=\;&\bigl(-7\,402\,950L_6L_4L_2-\frac{16\,923\,555}{4}L_7L_5-\frac{7\,473\,735}{2}L_8L_4-\frac{24\,424\,425}{4}L_9L_3\\
&-1\,260\,405L_{10}L_2+14\,784\,210L_{12}\bigr)\mathfrak{v}
\:.
\end{align*}
\end{proposition}

\begin{proof}
The number of linearly independent quasi-primary fields of conformal weight~$h$ in~${\mathfrak{V}}$ equals the coefficient of~$q^h$ in the series
\begin{displaymath}
\chi_{\mathfrak{V}}^{\qp}-1
\;=\;(1-q)(\chi_{\mathfrak{V}}-1)
\;=\;q^2+q^6+q^8+q^{10}+2q^{12}+2q^{14}+q^{15}+\ldots\:.
\end{displaymath}
The fields and their respective norm are obtained by direct computation. 
\end{proof}

The identity $L_1L_nv=(n-1)L_{n+1}v$ shows that $L_nv$ for any $n\geq 3$ correspond to iterated derivatives of $T$.
For the relatively prime integers ${m}=1\,275\,627\,234\,375$ and ${n}=2\,261$,
and for either choice of sign,
the linear combination
\begin{equation*}
v_{12,1}\sqrt{{m}}\pm v_{12,2}\sqrt{{n}}
\end{equation*}
is a null vector, 
\begin{equation*}
{n}\,\parallel v_{12,1}\parallel^2+{m}\,\parallel v_{12,2}\parallel^2
\;=\;0\:.
\end{equation*}
As such it is unique up to rescaling.

\begin{proposition}\label{proposition: quasi-primary fields and their squared norms in the W rep}
In the $(2,5)$ minimal model, to every conformal weight $h=h'-1/5$ with $0< h'\leq 11$, 
the space of quasi-primary fields in~${\mathfrak{W}}$ of weight $h$ is at most one-dimensional,
with the first quasi-primary field occurring at weight $19/5$.
An orthogonal basis is given by
\begin{center}
\begin{tabular}{lll}
$h'$&quasi-primary vector in ${\mathfrak{W}}$&squared norm\\
\hline
$4$&$\bigl(25L_3L_1-2L_4\bigr)\,\mathfrak{w}$&${5\,928}\parallel \mathfrak{w}\parallel^2/{5}$\\
$6$&$\bigl(21L_4L_2+7L_5L_1-\frac{43}{5}L_6\bigr)\,\mathfrak{w}$&${1\,634\,817}\parallel \mathfrak{w}\parallel^2/{125}$\\
$8$&$\bigl(7L_5L_3+2L_6L_2+\frac{9}{2}L_7L_1-\frac{27}{5}L_8\bigr)\,\mathfrak{w}$&$-{10\,759}\,c\,\parallel \mathfrak{w}\parallel^2/{5}$\\
$9$&$\bigl(7L_5L_3L_1+\frac{4}{5}L_6L_3+\frac{17}{10}L_7L_2-\frac{3}{2}L_8L_1-\frac{87}{50}L_9\bigr)\mathfrak{w}$&$-{4\,018\,443}\,c\,\parallel \mathfrak{w}\parallel^2/{3\,125}$\\
$10$&$\bigl(12L_6L_4+3L_7L_3+8L_8L_2+6L_9L_1-\frac{828}{55}L_{10}\bigr)\mathfrak{w}$&${356\,414\,616}\parallel\mathfrak{w}\parallel^2/{3\,025}$\\
$11$&$(3L_7L_4+7L_8L_3+14 L_9L_2-42 L_{10}L_1+60 L_6L_4L_1$&$-335\,699\,c^2\parallel\mathfrak{w}\parallel^2/10 $\\
&$\,+15 L_7L_3L_1-\frac{77}{5}L_{11})\,\mathfrak{w}$&\\
$12$&$w_{12,1}$&$\parallel w_{12,1}\parallel^2$\\
&$w_{12,2}$&$4\,577\,888\,353\,871\parallel\mathfrak{w}\parallel^2/3$\\
\end{tabular} \\ 
\label{table: quasi-primary fields in Fw}
\end{center}
where $\parallel w_{12,1}\parallel^2=-401\,304\,741\,575\,703\,261\,976\,819\,992\,213\,522\parallel\mathfrak{w}\parallel^2/5$
and
\begin{align*}
w_{12,1}
\;=\;&\bigl(\frac{490\,940\,248\,301\,793}{2}L_7L_5+331\,118\,144\,757\,253L_8L_4+\frac{750\,814\,671\,380\,107}{2}L_9L_3\\
&-6\,914\,283\,463\,318L_{10}L_2+\frac{1\,810\,769\,754\,094\,025}{2}L_{11}L_1+448\,490\,640\,452\,880L_6L_4L_2\\
&-280\,306\,650\,283\,050L_7L_4L_1-541\,926\,190\,547\,230L_8L_3L_1-1\,025\,605\,090\,206\,371L_{12}\bigr)\mathfrak{w}\:.\\
w_{12,2}
\;=\;&\bigl(-\frac{132\,275}{6}L_8L_4-16\,445L_9L_3+9\,880L_{10}L_2-\frac{122\,915}{2}L_{11}L_1-35\,750L_6L_4L_2\\
&+\frac{89\,375}{4}L_7L_4L_1+\frac{518\,375}{12} L_8L_3L_1+45\,959 L_{12}\bigr)\mathfrak{w}\:.
\end{align*}
\end{proposition}

\begin{proof}
The number of quasi-primary fields of conformal weight~$h=h'-1/5$ in~${\mathfrak{W}}$ is given by the coefficient of~$q^h$ in the series
\begin{displaymath}
\chi_{\mathfrak{W}}^{\qp}
\;=\;(1-q)\,\chi_{\mathfrak{W}}
\;=\;q^{-1/5}\,\bigl(1+q^4+q^6+q^8+q^9+q^{10}+q^{11}+2q^{12}+\ldots\bigr)
\:.
\end{displaymath} 
The fields and their respective norm are obtained by direct computation, using in particular eqs (\ref{eq: derived equation for N odd}) and (\ref{eq: derived equation for N even}).  
\end{proof}

For the relatively prime integers
$m=2\,775$ and $n=144\,904\,525\,088\,234\,638\,887\,846$, 
the linear combinations
\begin{equation*}
w_{12,1}\sqrt{{m}}\pm w_{12,2}\sqrt{{n}}
\end{equation*}
are null vectors.

For derivative fields, one has 

\begin{proposition}\label{proposition: squared Shapovalov norm of derivatives of T and of Phi}
For \hbox{$k\geq 0$} and the respective $k$\ts{th} derivative in the holomorphic coordinate, 
\begin{displaymath}
\frac{\parallel T^{(k)}\parallel^2}{c/2}
\;=\;\frac{k!(k+3)!}{3!}
\,\in\,\bigl\{1\,,\:4\,,\:40\,,\:720\,,\:20\,160\,,\ldots\bigr\}
\:,
\end{displaymath}
and
\begin{displaymath}
\frac{\parallel\Phi^{(k)}\parallel^2}{\parallel\Phi\parallel^2}
\;=\;k!\prod_{n=1}^{k}\biggl(k-n-\frac{2}{5}\biggr)
\,\in\,\Bigl\{1,-\frac{2}{5}\,,-\frac{12}{25}\,,-\frac{288}{125}\,,-\frac{14\,976}{625}\,,\ldots\Bigr\}
\:.
\end{displaymath} 
\end{proposition}

\begin{proof}
Let~$\phi$ be a field with the properties \hbox{$L_{-1}\phi=0$} and \hbox{$L_0\phi=h\phi$}.
For \hbox{$k\geq 1$},
\begin{equation*}
\begin{split}
\parallel\phi^{(k)}\parallel^2 
\;=\;\langle L_1^k\phi|L_1^k\phi\rangle
\;=&\;\langle L_1^{k-1}\phi|[L_{-1}L_1^k]\phi\rangle\\
\;=&\;\sum_{m=1}^k\langle L_1^{k-1}\phi|L_1^{m-1}[L_{-1}L_1]L_1^{k-m}\phi\rangle\\
\;=&\;2\sum_{m=1}^k(k-m+h)\langle L_1^{k-1}\phi|L_1^{k-1}\phi\rangle\\
\;=&\;k\,(k-1+2h)\parallel\partial^{k-1}\phi\parallel^2 
\:.
\end{split}
\end{equation*}
By induction,
\begin{displaymath}
\parallel\phi^{(k)}\parallel^2 
\;=\;k!\prod_{n=1}^{k}(k-n+2h)\parallel\phi\parallel^2
\:.
\end{displaymath}
\end{proof}

Now we specialise to genus \hbox{$g=1$}. 
Thus $q$ is identified with the nome~$e^{2\pi i\tau}$ for $\tau\in\mathfrak{H}_1$, the upper complex half plane.
In the character formulae above, 
we have $\chi_{\mathfrak{V}}=H(q)$ and $\chi_{\mathfrak{W}}=q^{-1/5}G(q)$
in terms of the Rogers-Ramanujan functions eq.~(\ref{def: Rogers-Ramanujan functions}).
The~zero-point functions differ from the corresponding characters by a factor of~$q^{-c/24}$.
Conventionally one numbers them in lexicographic order,
\begin{equation}\label{def: zero point functions in the (2,5) minimal model}
\begin{split}
\langle 1\rangle(\tau)^{g=1}_1
\;=&\;q^{-1/60}\,G(q)\:,\\
\langle 1\rangle(\tau)^{g=1}_2
\;=&\;q^{11/60}\,H(q)
\:.
\end{split}
\end{equation}
In accordance with Section \ref{subsub:minimal model in genus one},
the first and second correspond to the non-vacuum and vacuum sector, respectively,
and the partition function is given by eq.~(\ref{eq: g=1 partition function w.r.t. the flat metric}).
Together, the~zero-point functions form a vector valued modular function~$\begin{pmatrix}
 \langle 1\rangle_1^{g=1}\\
 \langle 1\rangle_2^{g=1}
 \end{pmatrix}$
for the full modular group w.r.t.\ a representation
\begin{displaymath}
\varrho_{\RR}:\quad\SL(2,\Z)\rechts M_2(\C) 
\end{displaymath}
that factors through $\SL(2,\Z/5\Z)$.
In particular, the genus one partition function eq.~(\ref{eq: g=1 partition function w.r.t. the flat metric})
is invariant under the full modular group.

For the remainder of this section, we drop the label $g=1$.
For $\ell\in\R$, 
let \hbox{$\D_{\ell}=d/d (2\pi i\tau)-(\ell/12)E_2(\tau)$} 
be the Serre-derivative operator.
(When the weight $\ell$ of a modular form $f$ is clear, 
we will often omit the index and write simply $\D f$ for $\D_{\ell} f$, $\D^2 f$ for $\D_{\ell+2}\D_{\ell} f$, etc.
For instance, both $\langle 1\rangle_1$ and $\langle 1\rangle_2$ are solutions of the Kaneko-Zagier differential equation
\begin{equation}\label{eq: Kaneko-Zagier ODE}
\D^2f
\;=\;\frac{\ell(\ell+2)}{144}E_4(\tau)f 
\end{equation}
which was originally studied for \hbox{$\ell=0$} or \hbox{$\ell\equiv 4\:(\text{mod}\:6)$} 
~\cite[\S8]{Z-K:1997}
but needs \hbox{$\ell=1/5$} here.)

For $N\geq 0$, 
the~$N$-point function of the 
Virasoro field~$T$ is the $N$-fold functional derivative of the~zero-point function.
In this sense the Virasoro field generates changes of the metric. 
For volume preserving metric changes, 
we have for $a=1,2$~\cite{EO:1987}\footnote{or preprint arXiv:1705.08294 for a direct proof}
\begin{equation}\label{eq: ODE relating <1> and <T>}
\frac{1}{2\pi i}\frac{d}{d\tau}\,\langle 1\rangle_a
\;=\;\frac{1}{(2\pi i)^2}\,\langle T\rangle_a
\:,
\end{equation}
where~$\langle T\rangle_a$ is the~one-point function of the field~$T$.
(By translational invariance of~$N$-point functions on the torus, 
$\langle T(z)\rangle_a$ is constant in position.)
As an aside, the OPE (\ref{OPE of T and T}) yields in addition
\begin{displaymath}
\D_2\,\langle T\rangle_a
\;=\;\frac{11}{3600}\,(2\pi i)^2E_4(\tau)\,\langle 1\rangle_a
\:,
\end{displaymath}
confirming eq.~(\ref{eq: Kaneko-Zagier ODE}).

For fixed \hbox{$\tau\in\mathfrak{H}_1$}, 
let \hbox{$\wp(z)\equiv\wp(z|\tau)$} and \hbox{$\zeta(z)\equiv\zeta(z|\tau)$} the Weierstrass~$\wp$ and~$\zeta$ function, respectively.
Denote their respective value at \hbox{$z_{ij}:=z_i-z_j$} (for \hbox{$z_i,z_j\in\C$}) by~$\wp_{ij}$ and~$\zeta_{ij}$.
We always set $z_0=0$ and omit the index in this case, writing simply $\wp_i$ for $\wp(z_i)$.

Now we calculate the~one-point function w.r.t.~the metric $|dz|^2$ of the field \hbox{$\Phi\in{R_2(\mathfrak{W})}$}
which corresponds to the lowest weight vector~$\mathfrak{w}\in\mathfrak{W}$.

\begin{proposition}
The one-point function of $\Phi$ satisfies
\begin{equation}\label{eq: DE for Phi}
\D_{-1/5}\langle\Phi\rangle
\;=\;0\:.
\end{equation}
\end{proposition}

\begin{proof}
By the OPE (\ref{OPE of T and Phi}), 
\begin{equation}\label{eq: 2-pt fct of T with Phi}
\frac{\langle T(z)\,\Phi(0)\rangle}{\langle\Phi(0)\rangle}
=\:h\,\wp(z) 
\:.
\end{equation}
(Note that $\langle\Phi''(0)\rangle=\frac{d^2}{d z^2}\langle\Phi(z)\rangle|_{z=0}=0$ by translational invariance.)
Moreover, 
\begin{equation}\label{eq: period integral over Weierstrass p}
\int_0^1\wp(z)dz
\;=\;-2\,\zeta(1/2)
\;=\;-\frac{\pi^2}{3}\,E_2(\tau)\:,
\end{equation}
so when the contour integral is taken along the real period, and \hbox{$\oint dz=1$},
\begin{displaymath}
\frac{1}{2\pi i}\,\frac{d}{d\tau}\langle\Phi(0)\rangle_a
\;=\;\oint\,\langle T(z)\,\Phi(0)\rangle_a\,\frac{dz}{(2\pi i)^2}\\
\;=\;-\frac{1}{60}\,E_2(\tau)\,\langle\Phi(0)\rangle_a\:.
\end{displaymath}
This shows that $\langle\Phi\rangle$ lies in the kernel of $\D_{-1/5}$.
\end{proof}

Since the kernel of $\D_{\ell}$ is spanned by $\eta(\tau)^{2\ell}$, 
where ~$\eta(\tau)$ is Dedekind~$\eta$-function, 
we have
\begin{equation}\label{eq: 1-point function of Phi}
\langle\Phi\rangle
\;=\;{\mu}_{\Phi}\,|\eta\,|^{-4/5} 
\:
\end{equation}
for some ${\mu}_{\Phi}\,\in\R^*$, 
which we compute in Proposition \ref{proposition: the normalisation constant of Phi}
(up to a sign that we choose positive).
We will use the factorisation 
$\langle\Phi\rangle=\langle\mathfrak{w}\rangle\langle\overline{\mathfrak{w}}\rangle$,
where $\langle\mathfrak{w}\rangle=\sqrt{{\mu}_{\Phi}\,}\,\eta^{-2/5}$ 
and $\langle\overline{\mathfrak{w}}\rangle=\sqrt{{\mu}_{\Phi}\,}\,\,\overline{\eta}^{-2/5}$


\begin{cor}\label{corollary: 2-pt fct of T with Phi}
We have
\begin{equation}\label{eq: 3-pt fct of T with Phi}
\frac{\langle T(z_1)T(z_2)\,\Phi(0)\rangle}{\langle\Phi(0)\rangle}
\;=\;\frac{c}{2}\,\wp_{12}^2
-\frac{1}{5}\,\wp_{12}\,\bigl(\wp_1+\wp_2\bigr)
+\frac{6}{25}\,\wp_1\,\wp_2+24\,G_4
\:.
\end{equation} 
\end{cor}

\begin{proof}
On the one hand, 
from the OPE (\ref{OPE of T and Phi}) for $T(z)$ and $\Phi(0)$, and eq.~(\ref{eq: 2-pt fct of T with Phi}),
\begin{displaymath}
\langle\Phi\rangle^{-1}\langle T(z)T(w)\,\Phi(0)\rangle
\;=\;\frac{h^2}{z^2}\,\wp(w)
-\frac{h}{z}\,\wp'(w)
+\text{terms that are regular for $z\rechts 0$}\:,
\end{displaymath}
where the occurring even and odd negative power of~$z$ can be replaced with~$\wp(z)$ and~$z\,\wp(z)$, respectively.
The latter expression is not elliptic. 
However, we may use 
\begin{displaymath}
-z\,\wp'(w)
\;=\;\wp(z-w)-\wp(w)+O(z^2) 
\:.
\end{displaymath}
Thus 
\begin{equation*}
\begin{split}
\langle\Phi\rangle^{-1}&\langle T(z)T(w)\,\Phi(0)\rangle\\
\;=&\;h\,\wp(z)\,\wp(z-w)+(h^2-h)\,\wp(z)\,\wp(w)
+\text{terms that are regular for $z\rechts 0$}\:.
\end{split}
\end{equation*}
On the other hand, 
the OPE (\ref{OPE of T and T}) for $T(z)$ and $T(w)$, and eq.~(\ref{eq: 2-pt fct of T with Phi})
yield
\begin{equation*}
\begin{split}
\langle\Phi\rangle^{-1}\langle T(z)T(w)\,\Phi(0)\rangle
\;=&\;\frac{c/2}{(z-w)^4}+\frac{h}{(z-w)^2}\{\wp(z)+\wp(w)\}-\frac{1}{5}\,h\,\wp''(w)+O(z-w)\\
\;=\;\frac{c}{2}\,\wp^2&(z-w)+h\,\wp(z-w)\{\wp(z)+\wp(w)\}-\frac{6}{5}\,h\,\wp(z)\,\wp(w)+\tC_{4,c}\:,
\end{split}
\end{equation*}
where~$\tC_{4,c}$ is constant in~$z$ and~$w$.
By comparison, we obtain
\begin{displaymath}
h\Bigl(h+\frac{1}{5}\Bigr)=\;0\:,
\quad
\tC_{4,c}
\;=\;-(c-2h)\,6\,G_4
\;=\;24\,G_4
\:,
\end{displaymath}
since $c=-22/5$.
\end{proof}


\begin{proposition}\label{proposition: 3rd order ODE for 3-pt fct of Phi in the analytic coordinate}
The two-point function of~$\Phi$ satisfies the ODE
\begin{equation}\label{3rd order ODE for 2-pt function of varphi in the analytic coordinate z}
\Bigl(\frac{25}{12}\frac{d^3}{d z^3}-\wp(z)\,\frac{d}{d z}+\frac{1}{5}\,\wp'(z)\Bigr)\,
\bigl\langle\Phi(z)\,\Phi(0)\bigr\rangle_A
\;=\;0
\:.
\end{equation}
\end{proposition}

\begin{proposition}\label{proposition: 3-point fct of T and two copies of Phi for g=1}
For $z_2\not=0$,  
\begin{equation*}
\begin{split}
\bigl\langle T(z_1)\,\Phi(z_2)\,\Phi(0)\bigr\rangle_A
\;=&\;-\frac{1}{5}\,\bigl(\wp_{12}+\wp_1-\wp_2\bigr)\,\bigl\langle\Phi(z_2)\,\Phi(0)\bigr\rangle_A\\
+&\:\bigl(\zeta_{12}-\zeta_1+\zeta_2\bigr)\,\bigl\langle\Phi'(z_2)\,\Phi(0)\bigr\rangle_A
+\frac{5}{2}\,\bigl\langle\Phi''(z_2)\,\Phi(0)\bigr\rangle_A
\:.
\end{split}
\end{equation*} 
\end{proposition}

\begin{proof}
By the OPE of $T(u)$ with $\Phi(z)$ and with $\Phi(0)$, 
respectively, 
\begin{equation*}
\begin{split}
\bigl\langle T(u)\,\Phi(z)\,&\Phi(0)\bigr\rangle_A\\
\;=&\;h\,\wp(u-z)\,\bigl\langle\Phi(z)\,\Phi(0)\bigr\rangle_A
+\zeta(u-z)\,\bigl\langle\Phi'(z)\,\Phi(0)\bigr\rangle_A
+\text{regular for $u\rechts z$}\\
\;=&\;h\,\wp(u)\,\bigl\langle\Phi(z)\,\Phi(0)\bigr\rangle_A
+\zeta(u)\,\bigl\langle\Phi(z)\,\Phi'(0)\bigr\rangle_A
+\text{regular for $u\rechts 0$}
\end{split}
\end{equation*}
By translational invariance,
\begin{displaymath}
\bigl\langle\Phi(z)\,\Phi'(0)\bigr\rangle_A
\;=\;-\,\bigl\langle\Phi'(z)\,\Phi(0)\bigr\rangle_A
\:.
\end{displaymath}
Considering all poles at once yields 
\begin{equation}\label{line: terms in the 3-pt fct of T varphi varphi which are singular in u}
\begin{split}
\bigl\langle T(u)&\,\Phi(z)\,\Phi(0)\bigr\rangle_A\\
\;=&\;h\bigl(\wp(z-u)+\wp(u)\bigr)\,\bigl\langle\Phi(z)\,\Phi(0)\bigr\rangle_A
+\bigl(\zeta(u-z)-\zeta(u)\bigr)\,\bigl\langle\Phi'(z)\,\Phi(0)\bigr\rangle_A \\
+&\:\text{terms that are constant in $u$,}
\end{split}
\end{equation}
by ellipticity of the three-point function. 
Comparison of the~$u^0$ terms on the r.h.s.\ of the first line of eq.~(\ref{line: terms in the 3-pt fct of T varphi varphi which are singular in u}) 
with the OPE (\ref{OPE of T and Phi}) for $T(u)$ and $\Phi(0)$ shows that the terms constant in~$u$ are equal to
\begin{displaymath}
-h\,\wp(z)\,\bigl\langle\Phi(z)\,\Phi(0)\bigr\rangle_A
+\zeta(z)\,\bigl\langle\Phi'(z)\,\Phi(0)\bigr\rangle_A
+\frac{5}{2}\;\bigl\langle\Phi''(z)\,\Phi(0)\bigr\rangle_A
\:
\end{displaymath}
since \hbox{$
\bigl\langle\Phi(z)\,\Phi''(0)\bigr\rangle_A
\;=\;
\bigl\langle\Phi''(z)\,\Phi(0)\bigr\rangle_A$}, by invariance under both translation and reflection.
\end{proof}

\begin{proof}[Proof of Proposition \ref{proposition: 3rd order ODE for 3-pt fct of Phi in the analytic coordinate}]
Eq.~(\ref{3rd order ODE for 2-pt function of varphi in the analytic coordinate z}) follows
by comparison of the terms in eq.~(\ref{line: terms in the 3-pt fct of T varphi varphi which are singular in u}) 
which are linear in~$u$, with the OPE (\ref{OPE of T and Phi}) for $T(u)$ and $\Phi(0)$, 
using that 
\begin{equation*}
\bigl\langle\Phi^{(3)}(z)\,\Phi(0)\bigr\rangle_a=-\bigl\langle\Phi^{(3)}(0)\,\Phi(z)\bigr\rangle_a
\:.
\end{equation*}
\end{proof}

\subsection{Graphical representation of correlation functions on the torus}\label{subsec: 2.2}

We now consider general rational conformal field theories on the torus with holomorphic differential $dz$ and periods $1,\tau$.
Throughout this section, all occurring correlation functions refer to one and the same holomorphic block
(we omit the lower index $A\in\mathfrak{A}$).


The following proposition is the reformulation of a known result~\cite{L:2013,L:PhD14} in terms of elliptic functions.  
Since the present version is somewhat simpler and uses an argument required for proving the related Proposition \ref{proposition: graphical representation of (N+1)-point functions of N copies of T and one copy of Phi}, which is new, we present it here.   

\begin{proposition}\label{proposition: graphical representation of N-point functions of T}
Let \hbox{$\langle 1\rangle$} be a holomorphic block in a sector of a CFT on the torus with central charge $c$.
For \hbox{$N\in\N_0$}, let \hbox{$S^{[1]}_N:=S(z_1,\ldots,z_N)$} be the set of oriented graphs with vertices \hbox{$z_1,\ldots,z_N$}
(which may or may not be connected), 
subject to the condition that every vertex has at most one ingoing and at most one outgoing line, and none is a tadpole
(with the line incoming to a vertex being identical to its outgoing line).
For \hbox{$n\in\N$}, there exist functions
\begin{displaymath}
{C}_{2n,c}:\quad\mathfrak{H}\:\rechts\:\C
\end{displaymath}
which depend on $c$, 
such that for the~$N$-point function of the Virasoro field, we have
\begin{equation}\label{graph rep}
\langle 1\rangle^{-1}\langle T(z_1)\ldots T(z_N)\rangle
\;=\;\sum_{\Gamma\in S^{[1]}_N}{\gamma}(\Gamma)\:,
\end{equation}
where for \hbox{$\Gamma\in S^{[1]}_N$},
\begin{displaymath}
{\gamma}(\Gamma)
\;:=\;\biggl(\frac{c}{2}\biggr)^{\sharp\text{loops}}
\,{C}_{2\cdot(N-\sharp\text{edges}),c}\,
\prod_{(z_i,z_j)\in\Gamma}\wp_{ij}\:
\end{displaymath}
Here \hbox{$(z_i,z_j)\in\Gamma$} is an oriented edge.
Moreover, for all \hbox{$n\in\N$, ${C}_{2n,c}$} 
is a modular form of weight~$2n$.
\end{proposition}

Note that the result holds true in general, and not just in the \hbox{$(2,5)$} minimal model.
For later use, here is the result for $N=3$ and for $N=4$, respectively:
\begin{equation}\label{eq: 3-pt function of T}
\begin{split}
\langle 1\rangle^{-1}\bigl\langle T(z_2)&T(z_1)T(z_3)\bigl\rangle \\
\;=\;&c\,\wp_{31}\wp_{12}\wp_{23}\,C_{0,c}\\
+&\frac{c}{2}\bigl(\wp_{13}^2+\wp_{23}^2+\wp_{12}^2\bigr)\,C_{2,c}
+2\,\bigl(\wp_{31}\wp_{12}+\wp_{13}\wp_{32}+\wp_{32}\wp_{21}\bigr)\,C_{2,c}\\
+&2\,\bigl(\wp_{31}+\wp_{12}+\wp_{23}\bigr)\,C_{4,c}
+C_{6,c}
\:,
\end{split}
\end{equation}
\begin{equation*}
\begin{split}
\langle 1\rangle^{-1}\bigl\langle T(z_2)&T(z_1)T(z_3)T(z_4)\bigl\rangle \\
\;=\;&\frac{c^2}{4}\,\bigl(
\wp_{12}^2\wp_{34}^2
+\wp_{13}^2\wp_{24}^2
+\wp_{14}^2\wp_{23}^2
\bigr)
\,C_{0,c}\\
+&c\,\bigl(
\wp_{12}\wp_{23}\wp_{34}\wp_{41}
+\wp_{13}\wp_{32}\wp_{24}\wp_{41}
+\wp_{12}\wp_{24}\wp_{43}\wp_{31}
\bigr)\,C_{0,c}\\
+&c\,\bigl(
\wp_{12}^2\wp_{34}+\wp_{13}^2\wp_{24}+\wp_{14}^2\wp_{23}+
+\wp_{23}^2\wp_{14}+\wp_{24}^2\wp_{13}
+\wp_{34}^2\wp_{12}
\bigr)
\,C_{2,c}\\
+&c\,\bigl(
\wp_{12}\wp_{23}\wp_{31}
+\wp_{23}\wp_{34}\wp_{42}
+\wp_{13}\wp_{34}\wp_{41}
+\wp_{12}\wp_{24}\wp_{41}
\bigr)
\,C_{2,c}\\
+&2\,\bigl(
\wp_{12}\wp_{23}\wp_{34}
+\wp_{23}\wp_{34}\wp_{41}
+\wp_{34}\wp_{41}\wp_{12}
+\wp_{41}\wp_{12}\wp_{23}\\
&\hspace{0.2cm}
+\wp_{42}\wp_{23}\wp_{31}
+\wp_{13}\wp_{34}\wp_{42}
+\wp_{24}\wp_{41}\wp_{13}
+\wp_{31}\wp_{12}\wp_{24}\\
&\hspace{0.2cm}
+\wp_{41}\wp_{13}\wp_{32}
+\wp_{14}\wp_{42}\wp_{23}
+\wp_{43}\wp_{31}\wp_{12}
+\wp_{34}\wp_{42}\wp_{21}
\bigr)
\,C_{2,c}\\
+&\frac{c}{2}\,\bigl(
\wp_{12}^2+\wp_{13}^2+\wp_{14}^2+\wp_{23}^2+\wp_{24}^2+\wp_{34}^2
\bigr)
\,C_{4,c}\\
+&2\,\bigl(
\wp_{41}\wp_{13}+\wp_{41}\wp_{12}+\wp_{31}\wp_{12}
+\wp_{12}\wp_{23}+\wp_{12}\wp_{24}+\wp_{42}\wp_{23}\\
&\hspace{0.2cm}
+\wp_{23}\wp_{34}+\wp_{23}\wp_{31}+\wp_{13}\wp_{34}
+\wp_{14}\wp_{42}+\wp_{14}\wp_{43}+\wp_{24}\wp_{43}
\bigr)
\,C_{4,c}\\
+&4\,
\bigl(
\wp_{41}\wp_{32}
+\wp_{12}\wp_{43}
+\wp_{13}\wp_{24}
\bigr)
\,C_{4,c}\\
+&2\,\bigl(
\wp_{12}+\wp_{13}+\wp_{14}+\wp_{23}+\wp_{24}+\wp_{34}
\bigr)
\,C_{6,c}\\
+&C_{8,c}
\:.
\end{split}
\end{equation*}
A proof of Proposition \ref{proposition: graphical representation of N-point functions of T} is given in Appendix \ref{appendix section: Proof of {proposition: graphical representation of N-point functions of T}}.

In order to compute higher order terms in the~${s}$-expansion of the fifth solution \hbox{$\langle 1\rangle_{\Phi}^{g=2}$}, 
we use the following result. 

\begin{proposition}\label{proposition: graphical representation of (N+1)-point functions of N copies of T and one copy of Phi}
Let $\Phi$ be a primary field of holomorphic conformal weight~$h$. 
Let \hbox{$\langle\Phi\rangle$} be the corresponding one-point function of a CFT on the torus
with central charge $c$.
For \hbox{$N\geq 0$}, let \hbox{$\tilde{S}^{[1]}_N:=\tilde{S}(0,z_1,\ldots,z_N)$} be the set of oriented graphs 
with vertices \hbox{$z_0=0,z_1,\ldots,z_N$} 
subject to the conditions
\begin{enumerate}
\item
 Every vertex different from~$0$ has at most one ingoing and at most one outgoing line, and none is a tadpole
 \item 
 No edge emanates from~$0$.
 \end{enumerate} 
 Let \hbox{$\lambda:\tilde{S}^{[1]}_N\rechts\{0,1,\ldots,N\}$} be the map that counts a graph's respective number of edges ending in~$0$. 
 For \hbox{$n\in\N$}, there exist functions
\begin{displaymath}
\tC_{2n,c}:\quad\mathfrak{H}\:\rechts\:\C
\end{displaymath}
which depend on $c$, 
such that for the \hbox{$(N+1)$}-point function of $N$ copies of the Virasoro field and one copy of~$\Phi$, we have
\begin{displaymath}
\langle\Phi\rangle^{-1}\bigl\langle T(z_1)\ldots T(z_N)\:\Phi(0)\bigr\rangle
\;=\;\sum_{\Gamma\in\tilde{S}(0,z_1,\ldots,z_N)}\widetilde{{\gamma}}(\Gamma)\:,
\end{displaymath}
where for \hbox{$\Gamma\in\tilde{S}(0,z_1,\ldots,z_N)$} with $\lambda=\lambda(\Gamma)$ and with number 
\hbox{$\sharp\text{edges}\geq\lambda$},
\begin{equation}\label{transcription of graphs representing the (N+1) pt fct of T with one Phi}
\widetilde{{\gamma}}(\Gamma)
\;:=\;\biggl(\frac{c}{2}\biggl)^{\sharp\text{loops}}\,
\;\tC_{2\cdot(N-\sharp\text{edges}),c}\,
\prod_{j=0}^{\lambda-1}(h-j)\:
\prod_{(z_i,z_j)\in\Gamma}\wp_{ij}\:
\:.
\end{equation}
Moreover, for all \hbox{$n\in\N$, $\tC_{2n,c}$} is a modular form of weight~\hbox{$2n$}.
\end{proposition}

Since holomorphic and anti-holomorphic variables do not give rise to mixed singularities 
and can be treated separately, 
we can write 
\begin{equation*}
\begin{split}
\bigl\langle T(z_1)\ldots T(z_N)\,\bar{T}(\bar{z}_1')&\ldots\bar{T}(\bar{z}_N')\,\Phi(0)\bigr\rangle\\
\;=&\;\bigl\langle T(z_1)\ldots T(z_N)\mathfrak{w}\bigr\rangle\,\bigl\langle\bar{T}(\bar{z}_1')\ldots\bar{T}(\bar{z}_N')\overline{\mathfrak{w}}\bigr\rangle 
\:,
\end{split}
\end{equation*}
where acording to eq.~(\ref{eq: 1-point function of Phi}), $\langle\mathfrak{w}\rangle=\sqrt{{\mu}_{\Phi}\,}\,\eta^{-2/5}$.
For illustration, 
Proposition \ref{proposition: graphical representation of (N+1)-point functions of N copies of T and one copy of Phi}
yields for $N=3$:
\begin{equation*}
\begin{split}
\langle\Phi\rangle^{-1}&\langle T(z_3)T(z_2)T(z_1)\Phi(z_0=0)\rangle \\
\;=\;&c\,\wp_{12}\wp_{23}\wp_{31}\,\tC_{0,c}
-\frac{c}{10}\,(\wp_{12}^2\wp_3+\wp_{31}^2\wp_2+\wp_{23}^2\wp_1)\,\tC_{0,c}
-\frac{66}{125}\,\wp_1\wp_2\wp_3\,\tC_{0,c}\\
-&\frac{1}{5}\,\bigl(\wp_{12}\wp_{23}(\wp_3+\wp_1)+\wp_{13}\wp_{32}(\wp_1+\wp_2)+\wp_{31}\wp_{12}(\wp_3+\wp_2)\bigr)\,\tC_{0,c}\\
+&\frac{6}{25}\,\bigl(\wp_{12}\wp_3(\wp_1+\wp_2)+\wp_{31}\wp_2(\wp_1+\wp_3)+\wp_{23}\wp_1(\wp_2+\wp_3)\bigr)\,\tC_{0,c}\\
+&2\,(\wp_{12}+\wp_{31}+\wp_{23})\,\tC_{4,c}
-\frac{1}{5}\,(\wp_1+\wp_2+\wp_3)\,\tC_{4,c}
+\tC_{6,c}\:.
\end{split}
\end{equation*}
Since $\tC_{2,c}=0$, the contribution of all graphs with $\sharp\text{edges}=2$,
\begin{equation*}
\begin{split}
&\frac{c}{2}\,(\wp_{12}^2+\wp_{31}^2+\wp_{23}^2)\,
+2\,(\wp_{12}\wp_{23}+\wp_{13}\wp_{32}+\wp_{31}\wp_{12})\,\\
-&\frac{1}{5}\,\bigl(\wp_{12}(\wp_1+\wp_2+2\wp_3)+\wp_{31}(\wp_1+2\wp_2+\wp_3)+\wp_{23}(2\wp_1+\wp_2+\wp_3)\bigr)\,\\
+&\frac{6}{25}\,(\wp_1\wp_2+\wp_3\wp_1+\wp_2\wp_3)\,\:,
\end{split}
\end{equation*}
drops out.
A proof of Proposition \ref{proposition: graphical representation of (N+1)-point functions of N copies of T and one copy of Phi} has been moved to Appendix \ref{appendix section: Proof of {proposition: graphical representation of (N+1)-point functions of N copies of T and one copy of Phi}}. 

We provide a machinery for computing successively, for \hbox{$N\geq 1$}, 
the modular forms \hbox{${C}_{2N,c}={\gamma}(\Gamma_0^N)$} and \hbox{$\tC_{2N,c}=\widetilde{{\gamma}}(\Gamma_0^N)$} 
from Proposition \ref{proposition: graphical representation of N-point functions of T} 
and Proposition \ref{proposition: graphical representation of (N+1)-point functions of N copies of T and one copy of Phi}, respectively.

Our method relies on a formula in Weinberg's book~\cite[p.~360]{Wein:1972}.
For the~zero-point 
and for the~one-point function of a primary field~$\Phi$ of holomorphic weight~$h$ of a CFT sector on the torus, 
we have
\begin{displaymath}
\frac{d^N\langle 1\rangle}{d\tau^N} 
\;=\;\oint\ldots\oint\bigl\langle T(z_1)\ldots T(z_N)\bigr\rangle\,\frac{dz_2\ldots dz_N}{(2\pi i)^N}
\end{displaymath}
and
\begin{displaymath}
\frac{d^N\langle\Phi\rangle}{d\tau^N} 
\;=\;\oint\ldots\oint\bigl\langle T(z_1)\ldots T(z_N)\,\Phi(z_0)\bigr\rangle\,\frac{dz_1\ldots dz_N}{(2\pi i)^N}
\end{displaymath}
respectively. In the present discussion, we integrate along the real period (using that the fields are holomorphic, and Cauchy's Theorem). 
This is particularly convenient for our purpose 
since the period integral over~$\wp$ is proportional to~$E_2$ by eq.~(\ref{eq: period integral over Weierstrass p}), 
and does not contribute to~${C}_{2N,c}$ and~$\tC_{2N,c}$. 
For \hbox{$N=1$}, we recover eq.~(\ref{eq: ODE relating <1> and <T>}) and (\ref{eq: DE for Phi}), respectively.

Let \hbox{$f,g:\C\rechts\C$} be two functions with period~$1$.
Suppose~$f$ is meromorphic on~$\C$, and~$g$ is meromorphic in a tubular neighbourhood of the real line.
The convolution of~$f$ and~$g$ along the real period is defined by
\begin{displaymath}
(f*g)(x)
\;:=\;\lim_{\sigma\searrow 0}\int_{-i\sigma}^{1-i\sigma}f(z)g(x-z)dz\:,\quad x\in\R\:. 
\end{displaymath}
One can show that \hbox{$f*g$} has a unique analytic continuation to complex arguments, 
which is regular on~$\R$, meromorphic on~$\C$, and which has period one.
Thus iterated convolutions can be considered.
The case of interest to us is when $f$ and $g$ equal the same function, or an iterated convolution of it.
For $k\geq 1$, we also write \hbox{$f*\ldots *f$} ($k$ factors) for the $k$-fold convolution of $f$, and we set $f^{*0}=1$.
For example, $(1*\wp)$ is given by eq.~(\ref{eq: period integral over Weierstrass p}).

\begin{proposition}\label{propos: the modular forms coming from loops in the graphical proposition}
Let \hbox{$\langle 1\rangle$} be a holomorphic block in a sector of a CFT on the torus with central charge $c$.
Let \hbox{$\langle\Phi\rangle$} be the~one-point function of a primary field~$\Phi$ of holomorphic weight~$h$ 
in the same sector.
For \hbox{$N\geq 1$}, let \hbox{$S_{\diamondsuit,N}\subset S^{[1]}_N$} and \hbox{$\tilde{S}_{\diamondsuit,N}\subset\tilde{S}^{[1]}_N$}, 
respectively, 
be the set of graphs whose connected components are all either isolated points, or loops.
We have
\begin{displaymath}
(2\pi i)^N\frac{d^N}{d\tau^N}\langle 1\rangle
-\oint\ldots\oint
\sum_{\Gamma\in S_{\diamondsuit,N}\setminus\Gamma_0^N}{\gamma}(\Gamma)\,dz_2\ldots dz_N
\;=\;{C}_{2N,c}
+\text{O($E_2$)}
\:
\end{displaymath}
and
\begin{displaymath}
(2\pi i)^N\frac{d^N}{d\tau^N}\langle\Phi\rangle
-\oint\ldots\oint
\sum_{\Gamma\in\tilde{S}_{\diamondsuit,N}\setminus\Gamma_0^N}\widetilde{{\gamma}}(\Gamma)\,dz_1\ldots dz_N
\;=\;\tC_{2N,c}
+\text{O($E_2$)}
\:,
\end{displaymath} 
respectively, where integration is performed along the real period.
\end{proposition}

Note that the meaning of ``$\text{O($E_2$)}$'' here is 
that the l.h.s.\ in each case is a quasimodular form (of weight~\hbox{$2N$}) which, 
when it is expressed as a polynomial in~$E_2$ with modular coefficients, 
has the constant term ${C}_{2N,c}$ or $\tC_{2N,c}$, respectively. 
Note that this determines the modular forms ${C}_{2N,c}$ and $\tC_{2N,c}$ uniquely.

\begin{proof}
In the graphical representation of \hbox{$\bigl\langle T(z_1)\ldots T(z_N)\bigr\rangle$} 
and of \hbox{$\bigl\langle T(z_1)\ldots T(z_N)\,\Phi(z_0)\bigr\rangle$},
for \hbox{$0\leq k\leq N$},
the iterated period integral over a connected component of~$k$ edges in a graph is proportional to the period integral over \hbox{$\wp^{*k}(z)$}.
For all graphs other than~$\Gamma^N_0$, we have \hbox{$k\geq 1$},
and for \hbox{$z\not=0$},
\begin{displaymath}
\oint\wp^{*k}(z)\,dz
\;=\;(1*\wp)^k
\:,
\end{displaymath}
where \hbox{$(1*\wp)$} is proportional to \hbox{$E_2(\tau)$} by eq.~(\ref{eq: period integral over Weierstrass p}).
Thus the only relevant convolutions are those at \hbox{$z=0$}. These correspond precisely to loops.  
\end{proof}

It turns out that it is useful to work with Eisenstein's zeta function \cite{E:1847,W:1976}. 
In modern terminology, it is defined as the modified Weierstrass zeta function
\begin{displaymath}
\mathcal{Z}(z)
\;:=\;\zeta(z)+z\,(1*\wp)
\:.
\end{displaymath}
For \hbox{$m,n\in\Z$}, the zeta function satisfies \cite{AS:1965}
\begin{equation*}
\zeta(z+m+n\tau)-\zeta(z)
\;=\;2m\zeta(1/2)+2n\:\zeta(\tau/2)
\:,
\end{equation*}
and Legendre's relation (together with eq.~(\ref{eq: period integral over Weierstrass p})) yields 
\begin{displaymath}
\mathcal{Z}(z+m+n\tau)-\mathcal{Z}(z)
\;=\;-2n\pi i
\:.
\end{displaymath}
Thus $1*\mathcal{Z}=-\pi i$.
Under differentiation, the convolution behaves according to $(f*g)'=f'*g=f*g'$,
and we have 
\begin{equation}\label{eq: relating convolutions of Weierstrass p to those of Eisenstein Z}
\wp^{*k}(z)
\;=\;(-1)^k\frac{d^n}{dz^k}\mathcal{Z}^{*k}(z)+(1*\wp)^k
\:.
\end{equation}
For convenience of the reader, we list the first few relevant terms:
\begin{equation*}
\begin{split}
\mathcal{Z}*\mathcal{Z}
\;=&\;-\frac{1}{2}\mathcal{Z}^2+\frac{1}{2}\,\wp+\frac{2\pi^2}{3}
\:,\\
\mathcal{Z}*\mathcal{Z}*\mathcal{Z}
\;=&\;\frac{1}{6}\mathcal{Z}^3-\frac{\pi i}{2}\mathcal{Z}^2-\frac{1}{2}\mathcal{Z}\,\wp-\frac{1}{6}\wp'+\frac{\pi i}{2}\wp
\:.
\end{split}
\end{equation*}
Eq.~(\ref{eq: relating convolutions of Weierstrass p to those of Eisenstein Z}) 
keeps being true if we replace the operation $*$ by a closely related commutative convolution $\ostar$,
for which iterated convolutions of $\mathcal{Z}$ are machine computable. 
$\wp^{*k}$ differs from $\wp^{\ostar k}$ by a residue \cite[cf.~Proposition 5]{L:2020}.


\begin{example}\label{example: modular coefficients}
We address the modular forms ${C}_{2N,c}$ and $\tC_{2N,c}$ 
from Proposition \ref{propos: the modular forms coming from loops in the graphical proposition}.
We have ${C}_{0,c}=1$ and ${C}_{2,c}=t$, where $t:=\langle 1\rangle^{-1}\langle T\rangle=d\log\langle 1\rangle/d(2\pi i\tau)$. 
For $N\leq 3$, the computation of at most $3$-fold convolutions of $\wp$ are required.
In the \hbox{$(2,5)$} minimal model,
earlier results \cite[and references therein]{L:PhD14} include
\begin{equation}\label{eq: <T(z)T(0)> for g=1 in the (2,5) minimal model}
\bigl\langle T(z)T(0)\bigl\rangle
\;=\;\frac{c}{2}\,\wp(z)^2\langle 1\rangle
+2\,\wp(z)\,\langle T\rangle
-6c\,G_4\,\langle 1\rangle
\:.
\end{equation}
Thus ${C}_{4,c}=-6c\,G_4$, in accordance with Proposition~\ref{propos: the modular forms coming from loops in the graphical proposition}. Moreover, we have
\begin{equation*}
\begin{split}
{C}_{6,c}
\;=&\;
-\frac{84}{5}\,\biggl(5c\,G_6-3\,G_4\,t\biggr)\\
{C}_{8,c}
\;=&\;
-\frac{16}{5}\,\biggl(367c\,G_4^2-336\,G_6\,t\biggr)
\:.
\end{split}
\end{equation*}
In Theorem~\ref{proposition: graphical representation of (N+1)-point functions of N copies of T and one copy of Phi},
\hbox{$\tC_{0,c}=1$}, \hbox{$\tC_{2,c}=0$},
and 
\begin{equation*}
\tC_{4,c}
\;=\;24\,G_4
\:,\hspace{1cm}
\tC_{6,c}
\;=\;336\,G_6\:.
\end{equation*}
Note that this recovers the modular form $\tC_{4,c}$ from Corollary \ref{corollary: 2-pt fct of T with Phi}.
\end{example}

\section{The genus \texorpdfstring{$2$}{Lg} partition function}\label{sec:2}

\subsection{The holomorphic blocks}\label{subsec: The holomorphic blocks}

The $(2,5)$ minimal model has the two sectors 
$ R_2(\mathfrak{V})$ and $R_2(\mathfrak{W})$.
In order to calculate, for $g\geq 1$, the partition function $\mathfrak{Z}_{\G}^g$ for $(\Sigma_g,\G)\in\mathscr{X}$,
one proceeds as follows:
\begin{enumerate}
 \item\label{item: cutting along g-1 homologically trivial cycles}  
 Cut $\Sigma_g$ along $g-1$ homologically trivial cycles $\gamma^s_i$, $i=1,\ldots,g-1$,
 so that $\Sigma_g$ decomposes into $g$ tori with boundary $\gamma^s_{i-1}\dot\cup\gamma^s_i$, 
where $\gamma^s_0$ and $\gamma^s_g$ are empty.
 \item\label{item: cutting along g homologically non-trivial cycles} 
 Cut the $i$th component ($i=1,\ldots,g$) along a homologically non-trivial cycle $\gamma^q_i$.
 This reduces its genus to zero. 
 \item\label{item: cutting along g-2 homologically non-trivial cycles} 
 For $i=2,\ldots,g-1$, cut once more along a separating cycle $\gamma^t_i$
 so that the connected components (pants) all have three boundary circles.
\end{enumerate}
Altogether we have cut along $3(g-1)$ cycles and obtained $2(g-1)$ pants (if $g\geq 2$). 
We encode the cutting scheme in the vector
\begin{equation}\label{expression: cutting scheme vector}
(\gamma^s_1,\ldots,\gamma^s_{g-1};\gamma^q_1,\ldots,\gamma^q_g;\gamma^t_1,\dots,\gamma^t_{g-2}) 
\end{equation}
whose components will be referred to as cutting cycles.
The calculations are performed by using the corresponding combination of ${s}$ and ${q}$ expansions.

Thus an $N$-point function on $M\in\mathscr{X}$ can be defined as a sum of products of three-point functions on a sphere.
The result does not depend on the cutting scheme
provided that the four-point functions on a sphere and the one-point functions on a torus are globally defined \cite{Sonoda:1988,Moore-Seiber:1989}.
It suffices to check this for the primary fields, and thus, in the $(2,5)$ minimal model, for the field $\Phi$.
For $\langle\Phi\rangle^{g=1}$, eq.~(\ref{eq: 1-point function of Phi}) is sufficient.
Moreover, $\langle\Phi\Phi\Phi\Phi\rangle^{g=0}$ can be explicitly calculated:
Here is a short derivation.
A comprehensive discussion is contained in \cite{Dotsenko:1988}.

$\langle T(z)\Phi(z_1)\ldots\Phi(z_N)\rangle^{g=0}_{\G}$ is a meromorphic function of $z$ 
whose poles are prescribed by the OPE (\ref{OPE of T and Phi}) for each of copy of $\Phi$.
By the Liouville proposition, for $N\geq 0$, we can write
\begin{equation}\label{eq: Mittag-Leffler for the 5-point function of T with 4 Phi, where the last is positioned in a different coordinate chart}
\begin{split}
\langle T(z)\Phi(z_1)\ldots\Phi(z_N)\rangle^{g=0}_{\G}
\;=&\;\langle\Phi(z_1)\ldots\Phi(z_N)\rangle^{g=0}_{\G}\,\sum_{i=1}^N\frac{-1/5}{(z-z_i)^2}\\
&+\sum_{i=1}^N\frac{1}{z-z_i}\,\frac{\partial}{\partial z_i}\,\langle\Phi(z_1)\ldots\Phi(z_N)\rangle^{g=0}_{\G}\:.
\end{split}
\end{equation}
According to eq.~(\ref{eq: transformation behaviour of T undr linear fractional transformation}),
for $z\rechts\infty$, \hbox{$\langle T(z)\Phi(z_1)\ldots\Phi(z_N)\rangle^{g=0}_{\G}=\text{O$(1/z^4)$}$}.
Thus there are no additional regular terms in eq.~(\ref{eq: Mittag-Leffler for the 5-point function of T with 4 Phi, where the last is positioned in a different coordinate chart}).
Moreover, the coefficient of $1/z^k$ for $k=1,2,3$ 
in eq.~(\ref{eq: Mittag-Leffler for the 5-point function of T with 4 Phi, where the last is positioned in a different coordinate chart}) must vanish,
\begin{equation*}
\sum_{i=1}^N
\biggl(
z_i^{k-1}\,\frac{\partial}{\partial z_i}
-\frac{1}{5}\,z_i^{k-2}(k-1)
\biggr)\langle\Phi(z_1)\ldots\Phi(z_N)\rangle^{g=0}_{\G}
\;=\;0\:,
\hspace{0.5cm} 
k=1,2,3\:.
\end{equation*}
Equivalently, $\langle\Phi(z_1)\ldots\Phi(z_4)\rangle^{g=0}_{\G}$ 
is invariant under translations, dilatations, and special conformal transformations.
Up to overall normalisation, 
this fully determines the solution for $N\leq 3$ and reduces the number of arguments from $N$ to $N-3$ otherwise.
For $N=2$, 
\begin{equation}\label{eq: 2-point function of Phi}
\langle\Phi(z)\,\Phi(0)\rangle^{g=0}_{\G}
\;=\;\eps_{\Phi}\,|z|^{4/5}\,\mathfrak{Z}^{g=0}_{\G}
\:,
\end{equation}
for the normalisation from eq.~(\ref{eq: normalisation of Phi}).
For $N=3$, 
\begin{equation}\label{eq: 3-point function of Phi}
\langle\Phi(z_1)\,\Phi(z_2)\,\Phi(z_3)\rangle^{g=0}_{\G}
\;=\;\lambda_{\Phi\Phi\Phi}\,|z_{12}\,z_{23}\,z_{31}|^{2/5}\mathfrak{Z}^{g=0}_{\G}
\:
\end{equation}
for some $\lambda_{\Phi\Phi\Phi}\in\R^*$. 
For $N=4$,
\begin{equation*}
\langle\Phi(z_1)\Phi(z_2)\Phi(z_3)\Phi(z_4)\rangle^{g=0}_{\G}
\;=\;\bigl(z_{32}\,z_{14}\bigr)^{2/5}\,f\Bigl(\frac{z_{12}}{z_{32}}\frac{z_{34}}{z_{14}}\Bigr)\,\mathfrak{Z}^{g=0}_{\G}
\:,
\end{equation*}
for some function $f$, which determine as follows:
By the OPE (\ref{OPE of T and Phi}) for $T$ and $\Phi$ about $z=z_1$,
\begin{equation}\label{eq: OPE for the 5-point function of T with 4 Phi}
\begin{split}
\langle T(z)\,\Phi(z_1)\ldots\Phi(z_N)\rangle^{g=0}_{\G}
\;=&\;\frac{-1/5}{(z-z_1)^2}\,\langle\Phi(z_1)\Phi(z_2)\ldots\Phi(z_N)\rangle^{g=0}_{\G}\\
&+\frac{1}{z-z_1}\,\langle\Phi'(z_1)\Phi(z_2)\ldots\Phi(z_N)\rangle^{g=0}_{\G}\\
&+\frac{5}{2}\,\langle\Phi''(z_1)\Phi(z_2)\ldots\Phi(z_N)\rangle^{g=0}_{\G}
+\text{O$(z-z_1)$}
\:.
\end{split}
\end{equation}
Eqs~(\ref{eq: Mittag-Leffler for the 5-point function of T with 4 Phi, where the last is positioned in a different coordinate chart})
and (\ref{eq: OPE for the 5-point function of T with 4 Phi}) together 
yield a 2\ts{nd} order differential equation for $\langle\Phi(z_1)\ldots\Phi(z_N)\rangle^{g=0}_{\G}$,
namely
\begin{equation}\label{eq: 2nd order ODE for N-pt fct of Phi}
\Bigl(\frac{\partial^2}{\partial z_1^2}
-\frac{2}{5}\,\sum_{i=2}^N\frac{1}{z_1-z_i}\,\frac{\partial}{\partial z_i} 
+\frac{2}{25}\,\sum_{i=2}^N\frac{1}{(z_1-z_i)^2}\Bigr)\langle\Phi(z_1)\ldots\Phi(z_N)\rangle^{g=0}_{\G}
\;=\;0
\:.
\end{equation}
(A correspoding ODE holds for the anti-holomorphic coordinate.)

For $N=4$, $(z_1,z_2,z_3)=(z,0,1)$ and $z_4\rechts\infty$, 
\begin{equation}\label{eq: 4-pt fct of Phi identificed with fct f}
\langle\Phi(z)\Phi(0)\Phi(1)\widetilde{\Phi}(0)\rangle^{g=0}_{\G}
\;=\;f(z)\,\mathfrak{Z}^{g=0}_{\G} 
\:,
\end{equation}
where $\widetilde{\Phi}$ denotes the field in the inverse coordinate.
Eq.~(\ref{eq: 2nd order ODE for N-pt fct of Phi}) is equivalent to a hypergeometric ODE for the function $f$.
Using the Frobenius ansatz $f(z)=|z\,(1-z)|^{4/5}\,\mathcal{F}(z)$, 
the ODE takes the standard form
\begin{align*}
z(1-z)\,{\mathcal{F}}''(z) 
+\frac{6}{5}\bigl(1-2z\bigr)\,{\mathcal{F}}'(z)
-\frac{12}{25}\,{\mathcal{F}}(z)
\;=\;0\:.
\end{align*}
Two fundamental solutions are \hbox{${\mathcal{F}}_1(z)=\, _2F_1\Bigl(\frac{4}{5},\frac{3}{5};\frac{6}{5};z\Bigr)$} and 
\hbox{${\mathcal{F}}_2(z)=z^{-1/5}\, _2F_1\Bigl(\frac{3}{5},\frac{2}{5};\frac{4}{5};z\Bigr)$}.
Since 
$f$ is real,
the full solution reads
\begin{equation*}
\mathcal{F}(z)
\;=\;{m}_1\,|{\mathcal{F}}_1(z)|^2+{m}_2\,|{\mathcal{F}}_2(z)|^2
\end{equation*}
for $m_1,m_2\in\R^*$.
The condition $\mathcal{F}(z)=\mathcal{F}(1-z)$ yields $m_2/m_1=\mathfrak{C}$, 
where 
\begin{equation}\label{eq: Cardy's constant}
\mathfrak{C}
\;=\;-\,\frac{\Gamma(6/5)^2}{\Gamma(4/5)^3}\frac{\Gamma(1/5)\,\Gamma(2/5)}{\Gamma(3/5)}
\;\approx\;-3.653116237
\:
\end{equation}
is Cardy's constant \cite{C:1985}.
In order to fix the constant $m_1$, we apply the ${s}$ formalism to \hbox{$(\CP^1,\G)=M_L\sqcup_{\gamma}M_R$}.
Thus by eqs (\ref{eq: Segal's s formula}) and (\ref{eq: 4-pt fct of Phi identificed with fct f}),
\begin{equation*}
\begin{split}
f(z)\,\mathfrak{Z}^{g=0}_{\G}
\;=\;
&\;\langle\Phi(z)\,\Phi(0)\rangle_{M_L\sqcup_{\gamma}D_R}\,\langle\Phi(1)\,\widetilde{\Phi}(0)\rangle_{D_L\sqcup_{\gamma}M_R}
+\ldots\\
+&\eps_{\Phi}\,\langle\Phi(z)\,\Phi(0)\Phi(z_R=0)\rangle_{M_L\sqcup_{\gamma}D_R}\,\langle\Phi(z_L=0)\Phi(1)\,\widetilde{\Phi}(0)\rangle_{D_L\sqcup_{\gamma}M_R}+\ldots,
\end{split}
\end{equation*}
where the summand in the first line corresponds to $1\in R_2(\mathfrak{V})$ and that in the second to $\Phi\in R_2(\mathfrak{W})$.
By eq.~(\ref{eq: 2-point function of Phi}) and eq.~(\ref{eq: 3-point function of Phi}), 
the lowest order term in $z$, for $z$ close to zero,
equals $\eps_{\Phi}^2|z|^{4/5}$ in the first and $\eps_{\Phi}\lambda_{\Phi\Phi\Phi}^2\,|z|^{2/5}$ in the second line, respectively.
By comparison with 
\begin{equation*}
f(z)
\;=\;|z\,(1-z)|^{4/5}\,{m}_1\biggl(\Bigl|\, _2F_1\Bigl(\frac{4}{5},\frac{3}{5};\frac{6}{5};z\Bigr)\Bigr|^2
+|z|^{-2/5}\,\mathfrak{C}\,\Bigl|\, _2F_1\Bigl(\frac{3}{5},\frac{2}{5};\frac{4}{5};z\Bigr)\Bigr|^2\biggr)
\:,
\end{equation*}
the first and the second summand 
correspond to the fields in $R_2(\mathfrak{V})$ and in $R_2(\mathfrak{W})$, respectively.
In particular, $m_1=1$ and $\mathfrak{C}=\eps_{\Phi}\lambda_{\Phi\Phi\Phi}^2$.
This implies 
\begin{equation}\label{eq: eps Phi}
\eps_{\Phi}
\;=\;-1
\:.
\end{equation}
Note that in Cardy's paper, $\Phi$ is an imaginary field and its squared norm is positive.

\begin{proposition}\label{proposition: the normalisation constant of Phi}
The positive normalisation factor of $\langle{\Phi}(z)\rangle\flatz^{g=1}$ in eq.~(\ref{eq: 1-point function of Phi}) 
equals
\begin{equation*}
{\mu}_{\Phi}\,
\;=\;\sqrt{|\mathfrak{C}|} 
\:,
\end{equation*}
where $\mathfrak{C}$ is Cardy's constant (\ref{eq: Cardy's constant}).
Moreover,
\begin{equation*}
\langle\Phi(z_L=0)\,\Phi(z=1)\,\Phi(z_R=0)\rangle_{\DD}
\;=\;-\sqrt{|\mathfrak{C}|}
\:.
\end{equation*}
\end{proposition}

\begin{proof}
By translational symmetry, 
we may set $z=1$.
In the ${q}$ formalism we let the flat torus of modulus $\tau=\log q/(2\pi i)$ for $0<|q|\ll 1$
degenerate into the cylinder $(\gamma\times[0,\ell])^{\theta}$ for $\ell=-\log|q|$ and $\theta=\arg(q)$.
According to eq.~(\ref{eq: N-point function for genus g in the q-formalism for arbitrary metrics}) 
for the single field $\Phi$ of weight $(h,\bar{h})=(-1/5,-1/5)$,
\begin{equation*}
\langle\Phi(z=1)\rangle^{g=1}\flatz
\;=\;
-|q|^{-c/12}\,|q|^{-2/5}\langle\Phi(z_L=0)\,\Phi(z=1)\,\Phi(z_R=0)\rangle^{g=0}_{\DD}
+\ldots
\:
\end{equation*}
Note that all contributions from the vacuum sector vanish since \hbox{$\langle\Phi(z)\rangle^{g=0}=0$}.
In the three-point function,
we perform the coordinate change $z_3=1/z_R$.
Thus
\begin{align*}
&\langle\Phi(z_L=0)\,\Phi(z=1)\,\Phi(z_R=0)\rangle_{\DD}\\
&\hspace{0.3cm}\;=\;\lim_{z_3\rechts\infty}|z_3|^{-4/5}\langle\Phi(z_L=0)\,\Phi(z=1)\,\widetilde{\Phi}(z_3)\rangle_{\DD}
\;=\;\lambda_{\Phi\Phi\Phi}
\:,
\end{align*}
by eqs~(\ref{eq: 3-point function of Phi}) and (\ref{eq: normalisation of partition function of lense space}).
Thus for $c=-22/5$,
\begin{equation*}
\langle\Phi(z=1)\rangle\flatz^{g=1}
\;=\;\;-\lambda_{\Phi\Phi\Phi}\,|q|^{-1/30}+\ldots
\end{equation*}
Comparison with eq.~(\ref{eq: 1-point function of Phi}) 
yields ${\mu}_{\Phi}\,=-\lambda_{\Phi\Phi\Phi}$,
as required.
\end{proof}

Because of Proposition \ref{proposition: the normalisation constant of Phi},
the partition function $\mathfrak{Z}_M$ is computable for $M\in\mathscr{X}$ of arbitrary genus,
in terms of an iterated infinite series.

Let $\mathfrak{A}$ the set of all maps from the cutting scheme (\ref{expression: cutting scheme vector}) into the set $\{\mathfrak{v},\mathfrak{w}\}$. 
Thus
an element $A\in\mathfrak{A}$ is an assignment of one of $\mathfrak{v}$ and $\mathfrak{w}$
to each component in (\ref{expression: cutting scheme vector}).
Thus 
$A$ specifies for each individual cutting cycle whether a pair of discs endowed with fields 
$e_i\in\mathfrak{V}\otimes\overline{\mathfrak{V}}$ 
or $e_j\in\mathfrak{W}\otimes\overline{\mathfrak{W}}$ is inserted.
Write
\begin{equation*}
\mathfrak{Z}_M
\;=\;\sum_{A\in\mathfrak{A}}\mathfrak{Z}_{A,M}
\:
\end{equation*}
for the corresponding decomposition.
The tensor product structure of the two sectors implies that
\begin{equation}\label{eq: partition function split into holomorphic and antiholomorphic part}
\mathfrak{Z}_{A,M}
\;=\;\eps_A\,{\mathcal{B}}_{A,M}\,\overline{\mathcal{B}_{A,M}}
\end{equation}
where ${\mathcal{B}}_{A,M}$ a holomorphic function and $\overline{\mathcal{B}_{A,M}}$ is its complex conjugate. 
The factor $\eps_A$ is a sign, which is due to 
\hbox{$\langle\overline{\mathfrak{w}}|\overline{\mathfrak{w}}\rangle=\eps_{\Phi}\,\langle\mathfrak{w}|\mathfrak{w}\rangle$}.

For $A\in\mathfrak{A}$ and $0\leq j\leq 3$, 
let $\pa_j(A):=$ the number of pants with exactly $j$ boundary cycles to which $\mathfrak{w}$ is assigned.
If $\pa_1(A)>0$ then ${\mathcal{B}}_{A,M}=0$, since \hbox{$\langle\Phi\rangle^{g=0}=0$}, for every metric.
(Also the correlation functions of all other fields in the non-vacuum sector are zero.)
Thus we exclude the corresponding assignment from $\mathfrak{A}$.

\begin{proposition}\label{proposition: the prefactor eps A}
In eq.~(\ref{eq: partition function split into holomorphic and antiholomorphic part}),
for $A\in\mathfrak{A}$,
\begin{equation*}
\eps_A
\;=\;{\eps_{\Phi}}^{\pa_2(A)-\cc(A)}
\end{equation*}
where $\cc(A):=$ number of cutting cycles that are assigned $\mathfrak{w}$ by $A$.
\end{proposition}

\begin{proof}
According to Segal's formulae for the partition function in the ${s}$ and the ${q}$ formalism,
the (nonzero) factor ${\mathcal{B}}_{A,M}$ is a product of 
three-point functions $\langle\psi_i\psi_j\psi_k\rangle^{g=0}$ and inverses of $\langle\psi_i\psi_i\rangle_{\DD}$
for fields $\psi_i,\psi_j,\psi_k\in R_2(\mathfrak{V})\oplus R_2(\mathfrak{W})$.
More precisely, for $m=0,2,3$,
$\pa_m(A)$ three-point functions have $m$ fields in $ R_2(\mathfrak{W})$ 
and $(3-m)$ fields in $ R_2(\mathfrak{V})$.
$\cc(A)$ factors $\langle\psi_i\psi_i\rangle_{\DD}$ correspond to $\psi_i\in R_2(\mathfrak{W})$.
The remaining $(3g-3-\cc(A))$ factors correspond to $\psi_i\in R_2(\mathfrak{V})$.
Thus we have a decomposition
Since  $\pa_1(A)=0$, $\pa_3(A)$ is even by construction.
Thus the numerator contributes $\eps_{\Phi}^{\pa_2(A)}$, while the denominator contributes $\eps_{\Phi}^{-\cc(A)}$,
as required.
\end{proof}

Different choices of cutting cycles induce unitary transformations of $({\mathcal{B}}_{A,M})_{A\in\mathfrak{A}}$.
A convenient notation for ${\mathcal{B}}_{A,M}$ is $\langle 1\rangle_{A,M}$ or $\langle 1\rangle_{A,\G}^g$
(in accordance with $\mathfrak{Z}_M$ and $\mathfrak{Z}_{\G}^g$), 
and it is called a holomorphic (or conformal) block (associated to $A\in\mathfrak{A}$) on $M$.
We also use the word zero-point function.

The cardinality $|\mathfrak{A}|$ of $\mathfrak{A}$
equals one for $g=0$, two for $g=1$, and $|\mathfrak{A}|\leq 2^{3g-3}$ for $g\geq 2$.
For hyperelliptic $M\in\mathscr{X}$, 
$|\mathfrak{A}|$ is a Fibonacci number \cite{LN:2017}. 
Here is the example we will use later:

\begin{proposition}\label{proposition: in genus g=2, there are 5 holomorphic blocks} 
In the $(2,5)$ minimal model for genus two, $|\mathfrak{A}|=5$.
\end{proposition}

\begin{proof}
The cutting scheme for $g=2$ results in two pants.
When $A\in\mathfrak{A}$ assigns to $\gamma^s$ the value $\mathfrak{v}$, then $\pa_0(A),\pa_2(A)\in\{0,1,2\}$ and $\pa_0(A)+\pa_2(A)=2$.
In the case where $\pa_0(A)=1=\pa_2(A)$, there is a choice for which of the pants has no $\mathfrak{w}$ assignment, while the other has two.
When the value $\mathfrak{w}$ is assigned to $\gamma^s$, only $p_3(A)$ does not vanish, so $p_3(A)=2$.
(The nonzero elements are listed in Table \ref{table: holomorphic blocks for g=2}.)
\end{proof}

$N$-point functions can be treated in the same way.
One can use additional circles with ${s}$ expansions 
to locate each field insertion in a separate component with two boundary circles only.

Since $T$ commutes with anti-holomorphic derivatives, 
application of $T(z)$ to the partition function in eq.~(\ref{eq: partition function split into holomorphic and antiholomorphic part})
yields an expression of the form
\begin{equation*}
T(z)\,\mathfrak{Z}_M
\;=\;\sum_{A\in\mathfrak{A}}\langle T(z)\rangle_{A,M}\,\eps_A\,\,\overline{\mathcal{B}_{A,M}}
\:.
\end{equation*}
For $A\in\mathfrak{A}$, 
this defines the corresponding $1$-point function $\langle T(z)\rangle_{A,M}$ on $M$.
Successive application of $T(z_1)\ldots T(z_N)$ to $\mathfrak{Z}_M$ gives rise to Virasoro $N$-point (or correlation) functions $\langle T(z_1)\ldots T(z_N)\rangle_{A,M}$
provided that $z_i\not=z_j$ for $1\leq i<j\leq N$.
The index $A$ will be omitted whenever possible.

\subsection{Degenerating hyperelliptic Riemann surfaces}\label{subsection: Hyperelliptic Riemann surfaces}

For $k\geq 1$ and for \hbox{$0\leq{g}\leq k-1$}, 
let \hbox{$(\CP^1,P_1,P_{i_1},\ldots,P_{i_{2{g}}},P_{2k})$} be the Riemann sphere
marked with $2g+2$ distinct points \hbox{$P_1,P_{i_1},\ldots,P_{i_{2{g}}},P_{2k}$},
and let 
\begin{equation*}
\Sigma_g=\Sigma(P_1,P_{i_1},\ldots,P_{i_{2{g}}},P_{2k}):\quad y^2=p(x) 
\end{equation*}
be its double cover, which is ramified precisely at these ${2{g}}+2$ marked points, 
namely the roots of the polynomial $p(x)$. 
A metric $\G_g$ on $\Sigma_g$ will be specified later.
For the purpose of this paper, $k=3$ and ${g}\in\{0,1,2\}$.
We shall give a perturbative expansion around the complex structure at which $\Sigma(P_1,\ldots,P_6)$ degenerates.
This is achieved by pinching a suitably chosen cycle $\gamma\subset\Sigma_2$, 
or by letting $m$ ramification points run together, where $m=2$ or $m=3$.
The parameter describing the degeneration can be introduced through the following procedure:
\begin{enumerate}
\item\label{item: cutting} 
Cut along the cycle $\gamma\subset\Sigma_2$ 
such that $\gamma$ is the inverse image of the curve with the equation $|x|=\const$ on $\CP^1$ 
that encloses precisely $m$ ramification points $P_1,\ldots,P_m$. 
\item 
Replace $x(P_i)$ for $i\leq m$ by $s x(P_i)$ for $0<|s|\ll 1$.
This is equivalent to cutting along $\gamma$ and inserting a cylinder with parameter $\log s=i\theta-\ell$.
\end{enumerate}
The perturbative expansion is a power series in $s$.

When $\gamma$ is homologically trivial (though nontrivial in the fundamental group), which is the case discussed in Section \ref{subsection: s formalism},
$m$ equals $3$, and the cutting results in two separate tori 
with a disc centred at $\infty$ and $0$, respectively, removed. 
When $\gamma$ is non-homologous to zero, which is the case addressed in Section \ref{subsection: q formalism}, 
$m=2$, and the cutting results in a single torus with two discs removed
and a cylinder. 
To distinguish the two cases
we shall refer to the first and second case as the ${q}$ and the ${s}$ formalism, respectively,
though both refer to the same geometric quantities $\ell,\theta$.
In the case of the ${q}$-formalism, the perturbation parameter will be named accordingly.

\subsection{The \texorpdfstring{$s$}{Lg} formalism in genus \texorpdfstring{$g=2$}{Lg} }\label{subsection: s formalism}

Let $(\Sigma_1,P)$ and $(\widehat{\Sigma}_1,\widehat{P})$ be two flat tori with a single puncture
and with holomorphic one-forms $dz$ and $d\hat{z}$,
normalised by period one along some cycle.
Let $D_L$ and $\widehat{D}_R$ be the flat unit disc centred at $P_5$ and $P_0$, respectively,
and containing no other ramification point.
Let $z$ and $\hat{z}$ be a corresponding pair of coordinates with $z(P)=0$ and $\hat{z}(\widehat{P})=0$.
By eq.~(\ref{eq: behaviour of partition fct under rescaling of metric for arbitrary g}),
we may w.l.o.g.~assume 
that the two discs lie inside the respective fundamental cell of the two tori.
If $M_L^{(1)}\subset\Sigma_1$ and $\widehat{M}_R^{(2)}\subset\widehat{\Sigma}_1$ are the left and right manifolds 
for $\gamma_1$ and $\gamma_2$, respectively, 
the connected sum
\begin{equation}\label{description: genus two Riemann surface with (unspecified) metric}
M_2({s})
:\;=\;M_L^{(1)}\sqcup_{\gamma_1}(\gamma\times[0,\ell])^{\theta}\sqcup_{\gamma_2}\widehat{M}_R^{(2)}
\end{equation}
along $\gamma_1$ and $\gamma_2$ defines an element of genus two in $\mathscr{X}$ (with the induced metric).
Here ${s}=\exp(i\theta-\ell)$.
We continue the coordinates $z$ and $\hat{z}$ to the cylinder so that
\begin{equation}\label{eq: z1z2=eps}
z\hat{z}
\;=\;{s}
\:.
\end{equation}
Since the sewing is compatible with the involutions $z\mapsto-z$, $\hat{z}\mapsto-\hat{z}$, 
the corresponding $\CP^1$ projections of the two tori are sewn and form a new $\CP^1$.
On this space, we obtain a pair of almost global coordinates:

\begin{proposition}\label{proposition: almost global coordinates on CP1 satisfying the epsilon condition}
Let $\Sigma_1,\widehat{\Sigma}_1$ be flat tori of modulus $\tau,\hat{\tau}\in\mathfrak{H}_1$ 
with a single puncture, and let $z$ and $\hat{z}$ be a corresponding pair of local analytic coordinates 
vanishing at the respective puncture and satisfying the sewing condition (\ref{eq: z1z2=eps}) as stated above.
There exists a triple $(X, \widehat{X},\xi)$ satisfying the following properties:

$X$ and $\widehat{X}$ define a pair of coordinates on $\CP^1$, 
such that $X$ is defined on $\CP^1\setminus\{\infty\}$, $\widehat{X}$ is defined on $\CP^1\setminus\{0\}$ 
and 
\begin{equation}\label{eq: X hat X= xi/(s squared)}
X\widehat{X}
={\xi}/{{s}^2}
\:
\end{equation}
holds on $\CP^1\setminus\{\infty,0\}$. 
Here ${\xi}\in\Q\bigl[G_4,\,\widehat{G_4},\,G_6,\,\widehat{G_6}\bigr][[{s}^2]]$
has an expansion for $0<|{s}|\ll 1$ which starts as follows:
\begin{align*}
\frac{{\xi}}{{s}^2}
\,=&\,
\frac{1}{s^2}
-144\,s^2\,G_4\widehat{G_4}
-320\,s^4\,G_6\widehat{G_6}
+18\,720\,G_4^2\widehat{G_4}^2\,s^6\\
&+\frac{10\,002\,240}{121}\,G_4G_6\,\widehat{G_4}\widehat{G_6}\,s^8
+O({s}^{10})
\:.
\end{align*}
Moreover, for
$x
=\wp(z\,|\,\tau)$
and 
$\hat{x}
=\wp(\hat{z}\,|\,\hat{\tau})$,
the terms of low order in ${s}$ read
\begin{align*}
X(x)
\;=\;x
\,\Bigl(
&1-6\,\widehat{G_4}\,s^4x^2
-10\,\widehat{G_6}\,s^6x\left(x^2-30\,G_4\right)\\
&+24\,\widehat{G_4}^2\,s^8x\left(x^3+9\,G_4\,x-25\,G_6\right)\\
&+\frac{60}{11}\,\widehat{G_4}\widehat{G_6}\,
s^{10}x 
\left(19\,x^4-402\,G_4x^2-200\,G_6x-3060\,G_4^2\right)\\
&-\frac{216}{143}\,
\widehat{G_4}^3s^{12}x 
\Bigl(
55\,x^5
-66\,G_4x^3
-4950\,G_6x^2
+33308\,G_4^2x
+71100\,G_4 G_6
\Bigr)\\
&-\frac{600}{143}\,
\widehat{G_6}^2
s^{12}x\Bigl(
-22x^5
+913\,G_4x^3
-165\,G_6 x^2
-9834\,G_4^2 x
+19790\,G_4G_6
\Bigr)\\
&+O({s}^{14})
\Bigr)
\:,
\end{align*}
resp.~$\widehat{X}(\hat{x})$ is obtained by replacing, in the above formula, 
$x$ by $\hat{x}$ and by exchanging $G_{2k}\leftrightarrow\widehat{G}_{2k}$ 
at every occurrence.
Here and in the following, 
\hbox{$G_{2k}=G_{2k}(\tau)$} and \hbox{$\widehat{G_{2k}}=G_{2k}(\hat{\tau})$} and likewise for other modular forms.
\end{proposition}


\begin{proof}
In a tubular neighbourhood of 
\hbox{$\{|z|=|\hat{z}|\}$},
which in this proof we refer to as the annulus, define a function $f_{{s}}$ by
\begin{equation*}
f_{{s}}(x)
:=\log x+\log\hat{x}
\:. 
\end{equation*}
By eq.~(\ref{eq: z1z2=eps}) for $0<{s}<1$,
\begin{equation*}
f_{{s}}(x)
=\log\biggl(\Bigl(\frac{z\hat{z}}{{s}}\Bigr)^2x\hat{x}\biggr)
=-2\log{s}+\log z^2x+\log\hat{z}^2\hat{x}
\:.
\end{equation*}
$\log z^2x$ and $\log\hat{z}^2\hat{x}$ are individually holomorphic in ${s}$.
Indeed, 
we have an expansion
\begin{equation*}
\log z^2x
=a_1 z^4
+a_2z^6
+\biggl(-\frac{a_1^2}{2}+a_3\biggr)\,z^8
+(-a_1a_2+a_4)\,z^{10}
+O(z^{12})
\:,
\end{equation*}
where for $m\geq 1$, $a_m\in\Q[G_{2m+2}]$, more specifically,
\begin{equation*}
a_m(\tau)
=2(2 m+1)\,G_{2m+2}(\tau)
\:.
\end{equation*}
According to eq.~(\ref{eq: z1z2=eps}), for $n\geq 1$,
we have $z^{2n}={s}^{2n}/\hat{z}^{2n}$, where
\begin{equation*}
\frac{1}{\hat{z}^{2n}}
=\frac{1}{(2n-1)!}\,\biggl(\wp^{(2n-2)}(\hat{z}\,|\,\hat{\tau})-\sum_{k=1}^{\infty}\hat{a}_k\,\frac{d^{2n-2}}{d\hat{z}^{2n-2}}\,\hat{z}^{2k}\biggr)
\:.
\end{equation*}
Here $\hat{a}_m=a_m(\hat{\tau})$,
and $\wp^{(2n-2)}(\hat{z}\,|\,\hat{\tau})$ denotes the $(2n-2)$\ts{nd} derivative w.r.t.\ $\hat{z}$.
For $n\geq 1$, it is a polynomial of degree $n$ in $\hat{x}$. For $n=1$ is equals $\hat{x}$, 
and the polynomials for higher values of $n$ are listed e.g.\ in~\cite[p.\ 640]{AS:1965} for $n\leq 8$.
As a result, \hbox{$\log z^2x=O({s}^4)$}, where the coefficient of ${s}^4$ equals
\begin{equation*}
a_1\hat{x}^2+a_1\hat{a}_1/3-10\,a_1\widehat{G}_4+O(\hat{z}^2) \:.
\end{equation*}
Using the corresponding expansion for \hbox{$\hat{z}^{2m}={s}^{2n}/z^{2m}$}
clears the coefficient of its $\hat{z}$ dependence.
Every such replacement introduces another positive power of ${s}$ 
and the alternating $k\geq 2$ fold application
yields an expansion such that the coefficient of ${s}^{2n}$ for $n\leq k$ is a linear combination of a polynomial in $x$ and one in $\hat{x}$,
respectively.
Thus we have a unique splitting
\begin{align}\label{eq: expansion of log z squared xi + log hat z squared hat xi}
\log(z^2x)+\log(\hat{z}^2\hat{x})
=A(x)
+B(\hat{x})
+C\:
\end{align}
where $A,B$ are expansions of the form
\begin{equation*}
A(x)
=\sum_{n=1}^{\infty}A_{n}x^{n}\:,
\quad
B(\hat{x})
=\sum_{n=1}^{\infty}B_{n}\hat{x}^{n}\:,
\end{equation*}
respectively, and \hbox{$C/{s}^4=2a_1\hat{a}_1/3-10(\hat{a}_1G_4+a_1\widehat{G_4})+O({s}^2)$}.
For $n\geq 1$, $A_n$ and $B_n$ are at least of order $O({s}^{2n})$, by construction. 
\hbox{$\log(z^2x)+\log(\hat{z}^2\hat{x})$} is holomorphic in both $x$ and in $\hat{x}$ on the annulus, 
so the series converge.
Moreover, 
$A,B$ have an analytic continuation to the outside of the annulus. 
Indeed, $A$ can be extended to small values of $|x|$ and thus to some open neighbourhood $U$ of \hbox{$|x|=0$}
containing the annulus,
while $B$ can be extended to some open neighbourhood $\widehat{U}$ of \hbox{$|\hat{x}|=0$}. 
In the notations of eq.~(\ref{eq: expansion of log z squared xi + log hat z squared hat xi}), 
we define for $x,\hat{x}\in\C$,
\begin{align*}
\log X
&:=\log x-A(x)
\:,\\
\log\widehat{X}
&:=\log\hat{x}-B(\hat{x})\:.
\end{align*}
$X$ and $\widehat{X}$ define coordinates on $\CP^1\setminus\{\infty\}$ and on $\CP^1\setminus\{0\}$, respectively,
which for ${\xi}=e^{C}$ satisfy
\begin{equation*}
X\widehat{X}
=x\hat{x}\,e^{-(A+B)}
=x\hat{x}e^{-f_{{s}}(x)-2\log{s}}{\xi}
={\xi}/{s}^2\:,
\end{equation*}
as required. 
\end{proof}

The triple $(\tau,\hat{\tau},s)$ of the moduli of the individual tori $\Sigma$ and $\widehat{\Sigma}$ and the local sewing parameter is related to the period matrix as follows: 

\begin{proposition}
The period matrix of $M_2({s})$ for the sewing condition (\ref{eq: z1z2=eps}) 
equals
\begin{equation*}
\Omega(\tau,\hat{\tau},s)
\;=\;
\begin{pmatrix}
\tau  & 0 \\
0 & \hat{\tau}  \\
\end{pmatrix}
\pm2\pi i{s}
\begin{pmatrix}
 0 & 1 \\
 1 & 0 \\
\end{pmatrix}  
+4\pi i{s}^2 
\begin{pmatrix}
 \widehat{G}_2 & 0 \\
 0 & G_2 \\
\end{pmatrix} 
+\text{O$(s^3)$}
\:.
\end{equation*} 
Here $G_2$ is the quasi-modular form of weight $2$.
\end{proposition}

The result is consistent with \cite{M-T:2006}.

\begin{proof}
For $j=2,3,4$, let $\theta_j=\theta_j(0;\tau)$ and $\widehat{\theta_j}=\theta_j(0;\hat{\tau})$ be the Jacobi theta constants. 
Set 
\begin{equation}\label{eqs: values of Weierstrass p at half periods put into an order}
x_1
\;=\;\frac{\pi^2}{3}\left(\theta_3^4+\theta_4^4\right)
\:,\quad
x_2
\;=\;\frac{\pi^2}{3}\left(\theta_2^4-\theta_4^4\right)
\:,\quad
x_3
\;=\;-\frac{\pi^2}{3}\left(\theta_2^4+\theta_3^4\right) 
\:,
\end{equation}
and define $\hat{x}_k$ for $k=1,2,3$ as a function of $\hat{\tau}$ accordingly.
Thus $x_1,x_2,x_3$ and $\hat{x}_1,\hat{x}_2,\hat{x}_3$ are the ramification points of $\Sigma$ and of $\widehat{\Sigma}$,
respectively \cite{AS:1965}.
For the pair of coordinates $X,\widehat{X}$ from Proposition \ref{proposition: almost global coordinates on CP1 satisfying the epsilon condition},
set
\begin{equation*}
X_k
\;=\;X(x_k)
\hspace{1cm}
\widehat{X}_{k}
\;=\;\widehat{X}(\hat{x}_{k})
\:,\hspace{1cm}k=1,2,3
\:.
\end{equation*}
The linear fractional transformation $x\mapsto\frac{x-x_2}{x-x_1}\Bigl(\frac{\theta_3}{\theta_2}\Bigr)^4$
maps $(x_1,x_2,x_3)$ to $(\infty,0,1)$,
and it induces for $k=4,5,6$ the association
\begin{equation}\label{map: Xk mapped to expansion in s}
X_k
\;\mapsto\;\Bigl(\frac{\theta_3}{\theta_2}(\tau)\Bigr)^4\,
\bigl(1+\pi^2s^2\theta_4^4\,\widehat{X}_k+\text{O$(s^4)$}\bigr)
\:,
\end{equation}
by eq.~(\ref{eq: X hat X= xi/(s squared)}) and since $X_1-X_2=\pi^2\theta_4^4+\text{O$(s^2)$}$.
On the other hand, 
set  
$\Omega(\tau,\hat{\tau},s)
=\begin{pmatrix}
\Omega_{11}&\nu\\        
\nu&\Omega_{22}
\end{pmatrix}$.
Let $X_{1,\Omega}=\infty$, $X_{2,\Omega}=0$, $X_{3,\Omega}=1$ and 
\begin{equation*}
X_{4,\Omega}
\;=\;\Theta_{3,2}(\Omega)\:,
\hspace{1cm}
X_{5,\Omega}
\;=\;\Theta_{2,4}(\Omega)\:,
\hspace{1cm}
X_{6,\Omega}
\;=\;\Theta_{3,4}(\Omega)
\:,
\end{equation*}
where for $k,\ell=2,3,4$, 
\begin{equation*}
\Theta_{k,\ell}(\Omega) 
\;=\;\Bigl(\frac{\theta_3}{\theta_2}(\Omega_{11})\Bigr)^4
\bigl(
1+
4\nu^2
\frac{d}{d\Omega_{11}}\log\frac{\theta_3}{\theta_2}(\Omega_{11})\,
\frac{d}{d\Omega_{22}}\bigl(\theta_{k}\theta_{\ell}\bigr)(\Omega_{22})+\text{O$(\nu^4)$}
\bigr)
\:.
\end{equation*}
is the Riemann theta function with rational characteristics.
We use the identities
\begin{equation}\label{eq: logarithmic derivative of theta 3/theta 2}
\bigl(\log\frac{\theta_3}{\theta_2}\bigr)' 
\;=\;-\frac{\pi i}{4}\,\theta_4^4
\:,
\end{equation}
where the dash indicates differentiation w.r.t.\ the modulus,
\begin{equation*}
\bigl(\theta_k\theta_{\ell}\bigr)'
\;=\;2\pi i\,\Bigl(\D\log\bigl(\theta_k\theta_{\ell}\bigr)+\frac{G_2}{2\pi^2}\Bigr)
\:
\end{equation*}
where $\D\log\bigl(\theta_k\theta_{\ell}\bigr)=\D\log\theta_k+\D\log\theta_{\ell}$ 
with
\begin{equation*}
\frac{\D\theta_2}{\theta_2}
\;=\;\frac{1}{24}\left(\theta_3^4+\theta_4^4\right)\:,
\quad
\frac{\D\theta_3}{\theta_3}
\;=\;\frac{1}{24}\left(\theta_2^4-\theta_4^4\right)\:,
\quad
\frac{\D\theta_4}{\theta_4}
\;=\;\frac{1}{24}\left(\theta_2^4+\theta_3^4\right)
\:.
\end{equation*}
By eqs~(\ref{eqs: values of Weierstrass p at half periods put into an order}), 
this yields for $k=1,2,3$,
\begin{equation}\label{eq: Riemann theta as expansion in nu}
X_{7-k,\Omega}
\;=\;\Bigl(\frac{\theta_3}{\theta_2}(\Omega_{11})\Bigr)^4
\bigl(
1
+\nu^2\,\theta_4^4(\Omega_{11})\bigl(G_2(\Omega_{22})-\frac{1}{4}\hat{x}_k(\Omega_{22})\bigr)+\text{O$(\nu^4)$}
\Bigr)\:.
\end{equation}
where $\hat{x}_k(\Omega_{22})$ is defined as a function of $\Omega_{22}$ 
by eqs~(\ref{eqs: values of Weierstrass p at half periods put into an order}).
Since the $X_{k,\Omega}$ are the ramification points of $\Sigma_2$ \cite[p.~56, eq.~91]{Rosenhain:1851}, 
the set $\{X_{k,\Omega}\}_{k=4,5,6}$ given by eq.~(\ref{eq: Riemann theta as expansion in nu})
must equal the image of the association (\ref{map: Xk mapped to expansion in s}).
By the fact that $\sum_k\hat{x}_k=0$,
it follows that
\begin{equation}\label{eq: comparison of vartheta(tau) and vartheta(Omega)}
\Bigl(\frac{\theta_3}{\theta_2}(\tau)\Bigr)^4
\;=\;\Bigl(\frac{\theta_3}{\theta_2}(\Omega_{11})\Bigr)^4
\Bigl(1
+\nu^2\theta_4^4(\Omega_{11})\,G_2(\Omega_{22})+\text{O$(\nu^4)$}\Bigr)
\:.
\end{equation}
Since $\theta_3/\theta_2$ is invertible, 
we find in particular that $\Omega_{11}-\tau=\text{O$(\nu^2)$}$.
Subtracting eq.~(\ref{eq: comparison of vartheta(tau) and vartheta(Omega)}) 
from eq.~(\ref{eq: Riemann theta as expansion in nu})
yields
\begin{equation*}
X_{7-k,\Omega}
-\Bigl(\frac{\theta_3}{\theta_2}(\tau)\Bigr)^4
\;=\;\Bigl(\frac{\theta_3}{\theta_2}(\Omega_{11})\Bigr)^4
\Bigl(
-\frac{\nu^2}{4}\,\theta_4^4\,\widehat{X}_k+\text{O$(\nu^4)$}
\Bigr)
\:,
\end{equation*}
and we conclude that
\begin{equation}\label{eq: nu squared in terms of s}
\nu^2
\;=\;
(2\pi i)^2s^2+\text{O$(s^4)$}
\:.
\end{equation}
By eqs~(\ref{eq: logarithmic derivative of theta 3/theta 2}), 
(\ref{eq: comparison of vartheta(tau) and vartheta(Omega)}) and (\ref{eq: nu squared in terms of s}),
\begin{equation*}
\Omega_{11}
\;=\;\tau+4\pi is^2\,\widehat{G_2}+\text{O$(s^4)$}
\:.
\end{equation*}
This yields the expansion of the period matrix to the required order.
\end{proof}

%

By eq.~(\ref{description: genus two Riemann surface with (unspecified) metric}),
$M_2({s})$ carries an induced metric given by
\begin{equation}\label{def: genus two metric obtained by inserting a cylinder between two flat tori}
\G\LN(s)
=
\begin{cases}
|dz|^2&\quad\text{$1\leq |z|$}\;,\\
|d\log z|^2&\quad\text{$\sqrt{|{s}|}\leq |z|\leq1$}\:,\\
|d\log \hat{z}|^2&\quad\text{$\sqrt{|{s}|}\leq |\hat{z}|\leq 1$}\:,\\
|d\hat{z}|^2&\quad\text{$1\leq |\hat{z}|$}\:,
\end{cases}
\end{equation}
where $z,\hat{z}$ lie in the fundamental cell of the respective torus and the pair satisfies eq.~(\ref{eq: z1z2=eps}).

This metric is continuous and the curvature is supported on the two circles $\gamma_1$ and $\gamma_2$.
Mark the latter by one point each and let
\begin{equation}\label{defs: M1 and hat M1(s)}
\begin{split}
M_1
\;=&\;M_L^{(1)}\sqcup_{\gamma_1}D_R^{(1)}\:, \\
\widehat{M}_1({s})
\;=&\;D_L^{(1)}\sqcup_{\gamma_1}(\gamma\times[0,\ell])^{\theta}\sqcup_{\gamma_2}\widehat{M}_R^{(2)}
\end{split}
\end{equation}
be the corresponding elements in $\mathscr{X}$ of genus one.
For ${s}=1$, we drop the argument and write $\widehat{M}_1$. 

\begin{table}[htb]
\centering
\begin{tabular}{lll|l|l|l}
$A(\gamma^s,$&$\gamma^{q},$&$\gamma^{\widehat{q}}\,)$&resp.~lowest weight terms in $g=0$&$\langle1\rangle({s},\tau,\hat{\tau})_A^{g=2}$&$\eps_A$\\
\hline
$(\mathfrak{v},$&$\mathfrak{v},$&$\mathfrak{v})$&$\langle1\rangle^{g=0}{\langle1\rangle^{g=0}}$&$\langle1\rangle({s},\tau,\hat{\tau})_{(2,2)}^{g=2}$&$\,1$\\
$(\mathfrak{v},$&$\mathfrak{v},$&$\mathfrak{w})$&$\langle1\rangle^{g=0}{\langle\Phi\Phi\rangle^{g=0}}$&$\langle1\rangle({s},\tau,\hat{\tau})_{(2,1)}^{g=2}$&$\,1$\\
$(\mathfrak{v},$&$\mathfrak{w},$&$\mathfrak{v})$&$\langle\Phi\Phi\rangle^{g=0}{\langle1\rangle^{g=0}}$&$\langle1\rangle({s},\tau,\hat{\tau})_{(1,2)}^{g=2}$&$\,1$\\
$(\mathfrak{v},$&$\mathfrak{w},$&$\mathfrak{w})$&$\langle\Phi\Phi\rangle^{g=0}{\langle\Phi\Phi\rangle^{g=0}}$&$\langle1\rangle({s},\tau,\hat{\tau})_{(1,1)}^{g=2}$&$\,1$\\
$(\mathfrak{w},$&$\mathfrak{w},$&$\mathfrak{w})$&$\langle\Phi\Phi\Phi\rangle^{g=0}{\langle\Phi\Phi\Phi\rangle^{g=0}}$&$\langle1\rangle({s},\tau,\hat{\tau})_{\Phi}^{g=2}$&$-1$\\
\end{tabular}\\
\vspace{0.3cm}
\caption{Holomorphic blocks in the $(2,5)$ minimal model on $M_2(s)$: 
We write \hbox{$q=\exp(2\pi i\tau)$} and \hbox{$\widehat{q}=\exp(2\pi i\hat{\tau})$}.
Using the $q$ formalism, $\langle1\rangle(\tau)_1^{g=1}=\langle\Phi\Phi\rangle^{g=0}+\ldots$ in the non-vacuum resp.\ $\langle1\rangle(\tau)_2^{g=1}=\langle1\rangle^{g=0}+\ldots$ in the vacuum sector on the flat torus. 
An additional holomorphic block is given by $\langle\Phi\rangle(\tau)^{g=1}=\langle\Phi\Phi\Phi\rangle^{g=0}+\ldots$
}
\label{table: holomorphic blocks for g=2}
\end{table}

By Proposition \ref{proposition: in genus g=2, there are 5 holomorphic blocks}, 
in the $(2,5)$ minimal model on $M_2(s)$, $|\mathfrak{A}|=5$.
As illustrated by Table \ref{table: holomorphic blocks for g=2},
it is convenient to use the description
\begin{equation}\label{eq: parametrisation of holomorphic blocks in genus two}
\mathfrak{A}
\;=\;\{(a,b),\Phi|\,a,b\in\{1,2\}\} 
\end{equation}
where $a=1$ and $a=2$ stand for the non-vacuum and the vacuum sector, respectively, 
in accordance with eqs (\ref{def: zero point functions in the (2,5) minimal model}).

Segal's formula (\ref{eq: Segal's s formula}) 
together with eq.~(\ref{eq: ket of twisted cylinder sewn with MR})
yield
the zero-point function associated to $(a,b)\in\mathfrak{A}$ 
\begin{equation}\label{eq: genus 2 partition function in the s formalism for arbitrary metric}
\langle 1\rangle_{(a,b),M_2({s})}
\;=\;\sum_{n}
\frac{\langle{\psi}_{n}(z_R=0)\rangle_{a,M_1}
\langle{\psi}_{n}(z_L=0)\rangle_{b,\widehat{M}_1({s})}}
{\langle{\psi}_{n}(z_L=0)\,{\psi}_{n}(z_R=0)\rangle_{\DD}}
\:,
\end{equation}
where $\{\psi_n\}_n$ be a standard orthogonal basis of $R_2({\mathfrak{V}})$.
(We indiscriminately write $\psi$ for the field in either coordinate.)
By eqs~(\ref{eq: Shapovalov form of ML and ei expressed through 1-point function}) and (\ref{eq: ket of twisted cylinder sewn with MR}),
\begin{equation}\label{eq: effect of reducing cylinder to lenth zero on a 1-pt fct in g=1}
\langle\psi_i(z_L=0)\rangle_{a,\widehat{M}_1({s})}
\;=\;{s}^{h_i-c/24}\,\langle e_i\,|\widehat{M}_R^{(2)}\rangle_a
\:,
\end{equation}
where $\langle e_i|\widehat{M}_R^{(2)}\rangle_a=\langle\widehat{\psi}_i(\widehat{z}=0)\rangle_{a,\widehat{M}_1}$.
Since all one-point functions in the analytic coordinate on $M_1$ and $\widehat{M}_1$ are constant in position,
we write (for $z_R=z$)
\begin{displaymath}
\langle\psi(z)\rangle_{a,M_1}
\;=\;\langle\psi\rangle_a\:,\quad
\langle\widehat{\psi}(\widehat{z})\rangle_{a,\widehat{M}_1}
\;=\;\reallywidehat{\langle\psi\rangle}_a
\:.
\end{displaymath}
By considering \hbox{${s}=-1$}, 
we see that $\reallywidehat{\langle\psi\rangle}_a=0$ for \hbox{$\psi\in {\mathfrak{V}}$} 
for which $h$ is odd.
Moreover, only the quasi-primary fields contribute to the sum in eq.~(\ref{eq: genus 2 partition function in the s formalism for arbitrary metric}).

\begin{proposition}\label{proposition: Taylor expansion in epsilon of the genus two 0-pt function obtained from the Rogers-Ramanujan functions and the one corresponding to the primary field Phi}
Let $M_1$ and $\widehat{M}_1(s)$ be the genus one Riemann surfaces defined by eqs (\ref{defs: M1 and hat M1(s)}).
Let $M_2(s)$ be the genus two Riemann surface from eq.~(\ref{description: genus two Riemann surface with (unspecified) metric}).
Let \hbox{$\langle 1\rangle_a=\langle 1\rangle_{a,M_1}$} 
and \hbox{$\widehat{\langle 1\rangle}_b=\langle 1\rangle_{b,\widehat{M}_1}$} be the holomorphic blocks of the $(2,5)$ minimal model
from eqs~(\ref{def: zero point functions in the (2,5) minimal model}).
Set $t_a:=d\log\langle 1\rangle_a/d(2\pi i\tau)$ and 
$\widehat{t_b}:=d\log\widehat{\langle 1\rangle}_b/d(2\pi i\hat{\tau})$ accordingly.
Associated to the set $\mathfrak{A}$ from eq.~(\ref{eq: parametrisation of holomorphic blocks in genus two})
are the five holomorphic blocks 
\begin{equation*}
\langle 1\rangle({s},\tau,\hat{\tau})^{g=2}_{(a,b)}
\;=\;{s}^{11/60}\langle 1\rangle_a\widehat{\langle 1\rangle}_b
\,\Xi^{g=2}_{(a,b)}\:,\quad a,b=1,2, 
\end{equation*}
and
\begin{equation}\label{def: fifth solution}
\langle 1\rangle({s},\tau,\hat{\tau})^{g=2}_{\Phi}
\;=\;{s}^{-1/60}\,\eps_{\Phi}\,{\mu}_{\Phi}\,(\eta\widehat{\eta})^{-2/5}
\,\Xi^{g=2}_{\Phi}
\:,
\end{equation}
where $\Xi^{g=2}_{A}$ has an expansion in modular forms as follows:
\begin{align*}
\Xi^{g=2}_{A}\;=\;\sum_{n\geq 0}\Xi^{g=2}_{A,n}\,s^{2n}
\:,
\end{align*}
where for $n\geq 0$, 
\hbox{$\Xi^{g=2}_{A,n}\in\Q\bigl[G_4,\,\widehat{G_4},G_6,\,\widehat{G_6},\,t_a,\widehat{t_b}\bigr]$} is a modular form 
of weight $2n$ both in $\tau$ and in $\hat{\tau}$.
It is linear in $t_a$ and in $\widehat{t_b}$ for $A=(a,b)$, and of degree zero for $A=\Phi$.
Exchanging $\tau\leftrightarrow\hat{\tau}$,
maps the matrix $\bigl(\Xi^{g=2}_{(a,b),n}\bigr)_{a,b=1,2}$ to its transpose,
while $\Xi^{g=2}_{\Phi,n}$ is left invariant.
\end{proposition}

\begin{proof}
The product structure and related symmetry between $\tau$ and $\hat{\tau}$ follow from Segal's formula~(\ref{eq: genus 2 partition function in the s formalism for arbitrary metric}).
We address the coefficients $\Xi^{g=2}_{A,n}$.
For $a=1,2$, the correct Laurent coefficient of the Virasoro $N$-point function \hbox{$\langle T(z_1)\ldots T(z_N)\rangle_a$}, 
or of one of its derivatives,
is of the form \hbox{$\bigl\langle L_{n_N}\ldots L_{n_1}\mathfrak{v}\bigr\rangle_a$},
thus, by  Proposition \ref{proposition: graphical representation of N-point functions of T},
it defines an element in the polynomial ring $\Q[G_4,G_6,t_a]$, 
which is linear in $t_a$.
Moreover, as mentioned previously, only one-point functions associated to even values of $h$ can be nonzero.
We address $A=\Phi$.
By Proposition \ref{proposition: graphical representation of (N+1)-point functions of N copies of T and one copy of Phi},
any \hbox{$\bigl\langle L_{n_N}\ldots L_{n_1}\mathfrak{w}\bigr\rangle$} is modular or a weakly holomorphic modular form.
Only powers of ${s}$ of the form \hbox{$h'-1/5$} occur with $h'$ even:
For $N\geq 0$,
while being constant in position,
\begin{equation*}
\langle L_{n_N}\ldots L_{n_1}\Phi(-z)\rangle
\;=\;(-1)^{h-\bar{h}}\langle L_{n_N}\ldots L_{n_1}\Phi(z)\rangle
\:,
\end{equation*}
since \hbox{$i(L_0-\bar{L}_0)$} generates rotations.
Here \hbox{$h=\sum_{i=1}^Nn_i-1/5$} and \hbox{$\bar{h}=-1/5$}.
Thus 
\begin{equation*}
\langle L_{n_N}\ldots L_{n_1}\Phi(z)\rangle
\;=\;0\hspace{1cm}\text{for $h'=h+1/5$ odd} 
\end{equation*}
Restricting to the holomorphic part $\Phi$ shows that, apart from a factor of ${s}^{-1/5}$, only even powers of ${s}$ occur.
\end{proof}  

The actual expansions of $\Xi^{g=2}_{(a,b)}$ and of $\Xi^{g=2}_{\Phi}$ are obtained from eq.~(\ref{eq: genus 2 partition function in the s formalism for arbitrary metric})
by direct computation and use of the list of quasi-primary fields from Proposition \ref{proposition: quasi-primary fields and their squared norms in the vacuum rep}.
For \hbox{$a,b\in\{1,2\}$}, 
\begin{equation*}
\begin{split}
\langle 1\rangle({s},\tau,\hat{\tau})^{g=2}_{(a,b)}
\;=\;&{s}^{11/60}\biggl(\langle 1\rangle_a\reallywidehat{\langle 1\rangle}_b
+{s}^2\,\frac{\langle T\rangle_a\reallywidehat{\langle T\rangle}_b}{c/2}
+{s}^6\,\frac{49\,\bigl\langle L_4L_2v\bigr\rangle_a\reallywidehat{\bigl\langle L_4L_2v\bigr\rangle}_b}{\parallel(7L_4L_2-2L_6)\,v\parallel^2}+\ldots\biggr)
.
\end{split}
\end{equation*} 
Recall from Section \ref{subsec: The fields of the (2,5) minimal model}
that $L_m\mathfrak{v}$ for $m\geq 3$ is a derivative field.
In the following, we drop the index $a,b\in\{1,2\}$.
By eq.~(\ref{eq: Virasoro field}),
\begin{equation*}
\bigl\langle T(z)T^{(k)}(0)\bigl\rangle
\;=\;k!\,\sum_{n\geq 2}z^{n-2}\bigl\langle L_nL_{k+2}\mathfrak{v}\bigl\rangle\:.
\end{equation*}
Comparison with eq.~(\ref{eq: <T(z)T(0)> for g=1 in the (2,5) minimal model}) yields:
For $h=6$,
\begin{equation*}
\bigl\langle L_4L_2\mathfrak{v}\bigl\rangle
\;=\;12\,G_4\langle T\rangle-44\,G_6\langle 1\rangle\\
\end{equation*}
for $h=8$,
\begin{equation*}
\begin{split}
\bigl\langle L_5L_3\mathfrak{v}\bigl\rangle
\;=&\;-4\,\bigl\langle L_6L_2\mathfrak{v}\bigl\rangle\\
\bigl\langle L_6L_2\mathfrak{v}\bigl\rangle
\;=&\;20\,G_6\langle T\rangle
-132\,G_4^2\langle 1\rangle
\:,
\end{split}
\end{equation*} 
for $h=10$,
\begin{equation*}
\begin{split}
\bigl\langle L_6L_4\mathfrak{v}\bigl\rangle
\;=&\;15\,\bigl\langle L_8L_2\mathfrak{v}\bigl\rangle\\
\bigl\langle L_7L_3\mathfrak{v}\bigl\rangle
\;=&\;-6\,\bigl\langle L_8L_2\mathfrak{v}\bigl\rangle\\
\bigl\langle L_8L_2\mathfrak{v}\bigl\rangle
\;=&\;24\,G_4^2\langle T\rangle
-336\,G_4G_6\langle 1\rangle\\
\:.
\end{split}
\end{equation*} 
In order to compute terms of order \hbox{$h\geq 12$}, 
~$N$-point functions for \hbox{$N\geq 3$} must be taken into account. 
Firstly, we have for $h=12$,
\begin{equation*}
\begin{split}
\bigl\langle L_7L_5\mathfrak{v}\bigl\rangle
\;=&\;-2\,\bigl\langle L_8L_4\mathfrak{v}\bigl\rangle\\
\bigl\langle L_8L_4\mathfrak{v}\bigl\rangle
\;=&\;-\frac{133\,056}{13}\,G_4^3\langle 1\rangle
+\frac{10\,080}{11}\,G_4G_6\langle T\rangle
-\frac{92\,400}{13}\,G_6^2\\
%
\bigl\langle L_9L_3\mathfrak{v}\bigl\rangle
\;=&\;-8\,\bigl\langle L_{10}L_2\mathfrak{v}\bigl\rangle\\
\bigl\langle L_{10}L_2\mathfrak{v}\bigl\rangle
\;=&\;-\frac{4\,752}{13}\,G_4^3
+\frac{360}{11}\,G_4G_6\langle T\rangle
-\frac{3\,300}{13}\,G_6^2\langle 1\rangle
\:.
\end{split}
\end{equation*} 
Moreover, by eq.~(\ref{eq: Virasoro field}),
\begin{equation*}
\bigl\langle T(z_2)T(z_1)T(0)\bigl\rangle
\;=\;\,\sum_{n_2\geq 2}\sum_{n_1\geq 2}z_2^{n_2-2}z_1^{n_1-2}\bigl\langle L_{n_2}L_{n_1}L_{2}\mathfrak{v}\bigl\rangle
\:.
\end{equation*}
Sorting out the coefficient of $z_2^4z_1^2z_0^0$ in eq.~(\ref{eq: 3-pt function of T})
with the correct modular coefficients from Example \ref{example: modular coefficients}
yields
\begin{equation*}
\bigl\langle L_6L_4L_2\mathfrak{v}\bigl\rangle
\;=\;-\frac{95\,040}{13}\,G_4^3
-\frac{7\,200}{11}\,G_4G_6\langle T\rangle
-\frac{100\,320}{13}\,G_6^2\langle 1\rangle
\:.
\end{equation*}
This yields an expansion of $\Xi^{g=2}_{(a,b)}$ 
up to terms of order $\text{O(${s}^{14}$)}$.

We address the fifth holomorphic block.
In analogy to eq.~(\ref{eq: genus 2 partition function in the s formalism for arbitrary metric}),
using Proposition \ref{proposition: quasi-primary fields and their squared norms in the W rep},
\begin{equation*}
\begin{split}
\langle 1\rangle({s},\tau,\hat{\tau})^{g=2}_{\Phi}
\;=\;&{s}^{-1/60}
\biggl(\frac{\langle\mathfrak{w}\rangle\,\reallywidehat{\langle\mathfrak{w}\rangle}}{\parallel\mathfrak{w}\parallel^2}
+{s}^4\,\frac{\bigl\langle (25L_3L_1-2L_4)\,\mathfrak{w}\bigr\rangle\reallywidehat{\bigl\langle (25L_3L_1-2L_4)\,\mathfrak{w}\bigr\rangle}}{\parallel(25L_3L_1-2L_4)\,\mathfrak{w}\parallel^2}+\ldots\biggr)
.
\end{split}
\end{equation*}
We have used eq.~(\ref{eq: 1-point function of Phi}) for the one-point function of $\Phi$ and the factorisation given there.
Put $h=h'-1/5$ with $h'\in\N_0$.
We have \hbox{$\langle T(z)\Phi(0)\rangle=\sum_{n}z^{n-2}\langle L_n\,\Phi(0)\rangle$},
so by eq.~(\ref{eq: 2-pt fct of T with Phi}),
\begin{align*}
\langle L_4\mathfrak{w}\rangle
\;=\;&-\frac{6}{5}\,G_4\,\langle\mathfrak{w}\rangle
\:,\\
\langle L_6\mathfrak{w}\rangle
\;=\;&-2\,G_6\,\langle\mathfrak{w}\rangle
\:,\\
\langle L_8\mathfrak{w}\rangle
\;=\;&-\frac{12}{5}\,G_4^2\,\langle\mathfrak{w}\rangle
\:,\\
\langle L_{10}\mathfrak{w}\rangle
\;=\;&-\frac{36}{11}\,G_4G_6\,\langle\mathfrak{w}\rangle
\:,\\
\langle L_{12}\mathfrak{w}\rangle
\;=\;&-\frac{4}{65}\,(36\,G_4^3+25\,G_6^2)\,\langle\mathfrak{w}\rangle
\:.
\end{align*}
We have \hbox{$\langle T(z_2)T(z_1)\Phi(0)\rangle=\sum_{n}z_2^{n_2-2}z_1^{n_1-2}\langle L_{n_2}L_{n_1}\,\Phi(0)\rangle$},
so by eq.~(\ref{eq: 3-pt fct of T with Phi}),
for $h'=4$,
\begin{align*}
\langle L_3L_1\mathfrak{w}\rangle
\;=\;&\frac{12}{5}\,G_4\,\langle\mathfrak{w}\rangle
\:,
\end{align*}
for $h'=6$,
\begin{align*}
\langle L_4L_2\mathfrak{w}\rangle
\;=\;&-60\,G_6\,\langle\mathfrak{w}\rangle
\:,\\
\langle L_5L_1\mathfrak{w}\rangle
\;=\;&8\,G_6\,\langle\mathfrak{w}\rangle
\:,
\end{align*}
for $h'=8$,
\begin{align*}
\langle L_5L_3\mathfrak{w}\rangle
\;=\;&600\,G_4^2\,\langle\mathfrak{w}\rangle
\:,\\
\langle L_6L_2\mathfrak{w}\rangle
\;=\;&-180\,G_4^2\,\langle\mathfrak{w}\rangle
\\
\langle L_7L_1\mathfrak{w}\rangle
\;=\;&\frac{72}{5}\,G_4^2\,\langle\mathfrak{w}\rangle
\:.
\end{align*}
For $h'=10$, 
\begin{align*}
\langle L_6L_4\mathfrak{w}\rangle
\;=\;&-\frac{299\,868}{55}\,G_4G_6\,\langle\mathfrak{w}\rangle
\:,\\
\langle L_7L_3\mathfrak{w}\rangle
\;=\;&\frac{25\,200}{11}\,G_4G_6\,\langle\mathfrak{w}\rangle
\\
\langle L_8L_2\mathfrak{w}\rangle
\;=\;&-\frac{5\,040}{11}\,G_4G_6\,\langle\mathfrak{w}\rangle
\:,\\
\langle L_9L_1\mathfrak{w}\rangle
\;=\;&\frac{288}{11}\,G_4G_6\,\langle\mathfrak{w}\rangle
\:. 
\end{align*}
In order to compute terms of order \hbox{$h'\geq 12$} in the non-vacuum sector, 
correlation functions with \hbox{$N\geq 3$} copies of $T$ and one copy of $\Phi$ must be taken into account.
These are provided by Proposition \ref{proposition: graphical representation of (N+1)-point functions of N copies of T and one copy of Phi}. Thus for $h'=12$,
\begin{align*}
\langle L_7L_5\mathfrak{w}\rangle
\;=\;& \frac{1\,407\,888}{65}\,G_4^3\,\langle\mathfrak{w}\rangle+\frac{193\,200}{13}\,G_6^2\,\langle\mathfrak{w}\rangle\\
\langle L_8L_4\mathfrak{w}\rangle
\;=\;& -\frac{3\,586\,104}{325}\,G_4^3\,\langle\mathfrak{w}\rangle-\frac{99\,120}{13}\,G_6^2\,\langle\mathfrak{w}\rangle\\
\langle L_9L_3\mathfrak{w}\rangle
\;=\;& \frac{43\,200}{13}\,G_4^3\,\langle\mathfrak{w}\rangle+\frac{30\,000}{13}\,G_6^2\,\langle\mathfrak{w}\rangle\\
\langle L_{10}L_2\mathfrak{w}\rangle
\;=\;& -\frac{6\,480}{13}\,G_4^3\,\langle\mathfrak{w}\rangle-\frac{4\,500}{13}\,G_6^2\,\langle\mathfrak{w}\rangle\\
\langle L_{11}L_1\mathfrak{w}\rangle
\;=\;& \frac{288}{13}\,G_4^3\,\langle\mathfrak{w}\rangle+\frac{200}{13}\,G_6^2\,\langle\mathfrak{w}\rangle
\end{align*}
and
\begin{align*}
\langle L_6L_4L_2\mathfrak{w}\rangle
\;=\;&-\frac{18\,551\,736}{325}\,G_4^3\,\langle\mathfrak{w}\rangle-\frac{541\,040}{13}\,G_6^2\,\langle\mathfrak{w}\rangle
\:,\\
\langle L_7L_4L_1\mathfrak{w}\rangle
\;=\;&-\frac{844\,560}{13}\,G_4^3\,\langle\mathfrak{w}\rangle-\frac{579\,480}{13}\,G_6^2\,\langle\mathfrak{w}\rangle
\:,\\
\langle L_8L_3L_1\mathfrak{w}\rangle
\;=\;&\frac{7\,177\,968}{325}\,G_4^3\,\langle\mathfrak{w}\rangle+\frac{198\,400}{13}\,G_6^2\,\langle\mathfrak{w}\rangle
\:.
\end{align*}
Together with the first of eqs~(\ref{eqs: squared Shapovalov norm of w and bar w}),
this yields the expansion for $\Xi^{g=2}_{\Phi}$ up to terms of order $\text{O(${s}^{14}$)}$.
More specifically, we have $\Xi^{g=2}_{(a,b)}
\;=\;\sum_{n\geq 0}\Xi^{g=2}_{(a,b),n}\,s^{2n}$, 
where
\begin{equation*}
\begin{split}
\Xi^{g=2}_{(a,b),0}
\;=&\;1\\
\Xi^{g=2}_{(a,b),1}
\;=&\;-\frac{5}{11}\,{t}_a\,\widehat{t_b}\:,\\
\Xi^{g=2}_{(a,b),2}
\;=&\;0\:,\\
\Xi^{g=2}_{(a,b),3}
\;=&\;\,\frac{280}{31}\,
\Bigl(\frac{9}{11}\,G_4\widehat{G_4}\,{t}_a\,\widehat{t_b}
-3\,\bigl(G_4\widehat{G_6}\,{t}_a+G_6\widehat{G_4}\,\widehat{t_b}\bigr)
+11\,G_6\widehat{G_6}\Bigr)\:,\\
\Xi^{g=2}_{(a,b),4}
\;=&\;\,\frac{624}{41}\,
\Bigl(
\frac{25}{11}\,G_6\widehat{G_6}\,{t}_a\,\widehat{t_b}
-15\,\bigl(G_4^2\widehat{G_6}\,\widehat{t_b}+G_6\widehat{G_4}^2\,{t}_a\bigr)
+99\,G_4^2\widehat{G_4}^2
\Bigr)\:,\\
\Xi^{g=2}_{(a,b),5}
\;=&\;\frac{4\,896}{31}
\Bigl(\,G_4^2\widehat{G_4}^2\,{t}_a\,\widehat{t_b}
-14\,\bigl(G_4^2\widehat{G_4}\widehat{G_6}\,{t}_a+G_4G_6\widehat{G_4}^2\,\widehat{t_b}\bigr)
+196\,G_4G_6\widehat{G_4}\widehat{G_6}\Bigr)\:,\\
\Xi^{g=2}_{(a,b),6}
\;=&-\frac{48}{61}\,
\Bigl(\frac{8\,817\,040\,260}{1\,691\,701}\,G_4G_6\widehat{G_4}\widehat{G_6}\,{t}_a\,\widehat{t_b}
+\frac{4\,994\,136}{451}\,\bigl(G_4^3\widehat{G_4}\widehat{G_6}\,\widehat{t_b}+G_4G_6\widehat{G_4}^3\,{t}_a\bigr)\\
&\hspace{0.9cm}
+\frac{6\,888\,210}{341}\,\bigl(G_4G_6\widehat{G_6}^2\,{t}_a+G_6^2\widehat{G_4}\widehat{G_6}\,\widehat{t_b}\bigr)
-\frac{529\,767\,216}{2\,665}\,G_4^3\widehat{G_4}^3\\
&\hspace{0.9cm}
-\frac{1\,970\,892}{13}\,\bigl(G_4^3\widehat{G_6}^2+G_6^2\widehat{G_4}^3\bigr)
-\frac{37\,043\,545}{403}\,G_6^2\widehat{G_6}^2\Bigr)
\:.
\end{split}
\end{equation*}
Moreover,
$\Xi^{g=2}_{\Phi}
\;=\;\sum_{n\geq 0}\Xi^{g=2}_{\Phi,n}\,s^{2n}$, where
\begin{equation*}
\begin{split}
\Xi^{g=2}_{\Phi,0}
\;=&\;1\:,\\
\Xi^{g=2}_{\Phi,1}
\;=&\;0\:,\\
\Xi^{g=2}_{\Phi,2}
\;=&\;\frac{312}{95}\,G_4\widehat{G_4}\:,\\
\Xi^{g=2}_{\Phi,3}
\;=&\;\frac{59\,340}{551}\,G_6\widehat{G_6}\:,\\
\Xi^{g=2}_{\Phi,4}
\;=&\;\frac{1\,175\,328}{725}\,G_4^2\,\widehat{G_4}^2\:,\\
\Xi^{g=2}_{\Phi,5}
\;=&\;\frac{75\,034\,656}{2\,299}\,G_4G_6\,\widehat{G_4}\widehat{G_6}\:,\\
\Xi^{g=2}_{\Phi,6}
\;=&\;\frac{32}{799\,833\,908\,590\,151\,400\,983\,568\,329\,597}\times\\
&\hspace{0.5cm}\Bigl(
{8\,336\,522\,269\,110\,502\,535\,732\,151\,408\,980\,286\,372}\,\frac{G_4^3\widehat{G_4}^3}{5}\\
&\hspace{0.6cm}+49\,126\,884\,084\,081\,132\,256\,334\,667\,534\,510\,924\,\bigl(G_4^3\widehat{G_6}^2+G_6^2\widehat{G_4}^3\bigr)\\
&\hspace{0.6cm}+{108\,570\,072\,190\,577\,226\,179\,433\,424\,864\,396\,525}\,\frac{G_6^2\widehat{G_6}^2}{3}
\Bigr)
\:.
\end{split}
\end{equation*}

Several terms in the $q$-expansions had been found previously by T.\ Gilroy and M.\ Tuite \cite{T:2015}
but these functions had not been given explicitly as modular forms.

\subsection{The \texorpdfstring{$q$}{Lg} formalism in genus \texorpdfstring{$g\geq 1$}{Lg}}\label{subsection: q formalism}

Let $M=(\Sigma,\G)\in\mathscr{X}$,
and let $M^{\theta}\subset M$ be the metric manifold obtained by cutting along a non-separating curve $\gamma\subset\Sigma$.
Let $\gamma_1,\gamma_2$ be the boundary curves of $M^{\theta}$.
Let $D_L^{(1)}$ and $D_R^{(2)}$ be the corresponding flat left and right unit discs
centred at $P_0$ and $P_5$, respectively, and containing no other ramification point.
We consider the case where 
\begin{equation*}
D_L^{(1)}\sqcup_{\gamma_1}M^{\theta}\sqcup_{\gamma_2}D_R^{(2)}
\:.
\end{equation*}
has genus zero or one.
Let $M^{\theta}=(S^1)^{\theta}$.
Let $z_L$ and $z_R$ be a pair of coordinates on $D_L^{(1)}$ and $D_R^{(2)}$
vanishing at the respective disc centre ($z=w$ and $z=-w$, respectively, for $w\not=0$) and satisfying $z_Lz_R=1$ on the boundary.
Now remove the discs (which leaves us with $S^1$ and an angle $\theta$) and insert the twisted cylinder bounded by 
$\gamma_1=\gamma\times\{0\}$ and $\gamma_2=\gamma\times\{\ell\}$. 
Thus $z(P)\sim\hat{q}\,z(P)$ and 
\begin{equation*}
z_L(P)\,z_R(P')
\;=\;\hat{q}
\:
\end{equation*}
for $\gamma$, for $\hat{q}=\exp(i\theta-\ell)$.  
We conclude that the self-sewn surface 
\begin{equation*}
\sqcup_{\gamma\times[0,\ell]}\,(S^1)^{\theta}
\;=\;\sqcup_{\gamma}\,(\gamma\times[0,\ell])^{\theta} 
\end{equation*}
is a torus $M_1(q)$ of modulus $\tau=\log\hat{q}/(2\pi i)$.
The latter carries the flat metric induced by that on the cylinder.
We write $\sqcup_{\gamma\times[0,\ell]}\,M^{\theta}$ 
when self-sewing is performed by inserting the flat cylinder of length $\ell$ 
(with the twist prescribed by the angle $\theta$). 

For later use, we note that for $N\geq 0$,
Segal's formula (\ref{eq: Segal's q formula}) 
together with eq.~(\ref{eq: ket of twisted cylinder sewn with MR})
yields for the $N$-point function of $\phi_1,\ldots,\phi_N\in R_2(\mathfrak{V})\oplus R_2(\mathfrak{W})$ associated to $A\in\mathfrak{A}$:
\begin{equation}
\begin{split}\label{eq: N-point function for genus g in the q-formalism for arbitrary metrics}
\bigl\langle\phi_1(z_1)&\ldots\phi_N(z_N)\bigr\rangle_{A,\sqcup_{\gamma\times[0,-\log|{q}|]}{M^{\theta}}}\\
\;=\;|\hat{q}|^{-c/12}&\sum_{n}\hat{q}^{h_n}\bar{\hat{q}}^{\bar{h}_n}\,
\frac{\langle{\psi}_{n}(z_L=0)\,\phi_1(z_1)\ldots\phi_N(z_N)\,{\psi}_{n}(z_R=0)\rangle_{A,D_L^{(1)}\sqcup_{\gamma_1}M^{\theta}\sqcup_{\gamma_2}D_R^{(2)}}}
{\langle{\psi}_{n}(z_L=0)\,{\psi}_{n}(z_R=0)\rangle_{{D_L}\sqcup_{\gamma}D_R}}
\:,
\end{split}
\end{equation}
provided $|\hat{q}|\ll 1$ and $\arg(\hat{q})=\theta$.
Here $\{\psi_n\}_n$ is a standard orthogonal basis of $R_2(\mathfrak{V})\oplus R_2(\mathfrak{W})$.
(We indiscriminately write $\psi$ for the field in either coordinate.)

Taking $N=0$ and restricting the summation in eq.~(\ref{eq: N-point function for genus g in the q-formalism for arbitrary metrics})
to a basis \hbox{$\{\psi_n\}_{n\geq 0}$} of~${\mathfrak{V}}$,
yields in particular
\begin{equation}\label{eq: general 0-pt fct for g=2 as expansion in rho}
\langle 1\rangle(\hat{q},w,\tau)_A^{g=2}
\;=\;\hat{q}^{11/60}\sum_{n\geq 0}\,\hat{q}^{h_n}\,\bigl\langle\psi_n(z_L=z-w)\,\psi_n(z_R=z+w)\bigr\rangle(\tau)_A^{g=1}
\:.
\end{equation}
Note that $\langle 1\rangle$ depends on $w$ rather than $z-w$ and $z+w$, by translational symmetry.

The period matrix can be computed in a way analogous to that in the $s$-formalism
and explicit formulae are provided by \cite{M-T:2006}.

For the $(2,5)$ minimal model, $|\mathfrak{A}|=2$, 
corresponding to the pair of Rogers-Ramanujan functions (for $n=0$).
For higher weight terms, derivative fields have to be taken into account as well.
In addition to two-point function for the quasi-primary fields listed in Proposition \ref{proposition: quasi-primary fields and their squared norms in the vacuum rep} for $h\geq 2$, 
there is one term corresponding to the field $T^{(h-2)}$ for each of $h=3,4,5$,
whose squared Shapovalov norm equals $\frac{c}{12}\,(h-2)!\,(h+1)!$
by Proposition \ref{proposition: squared Shapovalov norm of derivatives of T and of Phi}.
Note that for $h=4$, the normal ordered square of $T$ does not yield an additional field by eq.~(\ref{eq: normal ordered product of T in the (2,5) minimal model}).

The first interesting term in the series of eq.~(\ref{eq: general 0-pt fct for g=2 as expansion in rho}) occurs for weight \hbox{$h=6$}, which we provide by Proposition \ref{propos: the first interesting term in the rho expansion for h=6}. 
For the remainder of this section, 
all occurring correlation functions refer to the torus with holomorphic differential $dz$ and periods $1,\tau$.
We use the notation introduced subsequent to eq.~(\ref{eq: ODE relating <1> and <T>}).

\begin{proposition}\label{propos: the first interesting term in the rho expansion for h=6}
For $0\leq h\leq 6$ and $A\in\{1,2\}$, the summands in eq.~(\ref{eq: general 0-pt fct for g=2 as expansion in rho})
are the terms listed in Table \ref{table: the first few terms in eq. (19)}.
\begin{table}
\begin{tabular}{l|l}
$h$&coefficient of ${q}^h$ \\
\hline\noalign{\smallskip}
$0$&$\hspace{0.2cm}\langle 1\rangle_A$\\
$2$&$\biggl(\wp_{12}^2-12\,G_4\biggr)\,\langle 1\rangle_A
+4\wp_{12}\,\langle T\rangle_A/c$\\ 
$3$&$\biggl(
-5\wp_{12}^3
+90\,G_4\wp_{12}
+140\,G_6\biggr)\,\langle 1\rangle_A
+ \left(
-6\wp_{12}^2
+60\,G_4
\right)\,\langle T\rangle_A/c
$\\
$4$&$
\biggl(
21\wp_{12}^4
-504\,G_4\wp_{12}^2
-840\,G_6\wp_{12}
+540\,G_4^2
\biggr)\,\langle 1\rangle_A
+
\biggl(
12\wp_{12}^3
-216\,G_4\wp_{12}
-336\,G_6
\biggr)\,\langle T\rangle_A/c 
$\\
$5$&$\biggl(
-84\wp_{12}^5
+2520\,G_4\wp_{12}^3
+4200\,G_6\wp_{12}^2
-10\,080\,G_4^2\wp_{12}
-18\,480\,G_4G_6
\biggr)\,\langle 1\rangle_A$\\
&\hspace{0.5cm}
$+\biggl(
-28\wp_{12}^4
+672\,G_4\wp_{12}^2
+1\,120\,G_6\wp_{12}
-720\,G_4^2
\biggr)\,\langle T\rangle_A/c$\\
$6$
&$\biggl(\frac{162\,061}{31}\,\wp_{12}^6
-\frac{5\,834\,196}{31}\,G_4\wp_{12}^4
-\frac{9\,680\,300}{31}\,G_6\wp_{12}^3
+\frac{40\,948\,740}{31}\,G_4^2\wp_{12}^2
+\frac{100\,575\,720}{31}\,G_4G_6\wp_{12}$\\
&\hspace{0.5cm}
$-\frac{10\,618\,992}{31}\,G_4^3
+\frac{3\,301\,504\,920}{251}G_6^2
\biggr)\,\langle 1\rangle_A
$\\
&$+\biggl(
-\frac{66\,948}{31}\,\wp_{12}^5
+\frac{2\,060\,472}{31}\,G_4 \wp_{12}^3
+\frac{3\,420\,312}{31}\,G_6 \wp_{12}^2
-\frac{8\,786\,880}{31}\,G_4^2 \wp_{12}
-\frac{16\,258\,032}{31}\,G_4G_6
\biggr)\,\langle T\rangle_A/c$\\
\noalign{\smallskip}\hline
\end{tabular}\\
\vspace{0.3cm}
\caption{List of the terms for $c=-22/5$ and conformal holomorphic weight $0\leq h\leq 6$ in eq.~(\ref{eq: general 0-pt fct for g=2 as expansion in rho}). 
(The label $g=1$ is dropped.).
}
\label{table: the first few terms in eq. (19)}
\end{table}
\end{proposition}

\begin{proof}
It suffices to show that the term for $h=6$ is correct, and thus that 
\begin{equation*}
\begin{split}
\bigl\langle\bigl(&7L_4L_2-2L_6\bigr)(z_1)\bigl(7L_4L_2-2L_6\bigr)(z_2)\bigr\rangle_A\\
\;=\;&
\biggl(
-1\,062\,817\,\wp_{12}^6
+38\,261\,412\,G_4 \wp_{12}^4
+63\,465\,500\,G_6 \wp_{12}^3
-268\,595\,460\,G_4^2 \wp_{12}^2\\
&-659\,501\,640\,G_4G_6 \wp_{12}
+1\,512\Bigl(46\,062\,G_4^3
-176\,645\,G_6^2\Bigr)
\biggr)\,c\langle 1\rangle_A\\
+&\biggl(
121\,065\,\wp_{12}^5
-3\,723\,006\,G_4\wp_{12}^3
-6\,180\,846\,G_6\wp_{12}^2
+15\,845\,760\,G_4^2\wp_{12}\\
&+29\,310\,876\,G_4G_6
\biggr)\,\frac{\langle T\rangle_A}{4}\:.
\end{split}
\end{equation*}
We list the individual contributions:
\hbox{$\langle L_6(z_1)L_6(z_2)\rangle_A$} is the coefficient of \hbox{$(z-z_1)^4(w-z_2)^4$} in \hbox{$\langle T(z)T(w)\rangle_A$}
for \hbox{$|z_1-z_2|$} and \hbox{$|w-z_2|$} small.
\begin{equation*}
\begin{split}
\bigl\langle L_6(z_1)L_6(z_2)\bigr\rangle_A
=\biggl(&
5\,775\,\wp_{12}^6
-207\,900 \,G_4\,\wp_{12}^4
-346\,500 \,G_6\,\wp_{12}^3
+1\,455\,300\,G_4^2\,\,\wp_{12}^2\\
&+3\,591\,000\,G_4G_6\,\wp_{12}
-378\,000\,G_4^3
+1\,470\,000\,G_6^2
\biggr)\,c\langle 1\rangle_A\\
+\biggl(&
1\,260\,\wp_{12}^5
-\bigl(63\,000\,G_6+37800\,G_4\bigr)\,\wp_{12}^3
+151\,200\,G_4^2\,\,\wp_{12}\\
&+277\,200\,G_4G_6
\biggr)\,
\langle T\rangle_A
\:.
\end{split}
\end{equation*}
$\langle L_4L_2(z_1)L_6(z_2)\rangle_A$ is the coefficient of $(z-z_1)^2$ in $\langle T(z)T(z_1)T^{(4)}(z_2)\rangle_A$ for $|z-z_2|$ small.\begin{equation*}
\begin{split}
\bigl\langle L_4L_2(z_1)L_6(z_2)\bigr\rangle_A
\;=\;
\biggl(&
9\,878\,400\,G_6^2
-2\,592\,000\,G_4^3
+24\,494\,400\,G_4G_6\,\wp_{12}
+9\,979\,200\,G_4^2\,\,\wp_{12}^2\\
&-2\,368\,800\,G_6\,\wp_{12}^3
-1\,425\,600\,G_4\,\wp_{12}^4
+39\,600\,\,\wp_{12}^6
\biggr)\,c\langle 1\rangle_A\\
-
\biggl(&
4\,233\,600\,G_4G_6
+2\,332\,800\,G_4^2\,\,\wp_{12}
-907\,200\,G_6\,\wp_{12}^2
-552\,960\,G_4\,\wp_{12}^3\\
&+18\,144\,\wp_{12}^5
\biggr)\,
\langle T\rangle_A\,.
\end{split}
\end{equation*}
$\langle L_4L_2(z_1)L_4L_2(z_2)\rangle_A$ is the coefficient of $(z-z_1)^2(w-z_2)^2$ in $\langle T(z)T(z_1)T(w)T(z_2)\rangle_A$
for $|z-z_1|$ and $|w-z_2|$ small,
\begin{equation*}
\begin{split}
&\bigl\langle L_4L_2(z_1)L_4L_2(z_2)\bigr\rangle_A 
\;=\;\biggl(
74\,040\,G_6^2
-28\,944\,G_4^3
+244\,440\,G_4G_6\,\wp_{12}\\
&\qquad+102\,060\,G_4^2\,\,\wp_{12}^2
-30\,100\,G_6\,\wp_{12}^3
-16\,812\,G_4\,\wp_{12}^4
+467\,\wp_{12}^6
\biggr)\,c\langle 1\rangle_A\\
&\qquad+
\biggl(
49\,104\,G_4G_6
+51\,840\,G_4^2\,\,\wp_{12}
+18\,984\,G_6\,\wp_{12}^2
+15\,144\,G_4\,\wp_{12}^3\\
&\hspace{1cm}-588\,\wp_{12}^5
\biggr)\,
\langle T\rangle_A
\:.
\end{split}
\end{equation*}
This completes the proof.
\end{proof}

Since the overall number of holomorphic blocks in genus two is five
and in the ${q}$ formalism, only two of them are obtained from inserting the identity field
(corresponding to the pair of the Rogers-Ramanujan functions),
the remaining three solutions must be given by the two-point function of $\Phi$.
Indeed, as we have shown in Proposition \ref{proposition: 3rd order ODE for 3-pt fct of Phi in the analytic coordinate}, 
$\bigl\langle\Phi(z)\,\Phi(0)\bigr\rangle$ satisfies a $3$\ts{rd} order ODE.

Solving eq.~(\ref{3rd order ODE for 2-pt function of varphi in the analytic coordinate z}) 
will allow to compute the coefficients of \hbox{${q}^{k-1/5}/\parallel\Phi\parallel^2$} for \hbox{$k\geq 4$}
in the continuation of eq.~(\ref{eq: general 0-pt fct for g=2 as expansion in rho}) to \hbox{$A=3,4,5$}. 
For example, 
\hbox{$\bigl\langle L_4\Phi(z_1)\,L_1L_3\Phi(z_2)\bigr\rangle$} sorts out (in particular) the coefficient 
proportional to \hbox{$(z-z_1)^2(u-z_2)^{-1}(v-z_2)$} in the $5$-point function 
\hbox{$\bigl\langle T(z)T(u)T(v)\,\Phi(z_1)\,\Phi(z_2)\bigr\rangle$}.
A graphical proposition for correlation functions of $T$ and two copies of $\Phi$ is desirable. 

\subsection{The genus two partition function in the (2,5) minimal model}\label{subsec: The genus two partition function in the (2,5) minimal model}

Our methods provide the means to compute the genus two partition function to any order in the sewing parameter ${s}$:

\begin{theorem}\label{theorem: the full genus 2 partition fct}
For $0<|{s}|\ll 1$, 
let $M_2({s})$ be the genus two Riemann surface defined by eq.~(\ref{description: genus two Riemann surface with (unspecified) metric}) 
with metric $\G\LN(s)$ from eq.~(\ref{def: genus two metric obtained by inserting a cylinder between two flat tori}).
In the $(2,5)$ minimal model,
the partition function on $M_2({s})$ 
equals
\begin{equation*}
\mathfrak{Z}_{M_2({s})}
\;=\;|s|^{11/30}
\Bigl(\sum_{a,b=1,2}\bigl|\langle 1\rangle({s},\tau,\hat{\tau})_{(a,b)}^{g=2}\bigr|^2
-|s|^{-2/5}\,\bigl|\langle 1\rangle({s},\tau,\hat{\tau})_{\Phi}^{g=2}\bigr|^2\Bigr)
\:,
\end{equation*}
where the holomorphic blocks are those from Proposition \ref{proposition: Taylor expansion in epsilon of the genus two 0-pt function obtained from the Rogers-Ramanujan functions and the one corresponding to the primary field Phi}.
\end{theorem}

In particular, for small values of $|{s}|$, the partition function is negative.

\begin{proof}
For $g=2$, we have $p_2(A)=0$ and $\cc(A)=3$, so $\eps_A=-1$ according to Proposition \ref{proposition: the prefactor eps A},
since $\eps_{\Phi}=-1$ by eq.~(\ref{eq: eps Phi}).
So for $M_2=M_2({s}=1)$, Segal's formula (\ref{eq: Segal's s formula}) gives
\begin{align*}
\mathfrak{Z}_{M_2}
\;=\;&
\mathfrak{Z}\flatz^{g=1}\,\widehat{\mathfrak{Z}\flatz^{g=1}}
-\langle{\Phi}(z)\rangle\flatz^{g=1}
\reallywidehat{\langle{\Phi}(z)\rangle\flatz^{g=1}}
+\ldots
\:.
\end{align*}
Inserting the cylinder $(\gamma\times[0,-\log|s|])^{\arg s}$ yields the claimed identity, by eq.~(\ref{eq: effect of reducing cylinder to lenth zero on a 1-pt fct in g=1}).
\end{proof}

When dealing with general genus two surfaces, 
the metric (\ref{def: genus two metric obtained by inserting a cylinder between two flat tori}) is a rather unnatural choice.
We conclude the chapter by discussing an alternative approach, 
which uses the description of genus $g$ Riemann surfaces as branched coverings of the Riemann sphere.
The latter carries a metric that is induced by the natural distance function on $\C$.
This metric lifts to a metric on the genus $g$ surface 
which is singular at the branch points and flat outside.
Thus the metric surface resembles a polyhedron.
To regularise the metric, every ramification point $P$ on the surface is replaced by a flat disc centred at $P$,
with the amount of curvature spread evenly over the bounding circle. 
The radius $\varrho>0$ of the circle is chosen in such a way that for any pair of ramification points $P,P'$ on the surface,
the corresponding pair of closed discs has nonempty intersection iff $P=P'$.
The partition function associated with the regularised polyhedral metric will be called regularised partition function
and denoted by $\mathfrak{Z}^{g}\polyhedral$ (in genus $g$).
We compute the renormalised partition function in genus two, 
which is obtained from $\mathfrak{Z}^{g=2}\polyhedral$ by omitting the dependence on the radii of the regularisation.
(In practice, after application of a Weyl transformation if necessary, we set all radii equal to one.) 

Let $\Sigma_g$ be a hyperelliptic surface of genus $g\geq 0$,
which is unramified over the point at infinity.
Thus all $2g+2$ ramification points lie at finite distance from its opposite point,
the curvature distribution derives from proposition of Gauss-Bonnet according to 
\begin{equation*}
4\pi(1-g)
=\int_{\Sigma_g}\mathcal{K}
=\alpha(2g+2)+\beta
\:,
\end{equation*}
where $\alpha,\beta\in\R$.
Thus $\alpha g=-2\pi g$ and $2\alpha+\beta=-4\pi$,
where $\beta$ is the total amount of curvature that corresponds (in the way mentioned above) to the point at infinity,
while $\alpha$ is the curvature corresponding to a single ramification point.
In genus $g\geq 1$, $\alpha=-2\pi$, and $\beta=8\pi$.

Let $\Sigma_1$ and $\widehat{\Sigma}_1$ be flat tori of modulus $\tau$ and $\hat{\tau}$, respectively.
For $k=1,2,3$, let $x_k$ and $\hat{x}_k$ be the value of $\wp(z|\tau)$ and of $\wp(\hat{z}|\hat{\tau})$, 
respectively, at the half periods.
For the pair of coordinates $X,\widehat{X}$ from Proposition \ref{proposition: almost global coordinates on CP1 satisfying the epsilon condition},
let 
\begin{equation*}
X_k
\;=\;X(x_k)\:,
\hspace{0.7cm}
\widehat{X}_{k}
\;=\;\widehat{X}(\hat{x}_{k})\:,
\hspace{0.7cm}
X_{k+3}
\;:=\;\xi/(s^2\widehat{X}_k)\:,
\hspace{0.7cm}
k=1,2,3\:.
\end{equation*}
Let $\Sigma_2$ be the genus $g=2$ surface with ramification points $X_1,\ldots,X_6$. 
Let
\begin{equation*}
{Y}_k
:\;=\;({X}-{X}_k)^{1/2}
\:,
\hspace{0.7cm}
k=1,\ldots,6
\:,
\end{equation*}
be local coordinates on mutually disjoint discs in $\C$ of radius ${\varrho_k}>0$ about ${X}_k$,
and let $\widetilde{X}:={X}^{-1}$ be the coordinate (on both sheets) close to the point at infinity.
The regularised polyhedral metric on $\Sigma_2$
(associated to the numbers ${{\varrho_1}},{\varrho_2},{\varrho_3},{\varrho_4},{\varrho_5},{\varrho_6}$ and $\varrho_{\infty}$) 
is defined by
\begin{equation}\label{def: regularised polyhedral metric for genus two}
\G\polyhedral
=
\begin{cases}
4{\varrho_k}\,|d{Y}_k|^2&\quad\text{for $|{X}-{X}_k|\leq {\varrho_k}$\ for $k=1,\ldots,6$},\\
|d{X}|^2&\quad\text{for ${\varrho_k}\leq |{X}-{X}_k|<\varrho_{\infty}$, for $k=1,\ldots,6$}\:, \\
&\quad\text{and for $|{X}|\leq \varrho_{\infty}$}\:, \\
\varrho_{\infty}^4\,|d\widetilde{X}|^2&\quad\text{for $\varrho_{\infty}\leq |{X}|$}.
\end{cases}
\end{equation}
Note that the metric is everywhere continuous.

\begin{theorem}\label{theorem: renormalised partition function in genus two}
Let $c=-22/5$.
Let $x_1,x_2,x_3$ and $\hat{x}_1,\hat{x}_2,\hat{x}_3$ be the functions 
of the modulus $\tau$ of $\Sigma_1$ and $\hat{\tau}$ of $\widehat{\Sigma}_1$, respectively,
defined by eq.~(\ref{eqs: values of Weierstrass p at half periods put into an order}).
Let $X,\widehat{X}$ be the pair of almost global coordinates on $\CP^1$ from Proposition \ref{proposition: almost global coordinates on CP1 satisfying the epsilon condition}.
Let $\mathfrak{Z}_{M_2({s})}$ be the genus two partition function from Theorem \ref{theorem: the full genus 2 partition fct},
where $0<|{s}|\ll 1$. 
The renormalised genus two partition function w.r.t.\ the polyhedral metric (\ref{def: regularised polyhedral metric for genus two})
is given by
\begin{align*}
\Bigl(\mathfrak{Z}\polyhedral^{g=2}\Bigr)^{\text{renorm}}
=\;&2^{-c/3}
|s|^{c/6}
\,\Bigl|\frac{s^2}{\xi}\Bigr|^{11c/24}\,
\\
&|\Delta_{1,2,3}\,\widehat{\Delta}_{1,2,3}|^{-c/24}
|X'(x_1)X'(x_2)X'(x_3)\,\widehat{X}'(\hat{x}_1)\widehat{X}'(\hat{x}_2)\widehat{X}'(\hat{x}_3)|^{-c/24}\,\\
&|\widehat{X}_1\widehat{X}_2\widehat{X}_3|^{c/12}|\hat{x}_1\hat{x}_2\hat{x}_3|^{c/6}
\,\exp\Bigl(\frac{c}{6}\,\text{Res}_{z=0}\Bigl[\frac{1}{z}\log X'(x)\Bigr]\Bigr)\,\mathfrak{Z}_{M_2({s})}
\:,
\end{align*}
where $X'(x)=\frac{dX}{dx}$ etc, and
\begin{equation*}
\Delta_{1,2,3}
\;=\;\underset{j,k\in\{1,2,3\}}{\prod_{j\not=k}}|x_j-x_k|^2\:,
\hspace{0.5cm}
\widehat{\Delta}_{1,2,3}=\underset{j,k\in\{1,2,3\}}{\prod_{j\not=k}}|\hat{x}_j-\hat{x}_k|^2
\:.
\end{equation*}
\end{theorem}

We proof makes use of the following 

\begin{lemma}\label{lemma: integration on rotationally invariant region}
Let $P\in\C$, and let $U_P\subseteq\C$ be an open set containing $P$.
Let $\mathcal{K}_P$ be a two-form with support on $U_P$.
Suppose $U_P$ and $\mathcal{K}_P$ are invariant under rotation around $P$.
Let $f$ be a harmonic function on $U_P$ (i.e. $\partial_z\partial_{\bar{z}}f=0$ on $U_P$).
The following is true:
\begin{equation*}
\iint_{U_P}f(z)\,\mathcal{K}_P
\;=\;f(z(P))\;\iint_{U_P}\mathcal{K}_P
\:.
\end{equation*}
In particular,
for $n\in\Z$, 
\begin{equation*}
\frac{\iint_{U_P} z^n\,\mathcal{K}_P}{\iint_{U_P}\mathcal{K}_P}
\;=\;
\begin{cases}
\:z(P)^n&\text{in case $z(P)\not=0$}\:,\\
\:\delta_{n,0}&\text{in case $z(P)=0$}\;.
\end{cases}
\end{equation*}
\end{lemma}


\begin{proof}
By assumption on $f$, there exists a holomorphic function $F$ on $U_P$ such that $f=F+\overline{F}$.
Close to $z(P)$, $F$ admits a Taylor expansion in powers  of $(z-z(P))$. 
A rotation $z\mapsto z'$ around $P$ leaves $\mathcal{K}_P$ invariant,
but the only term $(z'-z(P))^n$ invariant under the coordinate change is the one for $n=0$.
\end{proof}

\begin{proof}[Proof of Theorem \ref{theorem: renormalised partition function in genus two}]
Let $dz$ and $d\hat{z}$ be the holomorphic one-form on the flat tori $\Sigma_1$ and $\widehat{\Sigma}_1$, respectively.
The proof of eq.~(\ref{eq: cylinder partition function})
shows that the partition function for the metric 
\begin{equation}\label{def: genus two metric obtained by sewing flat tori}
\G\MT(s)
=
\begin{cases}
|dz|^2&\quad\text{$\sqrt{|{s}|}\leq |z|$}\;,\\
|d\hat{z}|^2&\quad\text{$\sqrt{|{s}|}\leq |\hat{z}|$}\:,
\end{cases}
\end{equation}
on $\Sigma_2$ equals 
\begin{equation}\label{eq: MT partition function in genus two}
\mathfrak{Z}\MT^{g=2}
\;=\;
|{s}|^{c/12}\,\mathfrak{Z}_{M_2({s})}
\:.
\end{equation}
The metric (\ref{def: genus two metric obtained by sewing flat tori}) is continuous by eq.~(\ref{eq: z1z2=eps}).
Let $\G_A$ and $\G_B$ be the metric (\ref{def: genus two metric obtained by sewing flat tori})
and (\ref{def: regularised polyhedral metric for genus two}), respectively.
The curvature of the metric $\G_A$ from eq.~(\ref{def: genus two metric obtained by sewing flat tori}) 
is supported on the cycle 
\begin{equation}\label{sets: curvature circle of MT metric}
\{|z|=|{s}|^{1/2}\}
\;=\;\{|\hat{z}|=|{s}|^{1/2}\} 
\end{equation}
and has integral $-4\pi$ by Gauss-Bonnet.
On the other hand, all curvature of the regularised polyhedral metric $\G_B$ from eq.~(\ref{def: regularised polyhedral metric for genus two}) is spread over the circles bounding the discs centred at $X_1,\ldots,X_6$ and the point at infinity. 
Let $\alpha=-2\pi$ and $\beta=8\pi$ be the integrated curvature on the circle
\begin{equation}\label{set: curvature circle around k th finite ramification point}
\{|X-X_k|=\varrho_k\}
\:,
\end{equation}
for $k=1,\ldots,6$, and on 
\begin{equation}\label{set: curvature circle around the point at infinity}
\{|{X}|=\varrho_{\infty}\}\:, 
\end{equation}
respectively. 
Let $\mathcal{K}_k$ for $k=1,\ldots,6$ and $\mathcal{K}_{\infty}$ be the corresponding curvature two-form. 

Thus we have to compute the Weyl-factor on eight circles carrying curvature for one of the metrics $\G\initial$ and $\G\final$. 
On each of them the almost global coordinate $X$ from Proposition \ref{proposition: almost global coordinates on CP1 satisfying the epsilon condition} is defined. 
Namely, the respective Weyl factor is given as follows:
\begin{align*}
|dz|^2\mapsto|dX|^2
\;=\;&
\begin{cases}
\exp(\Delta\sigma\initial)\,|dz|^2 
&\text{on the cycle (\ref{sets: curvature circle of MT metric})},\\
\exp(\Delta\sigma\final)\,|dz|^2
&\text{on the circle (\ref{set: curvature circle around k th finite ramification point}) for $k=1,2,3$},
\end{cases}\\
|d\hat{z}|^2\,\mapsto\,|d{X}|^2
\;=\;&
\begin{cases}
\exp(\Delta\widehat{\sigma}\final)\,|d\hat{z}|^2
&\text{on the circle (\ref{set: curvature circle around k th finite ramification point}) for $k=4,5,6$},\\
\exp(\Delta\widehat{\sigma}_{\infty})\,|d\hat{z}|^2 
&\text{on the circle (\ref{set: curvature circle around the point at infinity})},
\end{cases}
\end{align*}
where 
\begin{align*}
\Delta\sigma\initial
\;=\;\Delta\sigma\final
\;=\;&\log\Bigl|\frac{dX}{dx}\Bigr|^2+\log\Bigl|\frac{dx}{dz}\Bigr|^2
\:,\\
\Delta\widehat{\sigma}\final
\;=\;\Delta\widehat{\sigma}_{\infty}
\;=\;&\log\Bigl|\frac{dX}{d\widehat{X}}\Bigr|^2
+\log\Bigl|\frac{d\widehat{X}}{d\hat{x}}\Bigr|^2
+\log\Bigl|\frac{d\hat{x}}{d\hat{z}}\Bigr|^2
\:.
\end{align*}
Note that by eq.~(\ref{eq: X hat X= xi/(s squared)}),
\begin{equation*}
\log\Bigl|\frac{dX}{d\widehat{X}}\Bigr| 
\;=\;\log|X|-\log|\widehat{X}|
\;=\;2\log|X|+\log\Bigl|\frac{s^2}{\xi}\Bigr|
\:.
\end{equation*}
Now we have the following list of integrals:
\begin{align}
\iint&\mathcal{K}\initial\log\Bigl|\frac{dx}{dz}\Bigr|^2
\;=\;-4\pi\Bigl(2\log2-3\log|s|\Bigr)
\:,\label{MT integral: log|dx/dz|} \\
\iint&\mathcal{K}\initial\log\frac{dX}{dx} 
\;=\;-4\pi\,\text{Res}_{z=0}\Bigl[\frac{1}{z}\log\frac{dX}{dx}\Bigr]
\:.\label{MT integral: log dX/dx}
\end{align}
For $k=1,2,3$,
\begin{align}
\iint\mathcal{K}_k\log\Bigl|\frac{dx}{dz}\Bigr|^2
\;=\;&-2\pi
\Bigl(2\log 2
+\log\varrho_k
+\sum_{j\not=k}\log|x_k-x_j|
-\log|X'(x_k)|
\Bigr)
\:,\label{k integral: log|dx/dz| squared} \\
\iint\mathcal{K}_k\log\Bigl|\frac{dX}{dx}\Bigr|^2
\;=\;&-2\pi\log|X'(x_k)|^2
\:,\label{k integral: log |dX/dx| squared}\\
\iint\mathcal{K}_{k+3}\log |X|
\;=\;&-2\pi\,\log|X_{k+3}|
\:.\label{k integral: log X}
\end{align}
and
\begin{align*}
\iint\mathcal{K}_{k+3}\log\Bigl|\frac{d\hat{x}}{dz}\Bigr|^2
\;=\;&-2\pi
\Bigl(2\log 2
+\log\widehat{\varrho_k}
+\sum_{j\not=k}\log|\hat{x}_k-\hat{x}_j|
-\log|\widehat{X}'(\hat{x}_k)|
\Bigr)
\:, \\
\iint\mathcal{K}_{k+3}\log\Bigl|\frac{d\widehat{X}}{d\hat{x}}\Bigr|^2 
\;=\;&-2\pi\log|\widehat{X}'(\hat{x}_k)|^2
\:,
\end{align*}
where for $k=1,2,3$,
\begin{equation*}
\log\widehat{\varrho_k}
\;=\;\log|\widehat{X}-\widehat{X}_k|
\;=\;\log\varrho_{k+3}-\log|X_{k+3}|+\log|\widehat{X}_k|
\:.
\end{equation*}
Moreover,
\begin{align}
\iint\mathcal{K}_{\infty}\log|X|
\;=\;&8\pi\,\log\varrho_{\infty}\:,\label{infty integral: log|X|}\\
\iint\mathcal{K}_{\infty}\log\frac{d\widehat{X}}{d\hat{x}}
\;=\;&0\:,\nn\\
\iint\mathcal{K}_{\infty}\log\Bigl|\frac{d\hat{x}}{d\hat{z}}\Bigr|^2
\;=\;&8\pi\Bigl(2\log2+\log|\hat{x}_1\hat{x}_2\hat{x}_3|\Bigr)
\:.\label{infty integral: log|d hat x/d hat z| squared}
\end{align}
We conclude that
\begin{align*}
\mathfrak{Z}\polyhedral^{g=2}/\mathfrak{Z}\MT^{g=2}
\;=\;&\exp\biggl\{
\frac{c}{48\pi}
\iint \Delta\sigma\initial\,\mathcal{K}\initial
+
\sum_{k=1}^3
\Bigl(\Delta\sigma\final\,\mathcal{K}_k
+\Delta\widehat{\sigma}\final\,\mathcal{K}_{k+3}\Bigr)
+\Delta\widehat{\sigma}_{\infty}\mathcal{K}_{\infty}
\biggr\}
\:,
\end{align*}
where
\begin{align*}
\exp\biggl\{
\frac{c}{48\pi}&
\iint \Delta\sigma\initial\,\mathcal{K}\initial
\biggr\}
\;=\;2^{-c/6}|s|^{c/4}\,\exp\Bigl(\frac{c}{6}\,\text{Res}_{z=0}\Bigl[\frac{1}{z}\log X'(x)\Bigr]\Bigr)
\:,
\end{align*}
and
\begin{align*}
\exp\biggl\{
&\frac{c}{48\pi}
\iint
\sum_{k=1}^3
\Bigl(\Delta\sigma\final\,\mathcal{K}_k
+\Delta\widehat{\sigma}\final\,\mathcal{K}_{k+3}\Bigr)
+\Delta\widehat{\sigma}_{\infty}\mathcal{K}_{\infty}
\biggr\}\\
\;=\;&
2^{-c/6}\,
(\varrho_1\ldots\varrho_6)^{-c/24}\,\varrho_{\infty}^{2c/3}
\,\Bigl|\frac{s^2}{\xi}\Bigr|^{11c/24}\,
|\Delta_{1,2,3}|^{-c/24}|\widehat{\Delta}_{1,2,3}|^{-c/24}\\
&|X'(x_1)X'(x_2)X'(x_3)|^{-c/24}\,|\widehat{X}'(\hat{x}_1)\widehat{X}'(\hat{x}_2)\widehat{X}'(\hat{x}_3)|^{-c/24}\,
|\widehat{X}_1\widehat{X}_2\widehat{X}_3|^{c/12}|\hat{x}_1\hat{x}_2\hat{x}_3|^{c/6}
\:.
\end{align*}
The renormalised partition function is obtained by setting $\varrho_1=\ldots=\varrho_6=\varrho_{\infty}=1$
and the claimed follows by taking eq.~(\ref{eq: MT partition function in genus two}) into account.
\end{proof}

Due to the choice of coordinate in the definition of the metric (\ref{def: regularised polyhedral metric for genus two}),
the renormalised partition function in Theorem \ref{theorem: renormalised partition function in genus two}
is not symmetric under exchange $\tau\leftrightarrow\hat{\tau}$.
Indeed, for $x_1,x_2,x_3$ from eq.~(\ref{eqs: values of Weierstrass p at half periods put into an order}),
\begin{equation*}
\sum_{i=1}^3x_i
\;=\;0\:,
\hspace{0.4cm}
\sum_{i=1}^3x_i^2
\;=\;60\,G_4\:,
\hspace{0.4cm}
\sum_{i=1}^3x_i^3
\;=\;210\,G_6\:,
\end{equation*}
and the corresponding set of equations for $\sum_{i=1}^3x_i^k$ with $4\leq k\leq 6$,
together with Proposition \ref{proposition: almost global coordinates on CP1 satisfying the epsilon condition}
yields
\begin{align*}
\text{Res}_{z=0}&\Bigl[\frac{1}{z}\log\frac{dX}{dx}\Bigr]\\
\;=\;&-216\,s^4\,G_4 \widehat{G_4}
-1\,200\,s^6\,G_6 \widehat{G_6}
-3\,312\,s^8\,G_4^2 \widehat{G_4}^2
-\frac{12\,614\,400}{121}\,s^{10}\,G_4 G_6 \widehat{G_4} \widehat{G_6}\\
&-\frac{384}{169}\,s^{12} 
\Bigl(
437\,157\,G_4^3\widehat{G_4}^3
+177\,075\,G_4^3\widehat{G_6}^2
+139\,050\,G_6^2\widehat{G_4}^3
+68\,125\,G_6^2\widehat{G_6}^2
\Bigr)\\
&+\frac{9\,845\,452\,800}{121}\,G_4^2G_6\,\widehat{G_4}^2\widehat{G_6}s^{14}
+\text{O$(s^{16})$} 
\:,
\end{align*}
and
\begin{align*}
\log&\Bigl|\frac{\widehat{X}_1\widehat{X}_2\widehat{X}_3}{\hat{x}_1\hat{x}_2\hat{x}_3}\Bigr|\\
\;=\;
&-360\,G_4\widehat{G_4}\,s^4
-2\,100\,G_6\widehat{G_6}\,s^6
+23\,760\,G_4^2\widehat{G_4}^2\,s^8
+\frac{3\,412\,800}{11}\,G_4G_6\widehat{G_4}\widehat{G_6}\,s^{10}\\
&+\frac{120}{13}s^{12}\Bigl(
119\,664\,G_4^3\widehat{G_4}^3-70\,560\,G_4^3\widehat{G_6}^2-61\,200\,G_6^2\widehat{G_4}^3-83\,125\,G_6^2\widehat{G_6}^2
\Bigr)\\
&-\frac{1\,364\,904\,000}{11}\,G_4^2G_6\,\widehat{G_4}^2\widehat{G_6}s^{14}
+\text{O$(s^{16})$}
\:.
\end{align*}
Thus the coefficient of $s^{12}$ in the expression 
$
\log\Bigl|\frac{\hat{x}_1\hat{x}_2\hat{x}_3}{\widehat{X}_1\widehat{X}_2\widehat{X}_3}\Bigr| 
+\text{Res}_{z=0}\Bigl[\frac{1}{z}\log\frac{dX}{dx}\Bigr]
$
reads
\begin{equation*}
\frac{-354\,544\,128}{169}\,G_4^3\widehat{G_4}^3+\frac{42\,076\,800}{169}\,G_4^3\widehat{G_6}^2+\frac{42\,076\,800}{169}\,G_6^2 \widehat{G_4}^3+\frac{103\,515\,000}{169}\,G_6^2\widehat{G_6}^2\:, 
\end{equation*}
so that
\begin{equation*}
\bigl(\mathfrak{Z}\polyhedral^{g=2}\bigr)^{\text{renorm}}
=\;|\widehat{X}_1\widehat{X}_2\widehat{X}_3|^{c/4}\times\bigl(\text{symmetric expression}+\text{O$(s^{16})$}\bigr)\:.
\end{equation*}
We explain the factor of $|\widehat{X}_1\widehat{X}_2\widehat{X}_3|^{c/4}$.
W.~Nahm suggests the following:

\begin{definition}
For $N\in\N$ and for $i=1,\ldots,N+1$, 
let $X_i\in\C\cup\{\infty\}$ with $X_i\not= X_j$ for $i\not=j$. 
For $A\in\SL(2,\C)$, set
\begin{equation*}
{\prod_{i<j}}'\bigl(A(X_i)-A(X_j)\bigr)^2 
\;=\;\prod_{i=1}^{N+1}\Gamma_A(X_i)^{-N}\,{\prod_{i<j}}'(X_i-X_j)^2
\:.
\end{equation*}
Here the dash means that infinite factors are omitted from the product.
\end{definition}
It follows immediately that the $\Gamma_A$ for $A\in\SL(2,\C)$ satisfy the cocycle condition,
namely for $A,A'\in\SL(2,\C)$, we have
\begin{equation*}
\Gamma_{AA'}(X)
\;=\;\Gamma_A(A'(X))\,\Gamma_{A'}(X)
\:.
\end{equation*}
Explicitly,
$A=
\begin{pmatrix}
    \alpha&\beta\\
    \gamma&\delta
\end{pmatrix}
$ acts on $X\in\C\cup\{\infty\}$ by 
$AX=\frac{\alpha X+\beta}{\gamma X+\delta}$,
so
\begin{equation*}
\Gamma_A(X)
\;=\;
\begin{cases}
(\gamma X+\delta)^2&\text{for $X\in\C$, provided $\gamma X+\delta\not=0$},\\
\gamma^{-2}&\text{for $X\in\C$ such that $\gamma X+\delta=0$},\\
\gamma^2&\text{for $X=\infty$ and $\gamma\not=0$},\\
\delta^{-2}&\text{for $X=\infty$ and $\gamma=0$}.
\end{cases}
\end{equation*}

Eq.~(\ref{eq: Nahm's formula}) implies
\begin{proposition}\label{proposition: effect of linear fractional trsf applied to ramification points on the partition function in genus g}
Let $\mathcal{X}=\{X_i\}_{i=1}^{2g+2}\subset\C\cup\{\infty\}$ be the set of ramification points
of a hyperelliptic genus $g$ Riemann surface with metric $|dX|^2$.
Let $\bigl(\mathfrak{Z}(\mathcal{X})\bigr)^{\text{renorm}}$
be the corresponding renormalised partition function.
For $A\in\SL(2,\C)$, we have
\begin{equation*}
\bigl(\mathfrak{Z}\bigl(A(\mathcal{X})\bigr)\bigr)^{\text{renorm}}
\;=\;{\prod_{i=1}^{2g+2}}|\Gamma_A(X_i)|^{c/4}\bigl(\mathfrak{Z}(\mathcal{X})\bigr)^{\text{renorm}}
\:.
\end{equation*}
\end{proposition}

\begin{example}
Let $A:\CP^1\rechts\CP^1$ be defined by $A(X)=\xi/(s^2X)=\widehat{X}$.
Proposition \ref{proposition: effect of linear fractional trsf applied to ramification points on the partition function in genus g} yields
\begin{equation*}
\bigl(\mathfrak{Z}\bigl(\{\widehat{X}_i\}_{i=1}^6\bigr)\bigr)^{\text{renorm}}
\;=\;\Bigl|\frac{s^2}{\xi}\Bigr|^{3c/4}\,{\prod_{i=1}^{6}}|X_i|^{c/4}\,
\bigl(\mathfrak{Z}\bigl(\{X_i\}_{i=1}^6\bigr)\bigr)^{\text{renorm}}
\:.
\end{equation*}
\end{example}

These observations can be used to define a universal partition function,
which does not depend on the specific set of ramification points of the genus $g$ Riemann surface,
but rather on the conformal class they define,
namely
\begin{equation*}
{\underset{i<j}{\prod_{i,j=1}^{2g+2}}}'\,|X_i-X_j|^{\frac{c}{4(2g+1)}}
\,\bigl(\mathfrak{Z}(\mathcal{X})\bigr)^{\text{renorm}}
\:.
\end{equation*}
%
%

\subsection{Discussion and outlook}

We have obtained a power series description of an object with good automorphic properties under the mapping class group in genus two.
This is provided by the $(2,5)$ minimal model studied above, apart from ramifications into questions of metric dependence.
We suggest using the polyhedral metric $\G\polyhedral$, which defines an almost natural metric on Riemann surfaces of arbitrary genus. 
A constant curvature metric seems preferable, and the corresponding partition function might be computable.
We would like to make contact with Siegel modular forms, but this requires the consideration of Dehn twists.

Dehn twists about simple closed curves $\gamma$, the type considered both in the ${s}$ and the ${q}$ sewing scheme,
define elements in the mapping class group $\Gamma_2$ (for genus two).
Those on separating simple closed (SSC) curves generate the Torelli group $\sheaf{I}_2$,
which is related to the Siegel modular group $\Sp(4,\Z)$ by the short exact sequence
\begin{equation*}
1\rechts \sheaf{I}_2\rechts\Gamma_2\rechts\Sp(4,\Z)\rechts 1 
\:.
\end{equation*}
Due to a fifth root singularity in ${s}$, 
the holomorphic block (\ref{def: fifth solution}) transforms trivially only under a quotient 
of $\Gamma_2$ by fifth powers of such Dehn twists.
This quotient is infinite for $g\geq 2$ \cite{Humphries:1991}.

Since ${s}=0$ is a regular interior point on the Siegel upper half plane $\mathfrak{H}_2$,
our holomorphic blocks do not form a Siegel modular form.
Taking fifth powers doesn't lift the singularity since the holomorphic blocks
transform into non-trivial linear combinations among themselves.
One should be able to construct a scalar representation of $\Gamma_2$, however, which can then be related to Siegel modular functions. 

As discussed in Section \ref{subsection: q formalism},
an alternatively way to compute $\mathfrak{Z}_{M_2}$ is by use of the ${q}$ formalism
though this requires knowledge of $\langle\Phi\Phi\rangle^{g=1}$.
The singular Riemann surface corresponding to ${q}=0$ defines a boundary point of the Siegel upper half plane $\mathfrak{H}_2$.
It is desirable to understand better the matching between the ${s}$ and the ${q}$ formalism,
since the partition function they deliver is the same.
A main task is to show that the solutions obtained through perturbative expansion
extend as a one-valued object to the full moduli space. This requires an understanding of the monodromy group.

The occurrence of linear differential equations of the  Kaneko-Zagier type \cite{Z-K:1997} in genus one,
and the $3$\ts{rd} order ODE from Corollary \ref{proposition: 3rd order ODE for 3-pt fct of Phi in the analytic coordinate} for $g=2$
is a general feature of holomorphic blocks in any genus. 
Since the use of algebraic coordinates allows to deal with (hyperelliptic) Riemann surfaces of all genera at once,
we provide here a reformulation of the ODE for $\langle\Phi\Phi\rangle^{g=1}$ in the description $\Sigma_1: y^2=p(x)$,
where
\begin{displaymath}
p(x)
\;=\;4\,\bigl(x^3-30\,G_4x-70\,G_6\bigl) 
\:.
\end{displaymath}

\begin{altcor}
$\Upsilon(x)$ defined by \hbox{$y^{-1/5}\Upsilon(x):=\bigl\langle\Phi(z)\,\Phi(0)\bigr\rangle$}
satisfies the ODE
\begin{equation}\label{3rd order ODE for 2-pt function of varphi in the algebraic coordinate x}
\bigl(p(x)\,\frac{d^3}{dx^3}
+f(x)\,\frac{d^2}{dx^2}
+g(x)\,\frac{d}{dx}
+h(x)\,\bigr)\,
\Upsilon(x)
\;=\;0\:, 
\end{equation}
where 
\begin{equation*}
\begin{split}
f
\;=&\;\frac{6}{5}\,p'\\
g
\;=&\;\frac{3}{100}\,\frac{[p']^2}{p}+\frac{9}{50}\,p''\\
h
\;=&\;-\frac{33}{500}\,\frac{[p']^3}{p^2}+\frac{33}{250}\,\frac{p'p''}{p}-\frac{288}{125}\:.
\end{split}
\end{equation*} 
\end{altcor}

In particular, the ODE has simple poles at the four ramification points.

\begin{proof} 
Set 
\begin{equation*}
L
\;=\;\frac{d^3}{dz^3}-\frac{12}{25}\,\wp(z)\,\frac{d}{dz}+\frac{12}{125}\,\wp'(z)
\:. 
\end{equation*}
Since \hbox{$d/dz=y\,d/dx$} and 
\begin{equation*}
\frac{d^3}{dz^3}
\;=\;y\,\Bigl(p(x)\frac{d^3}{dx^3}+\frac{3}{2}\,p'(x)\,\frac{d^2}{dx^2}+\frac{1}{2}\,p''(x)\,\frac{d}{dx}\Bigr)
\:,
\end{equation*}
we have in algebraic coordinates
\begin{align*}
L
\;=\;&y\,\Bigl(p(x)\,\frac{d^3}{dx^3}+\frac{3}{2}\,p'(x)\,\frac{d^2}{dx^2}+\frac{12}{25}\,p''(x)\,\frac{d}{dx}+\frac{12}{125}\Bigr)
\:.
\end{align*}
Thus
\begin{equation*}
L\Bigl(y^{-1/5}\,\Upsilon(x)\Bigr)
\;=\;y^{4/5}\biggl(
p(x)\,\frac{d^3}{dx^3}
+f(x)\,\frac{d^2}{dx^2}
+g(x)\,\frac{d}{dx}
+h(x)\,\biggr)\,
\Upsilon(x)\:,
\end{equation*}
where
\begin{align*}
f
\;=&\;y^{1/5}\,\Bigl(3p\,\frac{d}{dx}+\frac{3}{2}\,p'\Bigr)\,y^{-1/5}\:,\\
g
\;=&\;y^{1/5}\,\Bigl(3p\,\frac{d^2}{dx^2}+3\,p'\,\frac{d}{dx}+\frac{12}{25}\,p''\Bigr)\,y^{-1/5}\:,\\
h
\;=&\;y^{-4/5}\,L\,y^{-1/5}
\:.
\end{align*}
Computation of $d^ky^{-1/5}/dx^k$ for $k=1,2,3$ leads to the expressions for the functions $f,g,h$.
We conclude that the ODE~(\ref{3rd order ODE for 2-pt function of varphi in the algebraic coordinate x}) 
is equivalent to that of eq.~(\ref{3rd order ODE for 2-pt function of varphi in the analytic coordinate z}).
\end{proof}


By virtue of the Frobenius Ansatz \hbox{$\langle\Phi(z)\,\Phi(0)\rangle\sim z^{\alpha}$} times a function of $\tau$, 
the differential equation (\ref{3rd order ODE for 2-pt function of varphi in the analytic coordinate z}) imposes the condition
\begin{displaymath}
\frac{25}{12}\,\alpha(\alpha-1)(\alpha-2)
\;=\;\frac{2}{5}+\alpha
\:.
\end{displaymath}
on~$\alpha$, which produces the values~\hbox{$1/5,2/5$} and~$12/5$. 
On the other hand, 
using what is called power counting, 
and eqs~ (\ref{eq: 2-point function of Phi}) and (\ref{eq: 3-point function of Phi}) yield the OPE
\begin{equation*}
\Phi(z)\Phi(0)
\;=\;\eps_{\Phi}\,|z|^{4/5}\,1
+\eps_{\Phi}\,\lambda_{\Phi\Phi\Phi}\,|z|^{2/5}\,\Phi(0)+\ldots
\end{equation*}
The obvious solutions to the ODE are, to leading order,  
\begin{displaymath}
z^{2/5}\,\langle 1\rangle
\:,
\quad
z^{1/5}\,\langle\mathfrak{w}\rangle
\:,
\quad
z^{12/5}\,\langle T\rangle\:.
\end{displaymath}
The first two terms obviously correspond to the leading terms in the OPE,
but the third terms needs a better understanding.

The author is grateful to W.~Nahm for introducing her to the subject and for commenting on the manuscript.
Explicit computations in 
Propos.~\ref{proposition: almost global coordinates on CP1 satisfying the epsilon condition},
\ref{proposition: Taylor expansion in epsilon of the genus two 0-pt function obtained from the Rogers-Ramanujan functions and the one corresponding to the primary field Phi} and \ref{propos: the first interesting term in the rho expansion for h=6}
were performed using Mathematica.
The author found \cite{Headrick:2010} helpful for the calculations in Propos.~\ref{proposition: quasi-primary fields and their squared norms in the vacuum rep}
and \ref{proposition: quasi-primary fields and their squared norms in the W rep},
she thanks A.~Honecker for mentioning the resource to her.
This research was supported in part by the National Science Foundation under Grant No. NSF PHY-1748958.
The author acknowledges receipt of Higher Education Authority Support for a Covid-19 related Research Costed Extension of her IRC Government of Ireland Postdoctoral Award 2018/583 through Trinity College Dublin.

\appendix

\section{Proof of Proposition \ref{proposition: graphical representation of N-point functions of T}}\label{appendix section: Proof of {proposition: graphical representation of N-point functions of T}}

We use induction on~$N$. 
For $N\geq 0$, 
let \hbox{$\Gamma_0^N\in S^{[1]}_N$} be the graph whose vertices are all isolated.

For \hbox{$N=0$}, \hbox{$\langle 1\rangle={C}_{0,c}$} (corresponding to the empty graph).
For \hbox{$N=1$}, \hbox{$\Gamma_0^1(z)$} is the only graph, and 
\begin{displaymath}
\langle 1\rangle^{-1}\langle T\rangle
\;=\;\langle 1\rangle^{-1}\langle T(z)\rangle
\;=\;{\gamma}(\Gamma_0^1(z))
\;=\;{C}_{2,c}
\:
\end{displaymath}
is a modular form of weight~$2$.
For \hbox{$N=2$}, the admissible graphs form a closed loop,
a single line segment (with two possible orientations), and two isolated points.
Thus by eq.~(\ref{graph rep}),
\begin{displaymath}
{C}_{4,c}
\;=\;\langle 1\rangle^{-1}\bigl\langle T(z_1)T(z_2)\bigr\rangle
-\frac{c}{2}\,\wp_{12}^2\,{C}_{0,c}-2\,\wp_{12}{C}_{2,c}
\:.
\end{displaymath}
According to the OPE (\ref{OPE of T and T}),
${C}_{4,c}$ is regular, thus constant.
Suppose~${C}_{2k,c}$, for \hbox{$k\leq N$} has the required properties for \hbox{$k<N$}.
Let
\begin{displaymath}
E_N
\;:=\;\{1\leq j\leq N\:|\exists\:i\:\text{such that}\:(z_i,z_j)\in\Gamma\}\:, 
\end{displaymath}
and let~$E_N^c$ denote its complement in \hbox{$\{1,\ldots N\}$}.
We define 
\begin{displaymath}
\bigl\langle T(z_1)\ldots T(z_N)\bigr\rangle_r
\end{displaymath}
by
\begin{displaymath}
{\gamma}(\Gamma)
\;=\;\biggl(\frac{c}{2}\biggr)^{\sharp\text{loops}}
\langle 1\rangle^{-1}\Bigl\langle\bigotimes_{k\in E_N^c}T(z_k)\Bigr\rangle_r\,
\prod_{(z_i,z_j)\in\Gamma}\wp_{ij}\:
\:,
\end{displaymath}
and show first that 
$\bigl\langle T(z_1)\ldots T(z_N)\bigr\rangle_r$ is regular on any partial diagonal, thus constant.
In other words,
\hbox{$\sum_{\Gamma\not=\Gamma_0^N}{\gamma}(\Gamma)$} reproduces the correct singular part of the Virasoro~$N$-point function
as prescribed by the OPE (\ref{OPE of T and T}).
By symmetry, it suffices to verify this for \hbox{$\bigl\langle T(z_1)\ldots T(z_N)\bigr\rangle$} as a function of~$z_1$.

Since the~$z_i$ dependence is trivial, we write 
\hbox{$\bigl\langle T(z_1)\ldots T(z_n)\bigr\rangle_r=\langle 1\rangle {C}_{2n,c}$}.

Suppose the graphical representation for the~$k$-point function of the Virasoro field is correct for \hbox{$2\leq k\leq N-1$}.
For \hbox{$1\leq i\leq N$}, set 
\hbox{$S^{[i]}:=S(z_i,\ldots,z_N)$}.
For \hbox{$1\leq i,j\leq N$}, \hbox{$i\not=j$}, define
\begin{equation*}
\begin{split}
S_{(ij)}
\;:=&\;\{\Gamma\in S^{[1]}|\:(z_i,z_j),(z_j,z_i)\in\Gamma\}\:,\\
S_{(i,j)}
\;:=&\;\{\Gamma\in S^{[1]}|\:(z_i,z_j)\in\Gamma,(z_j,z_i)\notin\Gamma\}\:,\\
S_{(i)(j)}
\;:=&\;\{\Gamma\in S^{[1]}|\:(z_i,z_j),(z_j,z_i)\notin\Gamma\}\:.
\end{split}
\end{equation*}
$S^{[1]}$ decomposes as
\begin{displaymath}
S^{[1]}
\;=\;S_{(12)}\cup S_{(1,2)}\cup S_{(2,1)}\cup S_{(1)(2)}. 
\end{displaymath}
Since \hbox{$S_{(12)}\cong S^{[3]}$}, 
we have
\begin{displaymath}
\sum_{\Gamma\in S_{(12)}}{\gamma}(\Gamma) 
\;=\;\frac{c}{2}\,\wp_{12}^2\,\sum_{\Gamma\in\tilde{S}^{[3]}}{\gamma}(\Gamma)
\;=\;\frac{c}{2}\,\wp_{12}^2\,\langle 1\rangle^{-1}\bigl\langle T(z_3)\ldots T(z_N)\bigr\rangle
\end{displaymath}
by the induction hypothesis. 
Moreover,
\begin{displaymath}
\sum_{\Gamma\in S^{[1]}\setminus S_{(12)}}{\gamma}(\Gamma)
\;=\;\:2\,\wp_{12}\,\langle 1\rangle^{-1}\bigl\langle T(z_2)T(z_3)\ldots T(z_N)\bigr\rangle
+O\bigl((z_1-z_2)^{-1}\bigr)\:,
\end{displaymath}
since by induction hypothesis on~$S^{[2]}$, for \hbox{$\Gamma\in S^{[2]}$},
\begin{displaymath}
{\gamma}(\varphi_{12}^{-1}(\Gamma))+{\gamma}(\varphi_{21}^{-1}(\Gamma)) 
\;=\;2\,\wp_{12}\,{\gamma}(\Gamma)+O\bigl((z_1-z_2)^{-1}\bigr)\:.
\end{displaymath}
Here~$\varphi_{12}$ and~$\varphi_{21}$ are the isomorphisms
\begin{equation*}
\begin{split}
\varphi_{12}:\quad S_{(1,2)}&\rechts S^{[2]},\\
\varphi_{21}:\quad S_{(2,1)}&\rechts S^{[2]} 
\end{split}
\end{equation*}
given by contracting the link \hbox{$(z_1,z_2)$} resp.~\hbox{$(z_2,z_1)$} into the point~$z_2$ and leaving the graph unchanged otherwise:
Let \hbox{$\Gamma\in S_{(1,2)}$}. 
For \hbox{$j\not=2$},
we have  \hbox{$(z_i,z_j)\in\Gamma$} for \hbox{$j\not=1$} iff \hbox{$(z_i,z_j)\in\varphi_{12}(\Gamma)$},
and we have \hbox{$(z_i,z_1)\in\Gamma$} iff \hbox{$(z_i,z_2)\in\varphi_{12}(\Gamma)$}.
The case \hbox{$\Gamma\in S_{(2,1)}$} works analogously by changing the orientation.
One checks easily that either map defines an isomorphism.

We address the modularity property of~${C}_{2n,c}$.
From the transformation behaviour of~$T$ under conformal coordinate transformations \hbox{$z\mapsto\lambda z$} (with \hbox{$\lambda>0$})
follows that
\begin{displaymath}
\bigl\langle T(z_1)\ldots T(z_N)\bigr\rangle_{\Lambda} 
\;=\;\lambda^{2N}\bigl\langle T(\lambda z_1)\ldots T(\lambda z_N)\bigr\rangle_{\lambda\Lambda} 
\:.
\end{displaymath}
On the other hand, we have in eq.~(\ref{graph rep}), for every edge of a graph, 
\hbox{$\wp(z|\tau)=\lambda^2\,\wp(\lambda z|\lambda\tau)$}.
It follows that for \hbox{$1\leq n\leq N$, ${C}_{2n,c}$} transforms as a modular form of weight \hbox{$2n$},
\begin{displaymath}
{C}_{2n,c}(\Lambda)
\;=\;\lambda^{2n}{C}_{2n,c}(\lambda\Lambda)
\:.
\end{displaymath}
The~zero-point function is differentiable and generates the~$N$-point functions of~$T$ for \hbox{$N>0$}.
Thus the latter are also non-singular for finite~$\tau$, except possibly at the cusps. 
Since~$\mathfrak{Z}^{g=1}$ is modular on the full modular group,
it suffices to verify regularity at the cusp at infinity.
Let~$E_0$ be the lowest energy eigenvalue of~$L_0$ in the specific sector.
Let \hbox{$y=\IM \tau$}. 
For \hbox{$y\rechts\infty$}, 
we have
\begin{displaymath}
\frac{\bigl\langle T(z_1)\ldots T(z_n)\bigr\rangle_r}{\langle 1\rangle}
\sim\frac{O(e^{-2\pi E_0y})}{e^{-2\pi E_0y}}
\;=\;O(1)
\:.
\end{displaymath}
This completes the proof.

\section{Proof of Proposition \ref{proposition: graphical representation of (N+1)-point functions of N copies of T and one copy of Phi}}\label{appendix section: Proof of {proposition: graphical representation of (N+1)-point functions of N copies of T and one copy of Phi}}

We show that for \hbox{$\Gamma\in\tilde{S}(0,z_1,\ldots,z_N)$} and some polynomial $\varrho_{\lambda}(h)$,
\begin{equation*}
\widetilde{{\gamma}}(\Gamma)
\;:=\;\biggl(\frac{c}{2}\biggl)^{\sharp\text{loops}}\,
\;\tC_{2\cdot(N-\sharp\text{edges}),c}\hspace{0.2cm}
\varrho_{\lambda}(h)\,
\prod_{(z_i,z_j)\in\Gamma}\wp_{ij}\:
\:,
\end{equation*}
and we specify $\varrho_{\lambda}$.

For $N\geq 0$, 
let \hbox{$\Gamma_0^N\in\tilde{S}^{[1]}_N$} be the graph whose vertices are all isolated. 

For \hbox{$N=0$}, the only graph is~$\Gamma_0^0$, and we have \hbox{$\widetilde{{\gamma}}(\Gamma_0^0)=\tC_{0,c}=1$}.

For \hbox{$N=1$}, by lack of a modular form of weight~$2$, the only graph that contributes has one edge,
\begin{displaymath}
\langle\Phi\rangle^{-1}\,\bigl\langle T(z)\,\Phi(0)\bigr\rangle
\;=\;h\,\wp(z)\:\tC_{0,c}\:. 
\end{displaymath}
This is consistent with eq.~(\ref{eq: 2-pt fct of T with Phi}).

For \hbox{$1\leq i\leq N$}, set \hbox{$\tilde{S}^{[i]}:=\tilde{S}(z_0=0,z_i,\ldots,z_N)$}.
We define~$\tilde{S}_{(ij)}$,~$\tilde{S}_{(i,j)}$ and~$\tilde{S}_{(i)(j)}$ for \hbox{$1\leq i,j\leq N$} 
in the same way as we defined~$S_{(ij)}$,~$S_{(i,j)}$ and~$S_{(i)(j)}$ (proof of Proposition \ref{proposition: graphical representation of N-point functions of T})
but with~$\tilde{S}^{[1]}$ in  place of~$S^{[1]}$.
Moreover, for \hbox{$1\leq i\leq N$} we set 
\begin{displaymath}
\tilde{S}_{(i,0)}
\;:=\;\{\Gamma\in \tilde{S}^{[1]}|\:(z_i,0)\in\Gamma\}
\:.
\end{displaymath}
We show that
\begin{displaymath}
\langle\Phi\rangle^{-1}\langle T(z_1)\ldots T(z_N)\,\Phi(0)\rangle
-\sum_{\Gamma\in\tilde{S}^{[1]}\setminus\{\Gamma_0^N\}}\widetilde{{\gamma}}(\Gamma)
\;=\;\widetilde{{\gamma}}(\Gamma_0^N)
\:.
\end{displaymath}
The l.h.s.\ is well-defined: 
Every \hbox{$\Gamma\in\tilde{S}^{[1]}\setminus\{\Gamma_0^N\}$} has an edge \hbox{$(z_i,z_j)\in\Gamma$} 
with \hbox{$1\leq i\leq N$} and \hbox{$0\leq j\leq N$}.
So \hbox{$\widetilde{{\gamma}}(\Gamma)$} is proportional to~$\wp_{ij}$, 
according to eq.~(\ref{transcription of graphs representing the (N+1) pt fct of T with one Phi}),
and the induction hypothesis on~$\tilde{S}^{[2]}$ applies.  

The arguments employed previously in the proof of Proposition \ref{proposition: graphical representation of N-point functions of T}
show that
\begin{displaymath}
\sum_{\Gamma\in\tilde{S}_{(12)}}\widetilde{{\gamma}}(\Gamma)
\;=\;\frac{c}{2}\,\wp_{12}^2\bigl\langle T(z_3)\ldots T(z_N)\,\Phi(0)\bigr\rangle\:\langle\Phi\rangle^{-1}
\end{displaymath}
(recall our convention $\wp_{12}=\wp(z_1-z_2)$)
and 
\begin{displaymath}
\sum_{\Gamma\in\tilde{S}^{[1]}\setminus\tilde{S}_{(12)}}\widetilde{{\gamma}}(\Gamma)
\;=\;2\,\wp_{12}\,\bigl\langle T(z_2)\ldots T(z_N)\,\Phi(0)\bigr\rangle\:\langle\Phi\rangle^{-1}+O((z_1-z_2)^{-1})
\:.
\end{displaymath}
(Recall that $z_0=0$, $\wp_1=\wp_{10}$.)
Now we address graphs in~$\tilde{S}_{(1,0)}$.
By the OPE (\ref{OPE of T and Phi}) and the induction hypothesis on~$\tilde{S}^{[2]}$,
we have
\begin{displaymath}
\bigl\langle T(z_1)\ldots T(z_N)\,\Phi(0)\bigr\rangle\:\langle\Phi\rangle^{-1}
\;=\;h\,\wp_1\sum_{\Gamma'\in\tilde{S}^{[2]}}\widetilde{{\gamma}}(\Gamma')
+O(z_1^{-1})
\:.
\end{displaymath}
We show that 
\begin{displaymath}
h\,\wp_1\sum_{\Gamma'\in\tilde{S}^{[2]}}\widetilde{{\gamma}}(\Gamma')
-\sum_{\Gamma\in\tilde{S}_{(1,0)}}\widetilde{{\gamma}}(\Gamma)
\;=\;O(z_1^{-1})\:.
\end{displaymath}
Let 
\begin{displaymath}
\kappa:\quad\tilde{S}_{(1,0)}\rechts\tilde{S}^{[2]}
\end{displaymath}
be the map that contracts the edge \hbox{$(z_1,0)$} into~$0$.
We have the decomposition
\begin{displaymath}
\sum_{\Gamma\in\tilde{S}_{(1,0)}}\widetilde{{\gamma}}(\Gamma)
\;=\;\sum_{\Gamma'\in\tilde{S}^{[2]}}\underset{\kappa(\Gamma)=\Gamma'}{\sum_{\Gamma\in\tilde{S}_{(1,0)}}}\widetilde{{\gamma}}(\Gamma)
\:.
\end{displaymath}
Thus it remains to show that for every \hbox{$\Gamma'\in\tilde{S}^{[2]}$}, 
we have
\begin{equation}\label{eq: what remains to show, in terms of graphs}
h\,\wp_1\widetilde{{\gamma}}(\Gamma')
-\underset{\kappa(\Gamma)=\Gamma'}{\sum_{\Gamma\in\tilde{S}_{(1,0)}}}\widetilde{{\gamma}}(\Gamma)
\;=\;O(z_1^{-1})\:.
\end{equation}
Let \hbox{$\Gamma'\in\tilde{S}^{[2]}$} be such that \hbox{$(z_k,0)\in\Gamma'$} for~$\lambda$ values of~$k$, 
where \hbox{$2\leq k\leq N$}.
For each such value of~$k$ we obtain one graph \hbox{$\Gamma\in\tilde{S}_{(1,0)}$} with \hbox{$\kappa(\Gamma)=\Gamma'$}
by replacing, according to eq.~(\ref{transcription of graphs representing the (N+1) pt fct of T with one Phi}),
\begin{displaymath}
\wp_k
\mapsto
\wp_{k1}\,\wp_1\:.
\end{displaymath}
Different values of~$k$ lead to different graphs in~$\tilde{S}_{(1,0)}$.
One more graph \hbox{$\Gamma\in\tilde{S}_{(1,0)}$} is obtained by replacing
 \begin{displaymath}
\prod_{(z_i,z_j)\in\Gamma'}\wp_{ij}
\mapsto
\wp_1\prod_{(z_i,z_j)\in\Gamma'}\wp_{ij}
\:.
\end{displaymath}
This proves eq.~(\ref{eq: what remains to show, in terms of graphs}) 
provided that for the polynomial $\varrho_{\lambda}(h)$,
\begin{displaymath}
h\varrho_{\lambda}-(\lambda\varrho_{\lambda}+\varrho_{\lambda+1})
\;=\;0
\:
\end{displaymath}
with $\varrho_0(h)=1$.

An argument similar to that seen in the proof of Proposition \ref{proposition: graphical representation of N-point functions of T}
also shows that~$\tC_{2n,c}$ is a modular form of weight~\hbox{$2n$}.

\bibliographystyle{spbasic}      

\begin{thebibliography}{99}
\bibitem{AS:1965}
Abramowitz, M., and Stegun, I.: \textsl{Handbook of Mathematical Functions}, 
Dover Publications Inc., New York (1965); 
\bibitem{C:1985}
Cardy, J.L.: \textsl{Conformal Invariance and the Yang-Lee Edge Singularity in Two Dimensions},
Phys.\ Rev.\ Lett.\ \textbf{54}.13, 1354--1356 (1985);
\bibitem{Dotsenko:1988}
Dotsenko, Vl..S.: \textsl{Lectures on Conformal Field Theory}, Advanced Studies in Pure Mathematics 16 (1988),
Conformal Field Theory and Solvable Lattice Models, pp. 123--170;
\bibitem{E:1847}
Eisenstein, G: \textsl{Beitr\"age zur Theorie der elliptischen Funktionen}, 
Crelle, \textbf{35} (1847), part VI, 153--274,
reprinted in \textsl{Mathematische Abhandlungen besonders aus dem Gebiete der h\"oheren Arithmetik und der elliptischen Funktionen. Mit einer Vorrede von C.F. Gauss}, Georg Olms Verlagsbuchhandlung (1967), 213--334;
\bibitem{EO:1987}
Eguchi, T.\ and Ooguri, H.: \textsl{Conformal and current algebras on a general Riemann surface}, Nucl.\ Phys.~\textbf{B}282, 308--328 (1987);
\bibitem{Friedan-Shenker:1986}
Friedan, D.\ and Shenker, S.: \textsl{The analytic geometry of two-dimensional conformal field theory}, Nucl.\ Phys.~\textbf{B}281, 
(1987), 509--545;
\bibitem{Headrick:2010}
Headrick, M.: \textsl{Virasoro.nb} (Mathematica notebook), available at \href{http://people.brandeis.edu/~headrick/Mathematica/}{http://people.brandeis.edu/~headrick/Mathematica/};
\bibitem{H:1915}
Hilbert, D.: \textsl{Die Grundlagen der Physik, Erste Mitteilung, vorgelegt in der Sitzung vom 20.~November~1915},
Nachrichten von der Koeniglichen Gesellschaft der Wissenschaften zu Goettingen, Math-physik. Klasse, 1915, 395--407 (1915);
\bibitem{Humphries:1991}
Humphries, S.: \textsl{Normal closures of powers of Dehn twists in mapping class groups}, Glasgow Mathematical Journal, 34(3), (1992), 314--317;
\bibitem{L:2013}
Leitner, M.: \textsl{Virasoro correlation functions on hyperelliptic Riemann surfaces}, 
Lett.\ Math.\ Phys.\ 103.7, 701--728 (2013); 
\bibitem{L:PhD14}
Leitner, M.: \textsl{CFTs on Riemann Surfaces of genus $g\geq 1$}, 
PhD thesis, TCD (2014);
\bibitem{LN:2017}
Leitner, M., Nahm, W.: \textsl{Rational CFTs on Riemann surfaces}, in preparation, preprint arXiv:1705.07627;
\bibitem{L:2020}
Leitner, M.: \textsl{Convolutions on the complex torus}, Int. J. Number Theory doi.org/10.1142/S1793042121500391;
\bibitem{M-T:2006}
Mason, G., Tuite, M.: \textsl{On genus two Riemann surfaces formed from sewn tori}, Commun.\ Math.\ Phys.\ \textbf{270},  587--634 (2007);
\bibitem{Moore-Seiber:1989}
Moore, G.W., Seiberg, N.: \textsl{Classical and quantum conformal field theory}, Commun.\ Math.\ Phys. 123 (1989), 177;
\bibitem{Nahm:2020}
Nahm, W.: \textsl{Automorphic forms for $g\geq 1$}, Mathematisches Forschungsinstitut Oberwolfach, Report 38:2388, (2017), Oberwolfach reports; \textsl{Automorphic forms and quantum field theory}, talk given at the conference on \textsl{Integrable Systems and Automorphic Forms}, February 2020, Sirius Mathematics Center, Sochi;
\bibitem{P:1981}
Polyakov, A.M.: \textsl{Quantum Geometry of bosonic strings}; Phys.\ Lett.\ \textbf{B}103, 207--210 (1981);
\bibitem{Rosenhain:1851}
Rosenhain, G.: \textsl{Sur les fonctions de deux variables \`a quatre p\'eriodes, qui sont les inverses des int\'egrales ultra-elliptiques de la premi\`ere classe}, in M\'emoires pr\'esent\'es par divers savants, 2nd ser., 11 (1851), 361--468; also translated into German as \textsl{Abhandlung \"uber die Functionen zweier Variabler mit vier Perioden}, Ostwalds Klassiker der Exakten Wissenschaften, no. 65 (Leipzig, 1895), Acad\'emie des sciences (France). 
The citations in the text refer to the German translation;  
\bibitem{Segal:1988}
Segal, G.B.: \textsl{The Definition of Conformal Field Theory}, in: Bleuler K., Werner M.\ (eds) Differential Geometrical Methods in Theoretical Physics. NATO ASI Series (Series C: Mathematical and Physical Sciences), vol 250. Springer (1988), 165--171;
\bibitem{Sh:1972}
Shapovalov, N.N.: \textsl{On a bilinear form on the universal enveloping algebra of a complex semisimple Lie algebra},
Funktsional. Anal.\ i Prilozhen., 6:4 (1972), 65--70; Funct.\ Anal.\ Appl., 6:4 (1972), 307--312;
\bibitem{Sonoda:1988}
Sonoda, H.: \textsl{Sewing Conformal Field Theories I}, Nucl.Phys. \textbf{B}311, 401--416 (1988),
\textsl{Sewing Conformal Field Theories II}, ibid., 417--432;
\bibitem{T:2015}
Tuite, M., private communication (2015);
\bibitem{Vafa:1987}
Vafa, C.: \textsl{Conformal Theories and Punctured Surfaces}, Phys.\ Lett.\ \textbf{B}199, 195--202 (1987);
\bibitem{W:1976}
Weil, A.: \textsl{Elliptic functions according to Eisenstein and Kronecker}, Springer Verlag, New-York (1976); 
\bibitem{Wein:1972}
Weinberg, S.: \textsl{Gravitation and Cosmology},
John Wiley $\&$ Sons, New York-Chichester-Brisbane-Toronto-Singapore (1972);
\bibitem{Z-K:1997}
Zagier, D.\ and Kaneko, M.: \textsl{Supersingular j-invariants, hypergeometric series, and Atkin's orthogonal polynomials}
in Proceedings of the Conference on Computational Aspects of Number Theory, AMS/IP Studies in Advanced Math.\ 7, International Press, 
Cambridge (1997), 97--126;
\bibitem{Z:1-2-3}
Zagier, D.: \textsl{Elliptic modular forms and their applications}, 
in The 1-2-3 of Modular Forms: Lectures at a Summer School in Nordfjordeid, Norway, Universitext, 
Springer-Verlag, Berlin-Heidelberg-New York (2008), pp.\ 1--103.
\end{thebibliography}


\end{document}